\documentclass[npg,final]{copernicus}
\nolinenumbers

\usepackage{lscape}
\usepackage{latexsym}
\usepackage{amsmath}
\usepackage{amstext,amsfonts,epsf,epsfig,pifont}
\usepackage{graphics,graphicx,bm,amsmath}

\newlength{\intwidth}

\def\XXint#1#2#3{{\setbox0=\hbox{$#1{#2#3}{\int}$}
\vcenter{\hbox{$#2#3$}}\kern-.5\wd0}}

\usepackage{amssymb}

\usepackage{latexsym}
\usepackage{amstext,amsfonts,epsf,epsfig,pifont}
\usepackage{graphics,graphicx,bm,amsmath,multicol,rotating}

\begin{document}

\title{\vspace{-1.5in}\textmd{\normalsize{Lilly, Sykulski, Early, and Olhede (2017).  Fractional Brownian motion, the Mat\'ern process, and stochastic modeling of \\\vspace{-.125in} turbulent dispersion. \emph{Nonlin. Processes Geophys.} {\bf 24}, 481--514. \url{https://doi.org/10.5194/npg-24-481-2017}}}\\\vspace{-.15in}\rule{\textwidth}{1pt}\\
Fractional Brownian motion, the Mat\'ern process, and stochastic modeling of turbulent dispersion}

\Author[1]{Jonathan~M.}{Lilly}
\Author[2]{Adam M.}{Sykulski}
\Author[1]{Jeffrey J.}{Early}
\Author[3]{Sofia C.}{Olhede}

\affil[1]{NorthWest Research Associates, PO Box 3027, Bellevue, WA, USA }
\affil[2]{Data Science Institute / Department of Mathematics and Statistics, Lancaster University, Lancaster, UK}
\affil[2]{Department of Statistical Science, University College London, Gower Street, London, UK}

\runningtitle{The Mat\'ern process}

\runningauthor{Lilly, Sykulski, Early, and Olhede}

\correspondence{Jonathan Lilly (lilly@nwra.com)}

\received{}
\pubdiscuss{} 
\revised{}
\accepted{}
\published{}


\firstpage{1}

\maketitle

\begin{abstract}
Stochastic process exhibiting power-law slopes in the frequency domain are frequently well modeled by fractional Brownian motion (fBm), with the spectral slope at high frequencies being associated with the degree of small-scale roughness or fractal dimension.  However, a broad class of real-world signals have a high-frequency slope, like fBm, but a plateau in the vicinity of zero frequency.  This low-frequency plateau, it is shown, implies that the temporal integral of the process exhibits \emph{diffusive} behavior, dispersing from its initial location at a constant rate.  Such processes are not well modeled by fBm, which has a singularity at zero frequency corresponding to an unbounded rate of dispersion.  A more appropriate stochastic model is a much lesser-known random process called the Mat\'ern process, which is shown herein to be a \emph{damped} version of fractional Brownian motion.  This article first provides a thorough introduction to fractional Brownian motion, then examines the details of the Mat\'ern process and its relationship to fBm.  An algorithm for the simulation of the Mat\'ern process in $O(N \log N)$ operations is given.  Unlike fBm, the Mat\'ern process is found to provide an excellent match to modeling velocities from particle trajectories in an application to two-dimensional fluid turbulence.  
\end{abstract}

\tableofcontents
\introduction

Fractional Brownian motion (fBm), introduced by \citet{mandelbrot68-siam}, is a canonical stochastic process finding wide-ranging applications in fields as diverse as oceanography \citep{osborne89-tellus,sanderson90-tellus,sanderson91-tellus,summers02-npg}, geophysics \citep{molz97-wrr}, finance \citep{rogers97-mf}, and many others.  The essential features of this process are its \emph{self-similar} behavior---meaning that magnified and rescaled versions of the process appear statistically identical to the original---together with its \emph{nonstationarity}, implying a never-ending growth of variance with time.  Two other properties of fBm are its degree of small-scale roughness or \emph{fractal dimension} \citep[][Chapters 2 \& 3]{mandelbrot85-ps,falconer}, and the nature of its long-term \emph{memory} or \emph{long-range dependence} \citep{beran92-ss,beran}.  As pointed out by \citet{gneiting04-siam}, the self-similarity of fractional Brownian motion links the very small and the very large temporal scales behavior together, such that its memory, fractal dimension, and self-similarity aspect ratio are all controlled by the same parameter.  These, in turn, are all connected to the slope of the spectrum in the Fourier domain, in which fBm is found to exhibit a simple power-law behavior.

One important property that cannot be captured by fractional Brownian motion is the tendency for a process to \emph{diffuse}, or disperse from an initial location at a uniform rate.  In the fluid dynamics literature \citep[e.g.][]{davis83-jmr,lacasce08-pio}, it is known that the zero-frequency value of the spectrum of a process quantifies the dispersive tendency of the temporal integral of that process.  This recognition leads to a classification of processes, proposed here, based on their spectral value at zero frequency.  We refer to random processes as \emph{diffusive}, \emph{subdiffusive}, or \emph{superdiffusive}, depending on whether the spectral value is finite and nonzero, zero, or unbounded, respectively.  This quality of ``diffusiveness'' will be shown to be related to, but distinct from, the more familiar classification of processes as short-memory or long-memory depending on the long-time behaviors of their autocovariance functions \citep{beran92-ss,beran,gneiting04-siam}.  Fractional Brownian motion is found to be superdiffusive, and is associated with a diffusivity that tends to increase without bound.  

A particular application is the stochastic modeling of velocities obtained from particle trajectories in fluid flows.  In the field of oceanography, one of the main windows into studying the physics of the ocean circulation consists of position data from instruments that drift freely with the currents \citep{rupolo96-jpo,rossby07-lapcod,lumpkin07-lapcod}.  Similarly, numerical models of fluid systems are frequently analyzed by examining the motion of particles carried with the flow \citep{pasquero02-prl,veneziani05a-jmr,lilly11-grl}.  Such position records are known as \emph{Lagrangian trajectories}, on account of the moving frame of reference associated with the particles or instruments.   

One thread of research attempts to predict Lagrangian statistics based on dynamical assumptions \citep[e.g.][]{griffa96-smpo,majda99-pr,berloff02b-jpo,veneziani05a-jmr,majda13-ptrsla}.  Here, we instead try to identify the simplest stochastic model that can explain the major observed features, leaving the connection to the equations of motion to the future.   Velocities from Lagrangian trajectories are found \citep[e.g.][]{rupolo96-jpo} to exhibit power-law behaviors at high frequencies, and indeed fractional Brownian motion has been suggested as a stochastic model \citep{osborne89-tellus,sanderson90-tellus,sanderson91-tellus,summers02-npg}.  Yet a primary characteristic of these trajectories is their tendency to diffuse at a uniform rate at long times \citep{taylor21-plms,davis83-jmr,lacasce08-pio,koszalka10-od}, a feature that fBm cannot capture. 

A type of random process having a sloped spectrum that matches fBm at high frequencies, but that takes on a constant value in the vicinity of zero frequency, exists and is known as the \emph{Mat\'ern} process \citep{matern60-mss,guttorp06-biometrika}.  The same process has been referred to occasionally as the \emph{fractional Ornstein-Uhlenbeck} process \citep{wolpert05-sp,lim06-pla}, because it also generalizes the well-known Ornstein-Uhlenbeck process \citep{uhlenbeck30-pr} to fractional orders.  A multivariate version of the Mat\'ern process is broadly used for spatial statistics in various fields \citep{goff88-jgr,handcock93-technometrics,gneiting10-jasa,lindgren11-jrssb,schlather12-stmne}.  Yet despite the appeal of its generality, the Mat\'ern process appears in only a handful of papers in the time series literature \citep{wolpert05-sp,lim06-pla,li10-wtc,hartikainen10-mlsp,sykulski16-jrssc,sykulski17-itsp}.  In fluid dynamics, the only instances we are aware of is an application to wind tunnel data by \citet{vonkarman48-pnas}, pointed out by \citet{guttorp06-biometrika}, together with a more recent study by \citet{hedevang14-ijnsns}.
 
The purpose of this paper is to investigate the theoretical properties of the Mat\'ern process, in particular its relationship to fractional Brownian motion, and to establish the practical importance of this under-appreciated process for modeling time series that exhibit the fundamental phenomenon of diffusion.  On the theoretical side, the Mat\'ern process is seen to be a \emph{damped} version of fractional Brownian motion, in the same way that the Ornstein-Uhlenbeck process is a damped version of standard Brownian motion.  A simple generalization of the Mat\'ern process that incorporates a uniform rotation rate is shown to describe a \emph{forced/damped fractional oscillator}. By ``damped version'', we mean that the process is modified as would be expected if a physical damping were introduced into its stochastic differential or stochastic integral equation. This terminology, which draws upon intuition for damped and undamped oscillators from elementary physics, will be made more clear in Section~\ref{sie}.

On the practical side, we find the Mat\'ern process to be an excellent match for Lagrangian velocity spectra from a numerical simulation of two-dimensional turbulence, a classical system in fluid dynamics that has been the subject of a large number of studies,  \citep[e.g.][]{lin72-jas,mcwilliams90a-jfm,dritschel08-prl,bracco10-jfm,kadoch11-pre,scott13-jfm}.  The Mat\'ern process allows one to simultaneously vary the values of the three most important properties of Lagrangian trajectories: the kinetic energy, the degree of small-scale roughness or fractal dimension, and the long-time diffusive behavior.  Thus, it is arguably the simplest stochastic model that can capture the essential features of such data. 

A transition of the spectrum to constant values at sufficiently low frequencies is expected to be a common feature of many physical systems.  Systems are often characterized by a pressure to grow---represented by a forcing---together with some drag or resistance on that growth, represented by a damping.  After a sufficiently long time, the forcing and the damping equilibrate and one reaches a bounded state.  This leads to the speculation that many time series that are well described as fBm over relatively short timescales may be better matched by the Mat\'ern process over longer timescales.  More generally, the Mat\'ern process adds a third parameter (damping) to the two parameters (amplitude together with spectral slope or the Hurst parameter) of fBm, thus permitting a wider range of spectral forms to be accommodated. It is therefore reasonable to think that the Mat\'ern process could be of broad interest in many areas in which fBm has already proven itself useful.  

Many of the results herein may be found somewhere in the literature; the novelty and significance of this paper arise from placing these results in context.  The relevant literature is vast, and the results that form this narrative are widely distributed within disparate communities.  The concept of diffusivity discussed in Section~\ref{motivationsection} is well known within physics and fluid dynamics, but is largely unheard of in the time series literature.  The Mat\'ern process investigated in Section~\ref{maternsection} is well known in spatial statistics, but not in time series or in fluid dynamics.  That the Mat\'ern process is essentially damped fractional Brownian motion, one of our main points, has already been recognized by \citet{lim06-pla}, who, however, appear to have come upon the Mat\'ern form independently, without using this name and without referencing the existing literature.  Thus, the various results brought together here currently exist in such a dispersed state that the significance of combining them is not at all apparent.  

The main contributions of this work are: (i) to place the Mat\'ern process in context by understanding its relationship to fractional Brownian motion; (ii) to establish \emph{why} the Mat\'ern process is important for stochastic modeling of time series, geophysical time series in particular, which is its ability to simultaneously capture the effects of long-timescale diffusivity and small-scale fractal dimensionality; (iii) to demonstrate its performance with an application to a classical physical system; and (iv) to accomplish these goals in a way that is accessible to a general audience.  

This paper was inspired by the need to develop a stochastic model for a particular physical application.  As such, we are cognizant of the need to make stochastic modeling tools accessible to a broad audience.  We have therefore endeavored to present material in a manner that is grounded in concepts from signal analysis, as this is a common language shared by many fields.  A priority is placed on being self-contained, in order to avoid referring the reader repeatedly to the literature.  The use of stochastic differential equations, or other more mathematical tools, is avoided unless absolutely necessary.  At the same time, we are aware of the need to maintain rigor, and have therefore sought to carefully qualify any approximate or informal statements. New results are denoted as such.

The structure of the paper is as follows.  Section~\ref{motivationsection} introduces background material regarding the concept of \emph{diffusivity} and its relationship to the spectrum, and presents a preview of the application to turbulence as a motivation.  An introduction to fractional Brownian motion is presented in Section~\ref{Stochastic}.  The properties of the Mat\'ern process are then investigated in Section~\ref{maternsection}.  Section~\ref{Generation} presents a new algorithm for fast approximate numerical generation of the Mat\'ern process, and Section~\ref{Application} returns to the application with additional details.  The paper concludes with a discussion.  

All numerical software associated with this paper, including a script for figure generation, is distributed as a part of a freely available Matlab toolbox, as described in Appendix~\ref{jlabappendix}.  The paper includes two supplemental animations, \url{http://www.jmlilly.net/videos/dispersionmovie.mp4} and \url{http://www.jmlilly.net/videos/turbulencemovie.mp4}.

\section{Background and motivation}\label{motivationsection}

This section introduces background material on stochastic processes, and identifies the \emph{diffusivity} as a fundamental second-order stochastic quantity.  This importance of diffusivity is illustrated by briefly discussing an application to modeling particle velocities in fluid turbulence.

\subsection{Complex notation, continuous time}

In this paper, we will work with continuous-time, complex-valued processes, a choice that deserves comment.  The decision to use complex-valued processes stems from the fact that the main application, to fluid dynamics, consists of analyzing trajectories that may be regarded as positions on the complex plane.  For the most part, the results all apply equally well to real-valued processes.  The choice to work in continuous time reflects more than convenience, as physical phenomena are generally regarded as existing continuously in time.  A discrete time series arises when a process, such as a fluid flow, happens to be sampled at discrete intervals, owing to the constraints of measurements with real-world instruments.  For these reasons, we will work in continuous time, and discrete sampling effects will be addressed when relevant.  

\subsection{Autocovariance and spectrum}
 
Let $z(t)=u(t)+\mathrm{i} v(t)$ be a potentially nonstationary, complex-valued, zero-mean random process, where $\mathrm{i}\equiv\sqrt{-1}$.  For concreteness herein, $z(t)$ will be regarded as having units of velocity, with $u(t)$ and $v(t)$ giving eastward and northward velocity components, respectively.  The autocovariance function of $z(t)$ is defined as
\begin{equation}\label{autocovariance}
R_{zz} (t,\tau)\equiv\mathrm{E}\{z(t+\tau)\,z^*(t)\}
\end{equation}
where the asterisk denotes the complex conjugate; note this satisfies the symmetry $R_{zz}(t,\tau)=R_{zz}^*(t+\tau,-\tau)$.  If it is the case that $z(t)$ is \emph{second-order stationary}, meaning that its second-order statistics are independent of global time $t$, the autocovariance function is written as $R_{zz} (\tau)$.  In this case one finds $R_{zz}^* (-\tau)=R_{zz}(\tau)$, and thus the autocovariance function of a stationary complex-valued stochastic process has Hermitian symmetry.  Another useful property of $R_{zz}(\tau)$ is that it is \emph{rotationally invariant} in the $x$--$y$ plane: if one rotates the process counterclockwise through some some constant angle $\Theta$ by defining $\tilde z(t)\equiv e^{\mathrm{i} \Theta} z(t) $, we have $R_{\tilde z\tilde z}(\tau)=R_{zz}(\tau)$, and the autocovariance function remains unchanged.

It is well known that the autocovariance function of a complex-valued process does not completely characterize its second-order statistics \citep{mooers73-dsr,picinbono97a-itsp,schreier03b-itsp}.  Additional information is contained within a second covariance function
\begin{equation}\label{relation}
C_{zz} (t,\tau)\equiv\mathrm{E}\{z(t+\tau)\,z(t)\}
\end{equation}
which is the covariance between $z(t)$ and its own complex conjugate.\footnote{It is considered standard that the covariance between two zero-mean complex-valued time series $a(t)$ and $b(t)$ involves a conjugation of one of the two time series, e.g.  $R_{ab}(\tau)\equiv\mathrm{E}\{a(t+\tau)\,b^*(t)\}$.  This accounts for the conjugation in (\ref{autocovariance}) and the absence of conjugation in (\ref{relation}).  Thus, the quantity $C_{zz} (t,\tau)$ may be equivalently, but rather confusingly, denoted as $R_{zz^*} (t,\tau)$.  }
This quantity is variously known as the \emph{relation function} \citep{picinbono97a-itsp} or \emph{complementary autocovariance function} \citep{schreier03b-itsp} or \emph{pseudo-covariance} \citep{neeser93-itit} in the time series literature, and as the \emph{outer autocovariance} in oceanography and atmospheric science \citep{mooers73-dsr}.  Unlike the autocovariance function, the relation function changes with a coordinate rotation.   With $\tilde z(t)\equiv e^{\mathrm{i} \Theta} z(t)$ again being a rotated version the process, one finds $C_{\tilde z\tilde z}(\tau)=e^{\mathrm{i} 2\Theta}C_{zz}(\tau)$.  This shows that information regarding the directionality of variability must reside in $C_{zz} (t,\tau)$ and not in $R_{zz} (t,\tau)$.  If the process is \emph{isotropic}, meaning that its statistics are independent of the rotation angle $\Theta$, then clearly $C_{zz} (t,\tau)$ must vanish; the process is then said to be {\em proper} or {\em circular} or {\em circularly symmetric}. In the present paper we are concerned with isotropic processes, and we will therefore limit our attention to $R_{zz} (t,\tau)$.

The statistical information contained in the autocovariance function of a second-order stationary process, $R_{zz}(\tau)$, can be equivalently expressed in terms of its Fourier transform, the spectrum $S_{zz}(\omega)$, through the inverse Fourier relationship
\begin{equation}\label{fourierpair}
R_{zz} (\tau) = \frac{1}{2\pi}\int_{-\infty}^\infty e^{\mathrm{i} \omega \tau} S_{zz}(\omega) \, \mathrm{d}\omega.
\end{equation}
Rather than needing to deal separately with an eastward or $u$-velocity spectrum and a northward or $v$-velocity spectrum, the spectrum of the complex-valued velocity $z(t)=u(t)+\mathrm{i} v(t)$ compactly includes contributions due to positively-rotating circular motions $e^{\mathrm{i} |\omega| \tau} $ for $\omega>0$, and those due to negatively-rotating circular motions $e^{-\mathrm{i} |\omega| \tau} $ for $\omega<0$. For this reason $S_{zz}(\omega)$ is referred to as the \emph{rotary spectrum} in the oceanographic and atmospheric science literature \citep[][Chapter 5.4.4.2]{fofonoff69-dsr,gonella72-dsr,mooers73-dsr,emery}. Unlike the spectrum of a real-valued signal,  the rotary spectrum is in general not a symmetric function of $\omega$.  Because physical processes are generally better separated in the frequency domain than in the time domain, and because the spectrum is a more straightforward quantity to estimate than is the autocovariance, we will work with the spectrum rather than the autocovariance for stochastic modeling.  

\subsection{Diffusive processes}\label{diffusiveprocesssection}

The time integral of the velocity process $z(t)$ defines a complex-valued displacement or trajectory on the complex plane, denoted by
\begin{equation}\label{rintermsofz}
r(t)\equiv 
\int_{0}^t z(\tau)\, \mathrm{d} \tau
\end{equation}
where the integral is interpreted as $-\int_{t}^0 z(\tau) \,\mathrm{d} \tau$ for $t<0$.  This definition of $r(t)$ sets the initial condition $r(0)=0$.  Drawing on a key concept from physics we introduce the \emph{total} or \emph{isotropic diffusivity} as 
\begin{equation}\label{kappadef}
\kappa(t)\equiv\frac{1}{4} \frac{\mathrm{d}}{\mathrm{d} t}\,\mathrm{E}\left\{ |r(t)|^2\right\}
\end{equation}
which quantifies the expected rate at which the particles \emph{disperse}, or spread out, over time from an initial location. Here $\mathrm{E}\{\cdot\}$ is the expectation operator.   Note that $\kappa(t)$ is the defined as the {\em average} of the rates of dispersion in the $x$- and $y$-directions,  $\kappa_x(t)\equiv\frac{1}{2} \frac{\mathrm{d}}{\mathrm{d} t}\,\mathrm{E}\left\{ x^2(t)\right\}$ and $\kappa_y(t)\equiv\frac{1}{2} \frac{\mathrm{d}}{\mathrm{d} t}\,\mathrm{E}\left\{y^2(t)\right\}$. \footnote{Why $\kappa$ should be defined as the average of the component diffusivities $\kappa_x$ and $\kappa_y$, and not their sum, requires some comment.  Recall that the diffusion equation with constant diffusivity, but differing diffusivities in the $x$- and $y$-directions, is $\frac{\partial}{\partial t}\phi = \kappa_x\frac{\partial}{\partial x^2}\phi + \kappa_y\frac{\partial}{\partial y^2}\phi$ for some field $\phi(x,y,t)$. Under the assumption of isotropy, the definition $\kappa = \frac{1}{2}\left[\kappa_x+\kappa_y\right]$ leads to the usual form of the diffusion equation $\frac{\partial}{\partial t}\phi = \kappa \nabla^2\phi$ where $\nabla^2$ is the horizontal Laplacian.  Defining $\kappa$ instead as the {\em sum} of the component diffusivities would lead to a $\frac{1}{2}\kappa$ appearing in this equation, which is not standard.  This accounts for the factor of $1/4$  in (\ref{kappadef}), rather than the more familiar $1/2$ that is found in the definition of the component diffusivities $\kappa_x$ and $\kappa_y$.}
 
If an ensemble of particles exhibits a power-law dispersion near some time $t$ with
\begin{equation}\label{dispersiontlaw}
\mathrm{E}\left\{ |r(t)|^2\right\}\sim t^{\beta},\quad\quad \kappa(t) \sim t^{\beta-1}
\end{equation}
then the local behavior is said to be \emph{diffusive} if $\beta=1$, \emph{subdiffusive} if $\beta<1$, and \emph{superdiffusive} if $\beta>1$.  The same process may exhibit different diffusive regimes at different times, but if (\ref{dispersiontlaw}) holds in an asymptotic sense for large $t$, then the long-time limit of $\kappa(t)$ is given by
\begin{equation}\label{circles}
\kappa\equiv
\lim_{t\longrightarrow\infty}
 \frac{1}{4}\frac{\mathrm{d}}{\mathrm{d} t}\,\mathrm{E}\left\{ |r(t)|^2\right\}=
\left\{
\begin{array}{ccc}
0, && \beta<1\\
\mathrm{constant}, && \beta =1 \\
\infty, && \beta >1
\end{array}
\right.
\end{equation}
where the time-independent, asymptotic quantity $\kappa$ is conventionally known simply as \emph{the diffusivity}.  In the case that $\kappa$ is a nonzero constant, one has $\mathrm{E}\left\{ |r(t)|^2\right\}=4\kappa t$, and the \emph{expected area} enclosed by the particle ensemble grows linearly with time.  Thus $\kappa$ quantifies a tendency for random fluctuations to yield systematic outward or radial motion.  

The seminal work of \citet{taylor21-plms} applied the concept of diffusivity to study the random motions of macroscopic fluid particles, a usage that is now widespread in fluid dynamics \citep{lacasce08-pio}.  Here we employ the physical concept of diffusiveness to describe the long-term dispersive behavior of random processes in general, regardless of the system being represented.  

While the diffusivity is not a recognized quantity in time series analysis, we will show that is an essential second-order descriptor, on par with the variance.  If $z(t)$ is a zero-mean second-order stationary process with autocovariance function $R_{zz}(\tau)$ and Fourier spectral density $S_{zz}(\omega)$, and having variance $\sigma^2\equiv E\left\{|z(t)|^2\right\}$, one finds
\begin{align}\label{sacf}
\sigma^2 =
 R_{zz} (0) &= 
 \frac{1}{2\pi}\int_{-\infty}^\infty S_{zz}(\omega) \, \mathrm{d} \omega \\
 \kappa =
 \frac{1}{4} S_{zz} (0) &= 
 \frac{1}{4}\int_{-\infty}^\infty R_{zz}(\tau) \, \mathrm{d} \tau \label{kappaspectrum}
\end{align}
which shows that the variance $\sigma^2$ and diffusivity $\kappa$ may be seen as time- and frequency-domain analogues of one another.  The first of these relations is the well-known Parseval's theorem, while the second is shown in Appendix~\ref{diffusivityderivation}. Just as the variance $\sigma^2$ is given by the integral of the velocity spectrum, or the value of the autocovariance at zero, the diffusivity $\kappa$ is the integral of the autocovariance, or the value of the spectrum at zero.  As each is the zeroth-order moment in one of the two domains, they share a common footing as the two lowest-order and potentially most important second-order statistical properties of a stationary random process.  

The result that the diffusivity is the zero-frequency value of velocity spectrum is not entirely new.  It is implicit in a result of \citet{kampedeferiet39-assb}, see p.  527--528 of \citet{moninandyaglomII}.  It is also pointed out in \citet[p. 175]{davis83-jmr} and is mentioned in \citet{lacasce08-pio}.  However, this result does not appear widely appreciated in the ocean/atmosphere literature.  Within the time series literature, there does not appear to be a recognition of the potential importance of the zero-frequency value of the spectrum on account of its connection to dispersive behavior.  

Because the diffusivity appears as a second-order descriptor of the velocity process $z(t)$, it is useful to categorize $z(t)$ according to the associated diffusivity value.  For a given $z(t)$ we may \emph{define} $\kappa$ as in (\ref{kappaspectrum}) through the value of the spectrum at zero frequency, or equivalently, through the integral of the autocovariance.  We will refer to $z(t)$ as a \emph{diffusive process} if it is associated in this way with a non-zero and finite value of $\kappa$.  Processes associated with zero values of $\kappa$ will be said to be \emph{subdiffusive}, while those associated with unbounded values of $\kappa$ will be referred to as \emph{superdiffusive}.  Note that the diffusivity is a property that can be associated both with the velocity process $z(t)$, in the zero-frequency value of its spectrum, and the trajectory $r(t)$, in its rate of dispersion.  To avoid ambiguity, we will say that $z(t)$ is a diffusive \emph{process} whereas $r(t)$ is a diffusive \emph{trajectory}, and so forth for sub- and superdiffusive processes.\footnote{A \emph{diffusive process} in our terminology is distinct from the idea of a \emph{Markov diffusion process}, which is the solution to a particular type of first-order stochastic differential equation   \citep[e.g.][]{metzner07-thesis}.  As the latter usage appears to be somewhat restricted, we expect there to be little possibility of confusion.}

The classification of a stochastic process as diffusive, subdiffusive, or superdiffusive is related to a well-known property, the process \emph{memory}.  If the autocovariance of a finite-variance stationary process exhibits the long-term decay 
\begin{equation}\label{Rdecay}
R_{zz}(\tau) \sim |\tau|^{-\mu}, \quad\quad 0<\mu\le1 \quad\quad |\tau|\rightarrow \infty
\end{equation}
then the process is said to be a \emph{long-memory process} or to have \emph{long-range dependence} \citep{beran92-ss,beran,gneiting04-siam}.  A \emph{short-memory} process is one for which the autocovariance falls off more rapidly than $|1/\tau|$, in which case the autocovariance function will be absolutely integrable; note that the statement $R_{zz}(\tau) \sim |\tau|^{-\mu}$ means that the \emph{magnitude} of the autocovariance decays as $|\tau|^{-\mu}$. Thus, short-memory stationary processes are those for which the autocovariance function is absolutely integrable, and long-memory stationary processes are those for which it is not.

The process memory is therefore a classification based on the \emph{absolute integrability} of the autocovariance, whereas the diffusiveness is based on its \emph{integrability}, as seen in (\ref{kappaspectrum}).  From this one may establish that both short- and long-memory processes can be diffusive or subdiffusive, but only long-memory processes can be superdiffusive.  A long-memory process has an autocovariance that is not \emph{absolutely integrable}, whereas a diffusive process has an autocovariance that is \emph{integrable} and that integrates to a nonzero value.  A function can be integrable but not absolutely integrable, thus a diffusive process can be long-memory.  Similarly, both short-memory and long-memory processes could have autocovariances that integrate to zero, giving a subdiffusive process.  However, if a function is absolutely integrable then it is also integrable, thus a short-memory process cannot be superdiffusive.  For concreteness, examples of spectra of processes with different combinations of diffusiveness and memory are presented in Appendix~\ref{memorysection} based on modifications to the Mat\'ern process.
 
\subsection{Application to 2D turbulence}\label{applicationpreview}
In this paper, we will be concerned with an application to the stochastic modeling of particle trajectories, and the associated velocity time series, from a numerical simulation of fluid turbulence.  The system we will use, known as \emph{forced-dissipative two-dimensional turbulence}, see e.g. Chapter 8.3 of \citet{vallis}, generates temporally and spatially varying flows that exist purely in the horizontal plane.  This system is considered an idealized representation of turbulence in planetary fluid dynamics.  Details of the numerical model, including the model equations and parameter choices, are described in Section~\ref{modelsection}.  The simulation is carried out in a doubly periodic domain\footnote{A doubly periodic domain means that the $x$-axis is periodic, such that structures passing eastward across the eastern boundary return on the western boundary, and that the $y$-axis is similarly periodic.} having physical dimension of 2500 $\times$ 2500~km, and is integrated for three years.  The time series analyzed here are 512 particle trajectories taken from a total of 1024 that are tracked throughout this experiment, and that are initially uniformly distributed throughout the model grid at regular intervals.  

\begin{figure*}
	\hspace{-0.05in}\includegraphics[width=1.015\textwidth,angle=0]{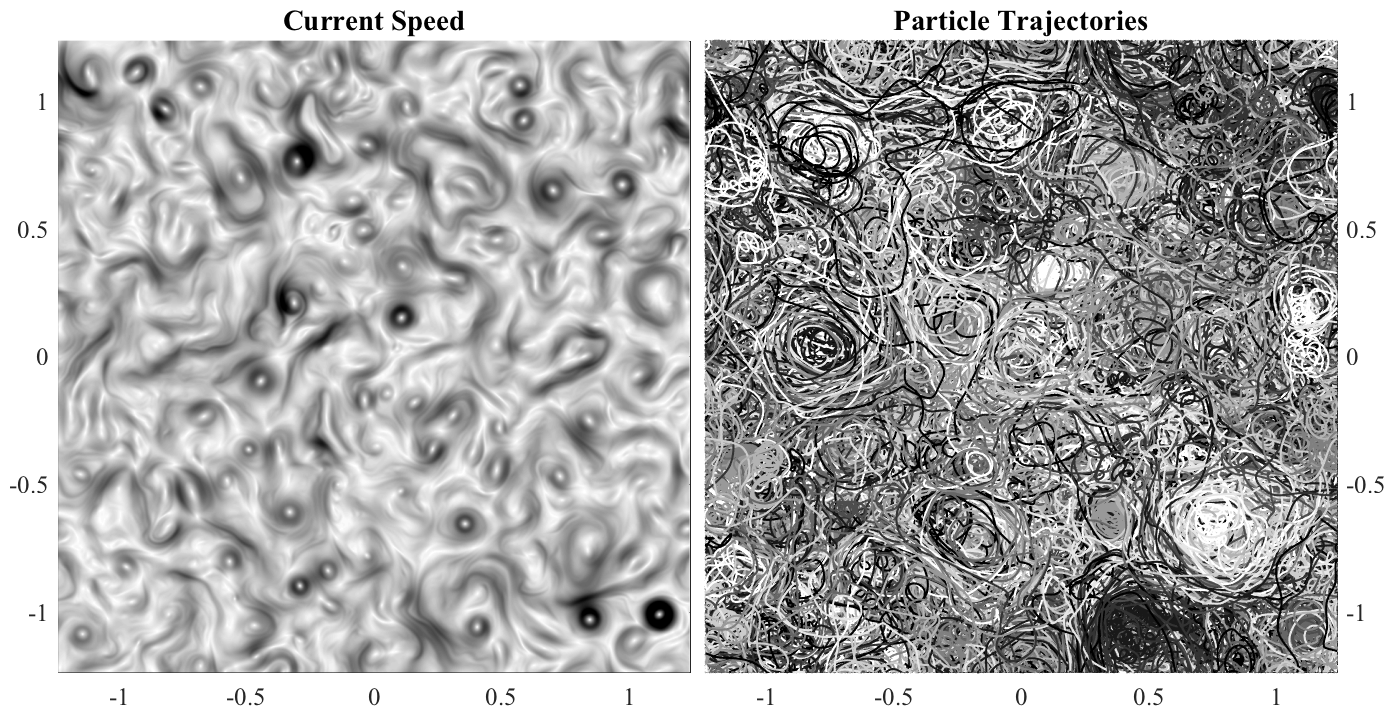}
	\caption{A snapshot of current speed from the turbulence simulation (left) together with 1024 particle trajectories (right).  In the left panel, shading is the speed $\sqrt{U^2(x,y,t)+V^2(x,y,t)}$ at each point, with white corresponding to zero velocity and black to 18~cm~s${^-1}$.  In the right-hand panel, different trajectories are represented by different shadings of gray.  The physical domain size is 2500 $\times$ 2500~km, with the $x$- and $y$-axes in this figure given in units of 1000~km. See turbulencemovie.mp4 for an animation of this figure, in which only the 512 particles to be analyzed are shown.}
	\label{snapshot}
\end{figure*}

A snapshot of the velocity field at the initial time, together with the particle trajectories from the entire simulation, is shown in Fig.~\ref{snapshot}.  The quantity plotted in the left-hand panel is the current speed $|U+\mathrm{i} V|=\sqrt{U^2+V^2}$ at time $t=0$, where $U=U(x,y,t)$ and $V=V(x,y,t)$ are the velocities at each point in the domain.   The roughly circular regions of high-speed currents correspond to long-lived swirling structures termed \emph{vortices} or \emph{eddies}. The emergence of vortices is one of the defining features of two-dimensional turbulence \citep[e.g.][]{mcwilliams90a-jfm}.  A method for their study based on trajectory data has been developed elsewhere \citep{lilly06-npg,lilly09-asilomar,lilly11-grl}. The focus here, however, is on trajectories {\em not} directly influenced by such structures.  For this reason, one-half of the trajectories are discarded in order to exclude those directly effected by vortices, using a criterion described in Section~\ref{modelsection}, leaving 512 trajectories that will be analyzed herein.  The supplementary animation turbulencemovie.mp4 presents the evolution of these 512 trajectories superimposed on the speed as in Fig.~\ref{snapshot}a.

These 512 ``eddy-free'' trajectories are also displayed in Fig.~\ref{dispersion}a.  Here, the position coordinates in the periodic domain have been unwrapped, and the resulting trajectories $r(t)$ offset in the horizontal so as to begin at the origin at time $t=0$.   Dispersion is then visualized by the circles, which have been drawn with radii
\begin{equation}\label{rtildeqn}
\tilde r_n\equiv\sqrt{\mathrm{E}\left\{ \left|r(n\Delta)\right|^2\right\}}\end{equation} at uniformly spaced time intervals $n\Delta t$, with $\Delta$ equal to six months and $n=1,2,\ldots 6$.  In this expression, the expectation operator is interpreted as the average over all 512 trajectories.  For constant diffusivity, one expects that $\tilde r_n^2 = 2\kappa n\Delta$ from~(\ref{kappadef}), such that the total enclosed area increases linearly, and the radius increases as the square root of time.  That the trajectories shown here are exhibiting diffusive behavior is thus indicated by the appearance of the circles in Fig.~\ref{dispersion}a, which become more closely spaced together as time increases.

\begin{figure*}[p!]
	\includegraphics[width=0.99\textwidth,angle=0]{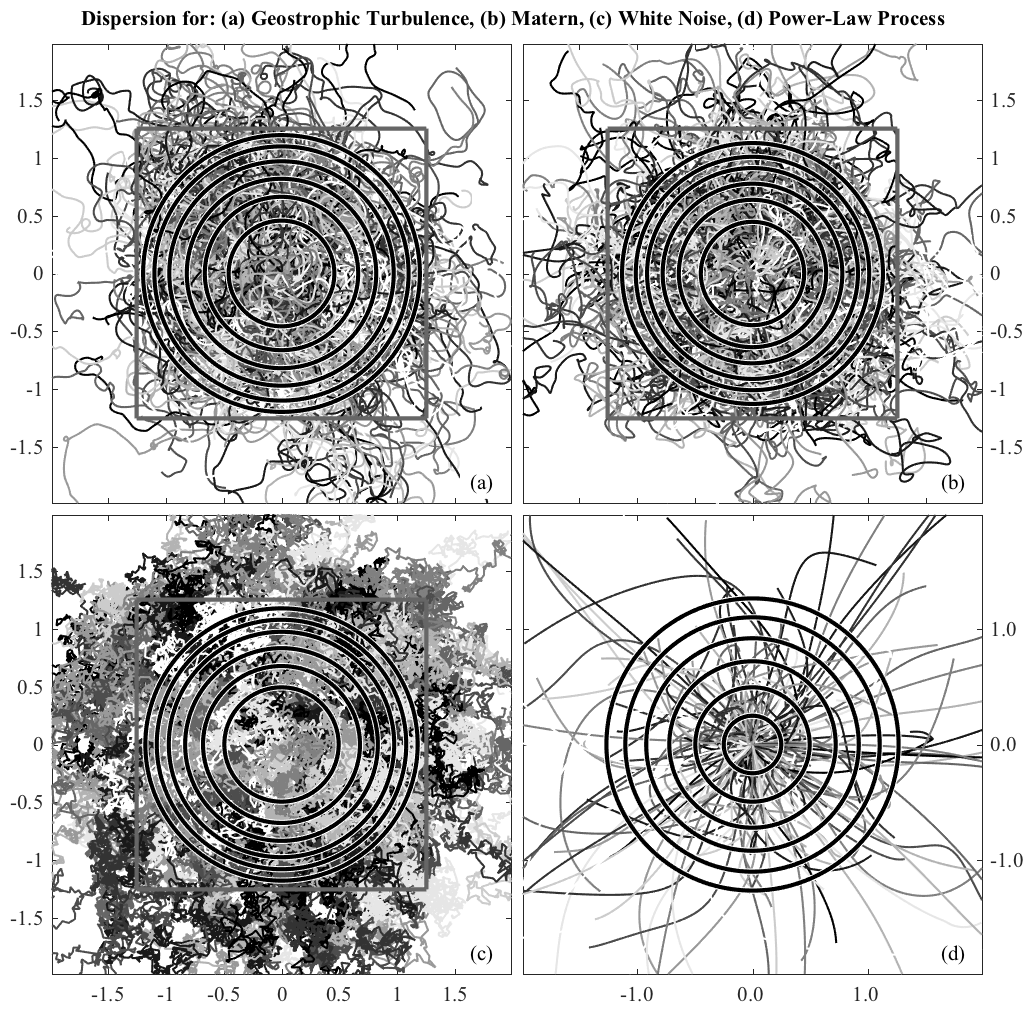}
	\caption{\small [See caption on next page]}
\end{figure*}

\setcounter{figure}{1}
\begin{figure*}[t!]
	\caption{\small [See figure on previous page] Dispersion curves for the three-year turbulence trajectories and the three different stochastic models discussed in Section~\ref{overview}.  Panel (a) shows 512 ``eddy-free'' trajectories, chosen from a larger set of 1024 as described later in Section~\ref{modelsection}.  All curves have been offset such that the initial points are located at the origin.  Panel (b) shows realizations of a Mat\'ern random process using parameters fit to the velocity spectra of each trajectory, and then cumulatively summed to produce a displacement, also with the initial condition at the origin.  Similarly, the lower two panels show trajectories corresponding to white noise velocities (c) and velocities for a power-law process (d), the latter approximated using a Mat\'ern process with very low damping.  The stochastic velocities in~(c) are chosen to match the low-frequency spectral levels of the turbulence trajectories, while in (d) they are chosen to match the high-frequency spectral slope.  All trajectories in the doubly-periodic domain have been unwrapped for presentational clarity, with the gray square in each panel showing the domain size.  Note that the x- and y-axes in panel (d) are a factor of one million times larger than those of the other panels, which is why the gray box is not visible.  In each panel, black circles show the root-mean-square distance from the origin $\tilde r_n$ defined as in (\ref{rtildeqn}).  Circles are drawn every six months, beginning at six months and ending at three years.  The circles in (d) do not become closer together with increasing radius, indicating superdiffusive behavior. See dispersionmovie.mp4 for an animation of the first two panels of this figure.}
	\label{dispersion}
\end{figure*}

The average estimated spectrum of the velocity signals $z(t)$ corresponding to these trajectories is shown as the heavy black curve in Fig.~\ref{spectra}.  Non-parametric estimates of the velocity spectra have been formed for each trajectory by tapering with a lowest-order Discrete Prolate Spheroidal Sequence or ``Slepian'' taper \citep[][Chapter 3.9]{slepian78-bell,thomson82-ieee,park87b-jgr,sapa} having a time-bandwidth product set to a value of 10, see p. 12,677 of \citet{park87b-jgr} for a definition of this parameter.  The spectra for all 512 velocity signals are averaged together, and because there is no expected difference between clockwise and anti-clockwise velocities, only spectra for positive frequencies are shown.

The velocity spectrum is observed to have three main features: an overall energy level, a high-frequency slope, and a low-frequency plateau.  As shown in the preceding section, the low-frequency plateau of the velocity signals is a reflection of the diffusive behavior of the trajectories.  The goal of this paper is to identify a stochastic model capable of reproducing these three features, and to thoroughly understand its properties.

\subsection{Overview of stochastic models}\label{overview}

Consider \mbox{one-,} \mbox{ two-,} and three-parameter frequency spectra having the forms
\begin{equation*}\label{three}
S_{zz}(\omega) = A^2, \quad
S_{zz}(\omega) = \frac{A^2}{|\omega|^{2\alpha}},\quad
S_{zz}(\omega) = \frac{A^2}{\left(\omega^2+\lambda^2\right)^{\alpha}}
\end{equation*} 
which are taken as models for the complex velocity time series $z(t)$ from the turbulence simulation.  The first type of spectrum corresponds to white noise.\footnote{For the sake of brevity, we here glossing over the fact that the spectrum of white noise is defined only up to the Nyquist frequency, whereas the other two spectra are defined for all frequencies.} The second is a power-law spectrum that arises for fractional Brownian motion \citep{mandelbrot68-siam} for $\alpha$, termed the {\em slope parameter}, in the range $1/2<\alpha<3/2$.  For the slope parameter $\alpha>1/2$, the third spectrum is that of a type of random process known as a Mat\'ern process \citep{matern60-mss,guttorp06-biometrika}, which we will show to be a \emph{damped} version of fractional Brownian motion, with $\lambda>0$ playing the role of an inverse damping timescale.  Note that these three models are formally nested within one another: choosing $\lambda=0$, the third becomes the second; and choose $\alpha=0$, the second becomes the first.

\begin{figure}[t!]\begin{center}
	\includegraphics[width=0.475\textwidth,angle=0]{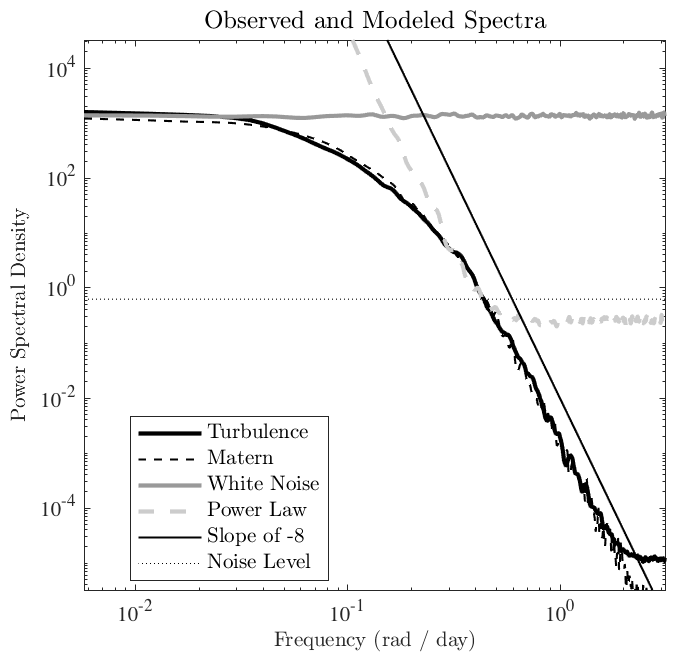}\end{center}
	\caption{\small Spectra for the trajectories shown in Fig.~\ref{dispersion}.  Estimated rotary spectra $S_{zz}(\omega)$ are shown for positive frequencies only, since the negative frequency side is statistically identical.  The first four curves show the mean values of the estimated spectra, formed as described in the text, of each of the four sets of trajectories shown in Fig.~\ref{dispersion}.  The fifth curve, indicated by a solid line, shows a slope of $-8$, corresponding to a slope parameter $\alpha=4$.  The dotted horizontal line marks the approximately limit of double numerical precision for the power law process, 15 orders of magnitude below its maximum; this numerical precision limit accounts for the flattening of gray dashed curve.} 
	\label{spectra}
\end{figure}

The form of the Mat\'ern spectrum is fit to the velocity spectra of the turbulence trajectories, in a way that will be described in Section~\ref{Application}, to generate best-fit values of the three Mat\'ern parameters ($A$, $\alpha$, and $\lambda$) for each of the 512 trajectories.  The low-frequency values from these fits are then used to match the white noise spectrum, while the parameters for the high-frequency slopes are used to match the power-law spectrum.  For each set of parameters, realizations of these three types of random processes are constructed from the best-fit parameters using the methods described in Section~\ref{Generation}.  The spectra of the simulated trajectories are then estimated in the same manner as for the original trajectories, and shown in Fig.~\ref{spectra}.  As expected due to their construction, the white noise and power-law process match only the low-frequency plateau or high-frequency spectral slope, respectively, of the original spectra.  

The Mat\'ern spectral form is seen to provide an excellent match to the observed Lagrangian velocity spectra over roughly eight decades of structure.  The high-frequency slope is seen to be roughly $|\omega|^{-8}$, a very steep slope.  We are not aware of any physical theory to account for this, nor for the value of the damping parameter $\lambda$.  Despite the fundamental role that the Eulerian {\em wavenumber} spectrum of velocity plays in turbulence theory, the Lagrangian {\em frequency} spectrum has received relatively little attention.  Attempting to connect the observed form of this spectrum to physical principles is, however, outside the scope of the present paper.

These three different sets of random processes for the velocity time series are then cumulatively summed to form trajectories, and are compared with the original trajectories in Fig.~\ref{dispersion}; note the axes limits in Fig.~\ref{dispersion}d are a factor of one million times larger than in the other panels, a consequence of the growth of the variance to enormous values.  The turbulence trajectories and the synthetic trajectories generated from the Mat\'ern model are observed to be virtually indistinguishable in character.  See the supplementary file dispersionmovie.mp4 for an animation of the upper two panels of Fig.~\ref{dispersion}, showing the good agreement between the Mat\'ern trajectories and the turbulence trajectories.
 
By contrast, the one-parameter and two-parameter spectral models provide poor fits to the observed trajectories, see Fig.~\ref{dispersion}c,d.  The trajectories associated with the white noise velocities match the dispersion curves closely, but the trajectories are far too rough in appearance.  When set to match the high-frequency spectral slope and thus the trajectory behavior at small scales, the power-law model for velocity spectra yields trajectories with a vastly incorrect range, too high a degree of smoothness at the large scale, and dispersion characteristic of a continually increasing diffusivity. 

Thus, the white noise model is able to correctly match the large scale, low-frequency component of the velocity spectra that accounts for the diffusive behavior of the trajectories.  The power-law model is able to correctly match the high-frequency component of the spectrum that sets the small-scale roughness.  The Mat\'ern spectrum allows one to match both.  This provides a compelling example that motivates examining the Mat\'ern process in more detail.

\section{Fractional Brownian motion}\label{Stochastic}

This section reviews the properties of fractional Brownian motion, focusing on the central importance of the spectrum.  With a few noted exceptions, this section presents material that is already known in the literature.  Readers already very familiar with this process may wish to skip to the description of the Mat\'ern process in the next section.

\subsection{Spectrum}\label{fbm}

As described in the Introduction, many real-world processes are found to exhibit power-law behavior over a broad range of frequencies.  For a range of spectral slopes, the power-law spectrum corresponds to that of a Gaussian random process\footnote{A Gaussian random process is one for which every finite linear combination of samples has a jointly Gaussian distribution.  For example, the distribution of the process at a fixed time is Gaussian, and the distribution between the process and itself at two different times is a jointly Gaussian function of two variables.} called \emph{fractional Brownian motion} (fBm), introduced by \citet{mandelbrot68-siam}.  While the spectrum of fBm is not defined in the usual sense due to its nonstationarity, an expanded version of the notion of a spectrum, discussed in Section~\ref{derivingthespectrum} and denoted as $\widetilde S_{zz}(\omega)$, is found to yield for fBm the form \citep{flandrin89-itit,solo92-sjam}
\begin{equation}
\widetilde S^{f\!Bm}_{zz}(\omega)=\frac{A^2}{|\omega|^{2\alpha}}, \quad\quad 1/2<\alpha < 3/2
\label{redspectrum}
\end{equation}
where $\alpha$ will be called the \emph{slope parameter}, and with $A$ setting the spectral level. Fractional Brownian motion is a generalization of classical Brownian motion---corresponding to the case $\alpha=1$ and therefore to an $\omega^{-2}$ spectrum---for which the slope parameter $\alpha$ can take a range of non-integral values.  It is clear that a process having a spectrum proportional to $|\omega|^{-2\alpha}$ for $\alpha>1/2$ will be singular at zero, and will integrate to an infinite value, thus possessing neither a finite diffusivity nor a finite variance.  Both the variance and the diffusivity of fBm will be found to increase without bound.  

\begin{figure*}[t!]
\includegraphics[width=0.99\textwidth]{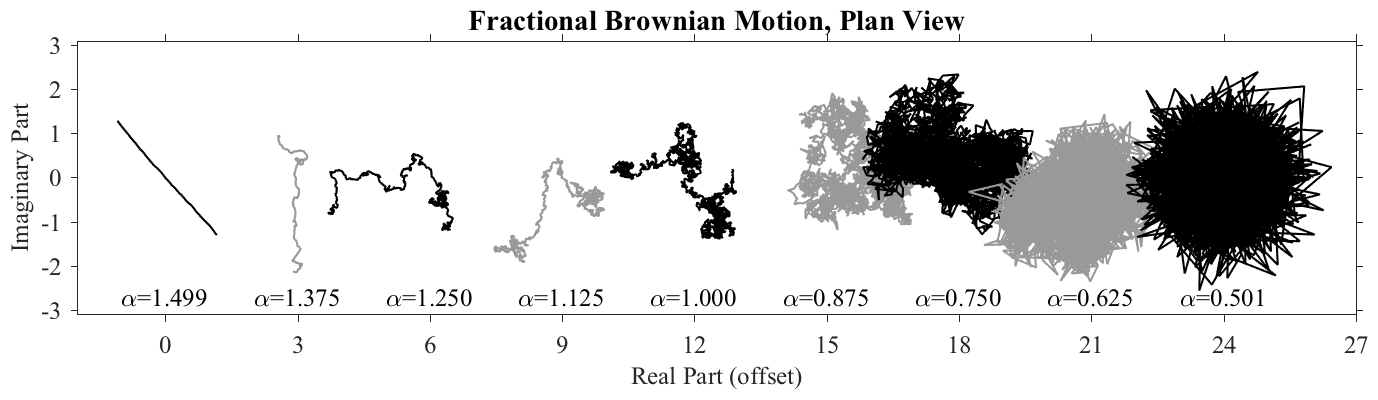}
\caption{\small Plan view of realizations of complex-valued fractional Brownian motion $z(t)$, for nine different values of the slope parameter $\alpha$, as indicated in the legend.  All nine time series have been set to have unit sample variance, and the real part of each curve is offset by a value of $-3$ from that for the next lower value of $\alpha$.  The smallest value of $\alpha$, corresponding to the least smooth process, is at the right.  The data aspect ratio is equal between the real and imaginary parts.  }
\label{randomwalk_brownianplan}
\end{figure*}

Examples of complex-valued fractional Brownian motion are shown in Fig.~\ref{randomwalk_brownianplan}.  Here nine curves are shown for nine different values of $\alpha$, varying from just greater than $1/2$ to just less than $3/2$.  The decrease in the degree of roughness as $\alpha$ increases, and the spectral slope becomes more steep, is readily apparent in the figure. This occurs due to the fact that larger values of $\alpha$ correspond to stronger degrees of `filtering', with steep spectral slopes removing high-frequency contributions to variance.  Because we are considering that $z(t)$ represents a velocity $z(t)=u(t)+\mathrm{i} v(t)$, this figure shows plots of $u(t)$ versus $v(t)$, as opposed to the trajectories that would arise from temporally integrating these quantities. 

The main goal of this section is to utilize fBm to understand the implications of the slope parameter $\alpha$.  It will be found that for fractional Brownian motion, $\alpha$ has several intuitively distinct but partly corresponding interpretations: it is directly linked to the temporal decay of the autocovariance function; it controls the aspect ratio of rescaling for self-similar behavior; it sets the \emph{fractal dimension} or \emph{degree of roughness}; and it determines the degree of \emph{persistence} or \emph{anti-persistence} of a differenced version of the process, see Appendix~\ref{fgnappendix}.  Note that in the fBm literature, the slope parameter $\alpha$ is conventionally replaced with $H=\alpha-1/2$, referred to as the \emph{Hurst parameter}, with $0<H< 1$, in terms of which the fBm spectrum is given by $\widetilde S^{f\!Bm}_{zz}(\omega)=A^2/|\omega|^{2H+1}$.  

There are compelling reasons to work with the slope parameter $\alpha$ rather than the Hurst parameter $H$.  While spectral slope could be characterized in the vicinity of any frequency, the Hurst parameter is, strictly speaking, a measure of the long-time process range or memory.  That is, $H$ is a \emph{limiting} quantity pertaining to the behavior of the process at very large timescales.  As pointed out by \citet{gneiting04-siam}, the self-similarity of fBm implies that the large-scale behavior (memory) and small-scale behavior (fractal dimension) must be linked.  However, for stochastic processes more generally, no such link between large and small scales is required.  The spectral slope is therefore more appropriate when showing the connection of fBm to its damped version, the Mat\'ern process, which is a short-memory process.  Furthermore, because the appearance of the Mat\'ern process as damped fractional Brownian motion is most clear in the frequency domain, it is sensible to work with a parameter that makes the spectral form simple.

\subsection{The fBm autocovariance function}\label{fbmcovsection}

Fractional Brownian motion is defined in terms of a stochastic integral equation, which will be presented later in this section.  This stochastic integral equation leads to a nonstationary autocovariance function given by \citep{mandelbrot68-siam}
\begin{multline}\label{fbmautocovariance}
R_{zz}^{f\!Bm}(t,\tau)=\mathrm{E}\{z(t+\tau)\,z^*(t)\}\\=\frac{V_\alpha}{2} \, A^2
\left[|t+\tau|^{2\alpha-1}+|t|^{2\alpha-1}-|\tau|^{2\alpha-1}\right]
\end{multline}
where $V_\alpha$ is a normalizing constant defined shortly.  The exponents take on values in the range $0 < 2\alpha- 1 < 2$ due to the fact that $1/2<\alpha<3/2$.  Thus the dependence of $R_{zz}^{f\!Bm}(t,\tau)$ on $t$ and $\tau$ varies from being relatively flat, near $\alpha=1/2$, to relatively steep, near $\alpha=3/2$.  

Observe that fractional Brownian motion is \emph{nonstationary}---its autocovariance is a function of ``global'' time $t$ as well as the time offset $\tau$.  Most significantly, the variance of fBm is 
\begin{equation}\label{fbmvariance}
\sigma^2(t)=\mathrm{E}\{|z(t)|^2\} = R_{zz}^{f\!Bm}(t,0)=V_\alpha \, A^2 |t|^{2\alpha-1}
\end{equation}
which increases without bound; the longer one waits, the larger the expected amplitude of variability becomes.  The time-varying fBm diffusivity is found to be 
\begin{equation}
\kappa(t) 
 = \frac{V_\alpha}{4} \frac{\alpha+1}{\alpha} \, A^2 |t|^{2\alpha},\quad\quad t\ge 0
\end{equation}
as we readily find by integrating the autocovariance as in (\ref{timevaryingkappa}).  Like the variance, the diffusivity tends to increase without bound, rather than taking on a constant value.  Note that the ratio of the diffusivity to the variance increases linearly with time, $\kappa(t)/\sigma^2(t)=\frac{1}{4}|t|(\alpha+1)/\alpha$.

The normalizing constant in fBm, conventionally denoted $V_\alpha$, is defined as the variance at time $t=1$ of an fBm process having the amplitude parameter $A$ set to unity, 
\begin{equation}
V_\alpha\equiv \mathrm{E}\{|z(1)|^2\} = R_{zz}^{f\!Bm}(1,0), \quad\quad A=1.
\end{equation}
Its value is found to be \citep{barton88-itit} 
\begin{equation}\label{valphastandard}
V_\alpha=\frac{\Gamma(2-2\alpha)\sin(\pi\alpha)}{\pi (\alpha-1/2)}
\end{equation}
where $\Gamma(x)$ is the gamma function.  We find in Appendix~\ref{coefficientappendix} that this constant can be cast in the more symmetric form
\begin{equation}\label{symmetricform}
V_\alpha=\frac{1}{\pi}\frac{\Gamma\left(\alpha-\frac{1}{2}\right)\Gamma\left(\frac{3}{2}-\alpha\right)}{ \Gamma(2\alpha)}
\end{equation}
which allows one to see behavior of this coefficient more clearly.  Recall that $\Gamma(x)$, while positive for positive $x$, is negative in the interval $(-1,0)$, as follows from the reflection formula $\Gamma(x)=\pi/\left[\sin(\pi x) \Gamma(1-x)\right]$.  Thus $V_\alpha$ is positive over the whole permitted range of $\alpha$, $1/2<\alpha<3/2$, but becomes unphysically negative as one passes outside of this range.  Because the gamma function has a singularity at zero, with $\Gamma(x)$ tending to positive infinity as $x$ approaches zero from above, $V_\alpha$ also tends to positive infinity as one approaches the two endpoints $\alpha=1/2$ and $\alpha=3/2$.  Finally, from $\Gamma(1/2)=\sqrt{\pi}$ and $\Gamma(2)=1$, the value of the coefficient for the Brownian case of $\alpha=1$ is found to be $V_1=1$. 


In addition to the autocovariance function, it is informative to also examine a related second-order statistical quantity,
\begin{multline}
 \gamma_{zz}(t,\tau)\equiv \frac{1}{2} \mathrm{E}\left\{ \left|z(t+\tau)-z(t)\right|^2\right\}\\= \frac{1}{2}\left[R_{zz}(t+\tau,0)+R_{zz}(t,0)
 -2\Re\left\{R_{zz}(t,\tau)\right\}\right]
\end{multline}
where $\Re\left\{\cdot\right\}$ denotes the real part.  This quantity is commonly known as the \emph{variogram} in time series analysis and geostatistics, following \citet{cressie88-jasa} and \citet{matheron63-eg}; in the turbulence literature, the same quantity is widely used and is known as the \emph{second-order structure function}, a term which dates back at least to the 1950's \citep{monin58-tpa}.  For a stationary random process, the variogram becomes simply $\gamma_{zz}(t,\tau)= \gamma_{zz}(\tau)= \sigma^2 -\Re\left\{R_{zz}(\tau)\right\}$.  Thus in the stationary case, the variogram merely repeats information already present in the autocovariance function.

For fractional Brownian motion, cancellations in the variogram occur and one obtains
\begin{equation}\label{fbmvariogram}
 \gamma_{zz}^{\,f\!Bm}(t,\tau)= \gamma_{zz}^{\,f\!Bm}(\tau)= \frac{V_\alpha}{2} A^2 |\tau|^{2\alpha-1}
\end{equation}
which is independent of global time $t$.  Thus unlike its autocovariance function, the variogram of fBm is stationary.  This equation states that the expected squared difference between fBm values at any two times is proportional to a power of the time difference, implying that the expected rate of growth of the fBm from its current value is independent of~$t$.  One might therefore say that fBm is nonstationary, but in a time-independent or stationary manner.  A process having a stationary variogram is said to be \emph{intrinsically stationary} \citep{ma04-jap}.  

\subsection{Linking the spectrum and autocovariance}\label{derivingthespectrum}
Owing to its nonstationarity, the fBm autocovariance cannot be Fourier transformed in the usual way to yield a spectrum that is independent of global time $t$.  Evidently the notion of what it means to be a Fourier transform pair must be generalized to accommodate the time-dependent autocovariance.  That the spectrum of fractional Brownian motion should be a power law of the form $|\omega|^{-2\alpha}$ was already conjectured by \citet{mandelbrot68-siam}, based on earlier work by \citet{hunt51-tams} on the spectrum of its increments.  Proving that this should be the case was accomplished by  \citet{solo92-sjam} using one approach, and by \citet{flandrin89-itit} and \citet{oigard06-pre} using two variants of a different approach.  Here, we essentially follow the latter paper, incorporating some additional details.

In general, the Fourier transform with respect to $\tau$ of a nonstationary autocovariance function $R_{zz}(t,\tau)$ defines a time-varying relative of the spectrum
\begin{equation}\label{fbmvariogramtransformgeneral}
 S_{zz}(t,\omega)\equiv  \int_{-\infty}^\infty R_{zz}(t,\tau) e^{-\mathrm{i} \omega \tau} \,\mathrm{d} \tau
\end{equation}
which, provided the integral on the right-hand side is well defined, is  known as the Rihaczek \citep[p. 60--62]{rihaczek68-itit,flandrin} or Kirkwood-Rihaczek  \citep{kirkwood33-pr,hindberg07-itsp,oigard06-pre} distribution, or alternatively as the \emph{time-frequency spectral density} \citep{hanssen03-itsp}. If one averages the Rihaczek distribution across global time in moving windows of length~$T$, then takes the limit as $T$ approach infinity, one obtains
\begin{align}\label{rihaczekaverage}
\overline {S}_{zz}(t,\omega;T) &\equiv \frac{1}{T}\int_{t-T/2}^{t+T/2} S_{zz}(u,\omega) \,\mathrm{d} u\\
 \widetilde S_{zz}(\omega)&\equiv\lim_{T\longrightarrow\infty}\overline {S}_{zz}(t,\omega;T)
\end{align}
where $\widetilde S_{zz}(\omega)$ is a {\em time-averaged} spectrum of a potentially nonstationary process.  Observe that for stationary processes, $R_{zz}(t,\tau)$ and therefore $ S_{zz}(t,\omega)$ are independent of the global time $t$.  In this case,  $ S_{zz}(t,\omega)$ reduces to the usual Fourier spectrum $S_{zz}(\omega)$, the time average in (\ref{rihaczekaverage}) has no effect, and $ \widetilde S_{zz}(\omega)$ is therefore also identical to the usual Fourier spectrum $S_{zz}(\omega)$.   Thus $ \widetilde S_{zz}(\omega)$ is a generalization of the usual Fourier spectrum, to which it reduces in the stationary case, that may be used to describe nonstationary processes.

For fractional Brownian motion, the Rihaczek distribution was stated by \citet{oigard06-pre} to be
\begin{align}\label{fbmvariogramtransformdefinition}
 S_{zz}^{f\!Bm}(t,\omega) &\equiv \int_{-\infty}^\infty R_{zz}^{f\!Bm}(t,\tau) e^{-\mathrm{i} \omega \tau} \,\mathrm{d} \tau\\&=
\frac{A^2}{|\omega|^{2\alpha}} \left(1-e^{\mathrm{i}\omega t}\right)+V_\alpha A^2 \pi|t|^{2\alpha-1}\delta(\omega)\label{fbmvariogramtransform}
\end{align}
with $\delta(t)$ being the Dirac delta function; see Appendix~\ref{fbmspectrumappendix} for details of the derivation.  The time-averaged version of the fBm  Rihaczek distribution, defined as in  (\ref{rihaczekaverage}), is  given by 
\begin{multline}\label{fbmlimit}
\overline {S}^{f\!Bm}_{zz}(t,\omega;T) = 
\frac{A^2}{|\omega|^{2\alpha}} \left[1-e^{\mathrm{i}\omega t} \frac{\sin(\omega T/2)}{\omega T/2}\right]\\
+\frac{V_\alpha}{2\alpha T} A^2 \pi \left[\left|t+\frac{T}{2}\right|^{2\alpha} -\left|t-\frac{T}{2}\right|^{2\alpha} \right] \delta(\omega)
\end{multline}
and in the limit as the averaging time $T$ approaches infinity, we have the time-averaged nonstationary spectrum for fBm,
\begin{equation}\label{limitingoperator}
 {\widetilde S^{f\!Bm}_{zz}}(\omega)\equiv \lim_{T\longrightarrow\infty}\overline {S}^{f\!Bm}_{zz}(t,\omega;T)=
\frac{A^2}{|\omega|^{2\alpha}},
\end{equation}
where all terms dependent on global time~$t$ are found to vanish.  This determines a sense in which the power-law form is the correct spectrum to associate with nonstationary fractional Brownian motion.   In the approach of \citet{flandrin89-itit}, a different, but related, time-varying generalization of the spectrum is used instead of the Rihaczek distribution, but leading to the same power-law form for the time-averaged spectrum. 

This approach to proving that the power-law form is the correct spectrum to associate with fBm may be critiqued on the grounds that taking the limit of an average of the time-frequency spectral density, while mathematically sensible, does not correspond well with a limiting action that occurs in actual practice.  \citet{solo92-sjam} took a different approach, and found that if the expected autocovariance and spectrum are estimated from a sample over a \emph{finite} time interval, the power-law form again emerges in the limit as that the time interval tends to infinity.  That proof therefore has a strong intuitive appeal, but is more involved than the argument presented here.  

\subsection{Self-similarity}\label{selfsimilarsection}

The most striking feature of fBm is that it is statistically identical to rescaled versions of itself.  To show this, we define a time- and amplitude-rescaled version of $z(t)$ as
\begin{equation}\label{selfaffinity}
\tilde z(t)\equiv \beta^{\alpha-1/2}\,z(t/\beta)
\end{equation}
where the amplitude rescaling has been chosen to depend upon $\beta$ as well as the slope parameter $\alpha$.  From (\ref{fbmautocovariance}), one finds 
\begin{multline}
R_{\tilde z\tilde z}^{f\!Bm}(t,\tau)=\beta^{2\alpha-1} R_{zz}^{f\!Bm}(t/\beta,\tau/\beta )\\=
\frac{ V_\alpha}{2} \, A^2\beta^{2\alpha-1}
\left[|(t+\tau)/\beta|^{2\alpha -1}+|t/\beta|^{2\alpha -1}-|\tau/\beta|^{2\alpha -1}\right]\\=
\frac{ V_\alpha}{2} \, A^2
\left[|t+\tau|^{2\alpha -1}+|t|^{2\alpha -1}-|\tau|^{2\alpha -1}\right] \\= R_{zz}^{f\!Bm}(t,\tau)\label{originalform}
\end{multline}
and the autocovariance function of the rescaled process is determined to be the same as that of the original process.

Because the original, unrescaled fBm process is Gaussian as well as zero mean, its statistical behavior is completely characterized by its autocovariance function.  Thus fBm is statistically identical to itself when we ``zoom in'' in time, provided we also magnify the amplitude appropriately.  This property was referred to as \emph{self-similarity} in the original work of \citet{mandelbrot68-siam}; although later the term \emph{self-affinity} was suggested as a substitute \citep{mandelbrot85-ps}, the original term appears to be in more widespread use.

The positive constant $\beta$ can be seen as a temporal zoom factor, while the coefficient $\beta^{\alpha -1/2}$ describes how the amplitude is to be rescaled.  Choosing $\beta>1$ corresponds to zooming in in time, since then the interval from zero to $\beta$ in the new process $\tilde z(t)$ is drawn from the smaller interval zero to one in $z(t/\beta)$.  Similarly, $\beta^{\alpha -1/2}$ with $\alpha>1/2$ is greater than one, implying the amplitude must also be magnified.  The required degree of amplitude magnification increases with $\alpha$ from a minimum value of unity at $\alpha=1/2$ to a value of $\beta$ at $\alpha=3/2$.  The slope parameter $\alpha$ therefore governs the \emph{aspect ratio} of rescaling for this self-similar behavior.

\begin{figure*}[t!]
\includegraphics[width=0.99\textwidth]{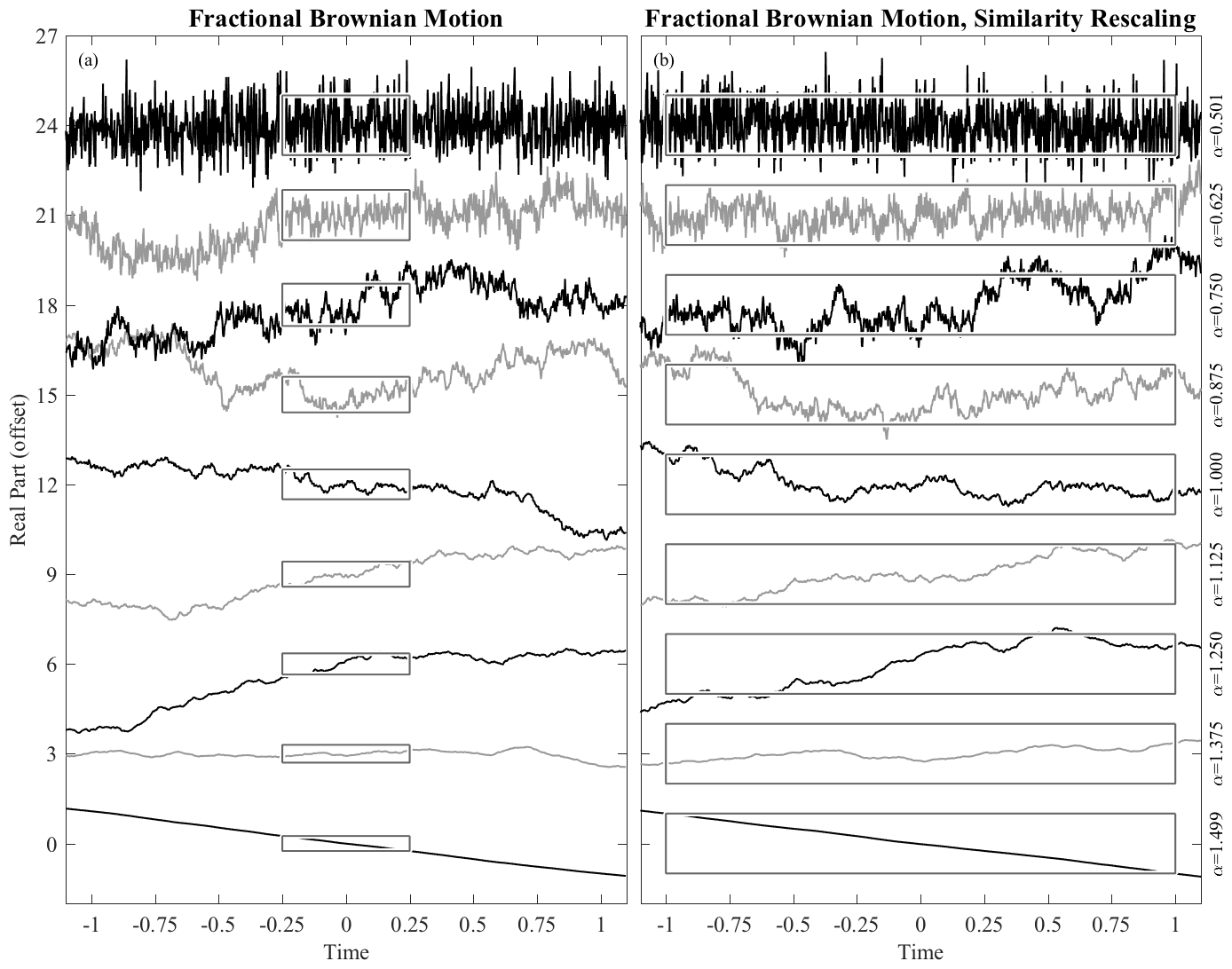}
\caption{\small A demonstration of self-similarity for fractional Brownian motion, using the realizations presented in Fig.~\ref{randomwalk_brownianplan}.  The real part of each process is shown, with the $y$-axes in this figure corresponding exactly to the $x$-axis in Fig.~\ref{randomwalk_brownianplan}.  The gray boxes in panel (a) illustrate the different scaling behaviors, as described by (\ref{selfaffinity}).  When each process is rescaled such that the boxes in (a) are transformed to the boxes in (b), the resulting time series are statistically identical to the originals.  Thus the rescaled curves in (b) present the same degree of roughness as the corresponding curves in (a).  The temporal magnification factor is $\beta=4$, while the amplitude magnification factor $\beta^{\alpha-1/2}$ varies from 1 at $\alpha=1/2$ to $4$ at $\alpha=3/2$.  In order to avoid the appearance of additional roughness in (a) due only to numerical resolution, only every fourth point in (a) is shown; thus the curves in (a) and (b) consist of the same number of points.  }
\label{randomwalk_brownian}
\end{figure*}

An illustration of self-similarity is presented in Fig.~\ref{randomwalk_brownian}, using the real parts of the nine realizations shown in Fig.~\ref{randomwalk_brownianplan}.  The two panels show the effects of the self-similar rescaling (\ref{selfaffinity}) on each time series with a zoom factor $\beta=4$, with the zooming represented by the gray boxes.  The boxes on the left, of different aspect ratios, are rescaled according to the law (\ref{selfaffinity}) to have the same aspect ratios, as shown on the right.  It is clear that each of the nine curves presents the same degree of roughness, and same amplitude of variability, on the left as on the right.  This demonstrates what is meant by statistical self-similarity, and shows how $\alpha$ controls the aspect ratio.  A distinguishing feature of fractional Brownian motion is that this zooming may be continued indefinitely in either direction.  

For stationary processes, self-similarity may also be seen in the frequency domain.  Apply the rescaling (\ref{selfaffinity}) to some process $z(t)$, which is now assumed to be stationary.  From the Fourier representation of the autocovariance, one finds
\begin{multline}
R_{\tilde z\tilde z}(\tau) = \beta^{2\alpha-1} \frac{1}{2\pi} \int_{-\infty}^\infty S_{zz}(\omega)\, e^{\mathrm{i} \omega \tau/\beta}\mathrm{d} \omega \\=
 \beta^{2\alpha} \frac{1}{2\pi} \int_{-\infty}^\infty S_{zz}(\beta\omega) \,e^{\mathrm{i} \omega \tau}\mathrm{d} \omega 
\end{multline}
after employing the change of variables $\omega/\beta\mapsto\omega$.  Thus, in order for the process to be self-similar, one must have
\begin{equation}\label{frequencyscaling}
S_{zz}(\omega)=\beta^{2\alpha} S_{zz}(\beta\omega)
\end{equation}
in the spectral domain.  This would clearly be the case for the power-law spectrum $S_{zz}(\omega)=A^2|\omega|^{-2\alpha}$, if a stationary process with such a spectrum were to exist.  More generally, if a process has an \emph{approximately} power-law spectrum over a range of frequencies, then the self-similarity condition (\ref{frequencyscaling}) is expected to be \emph{approximately} satisfied over that range.  In this sense a power-law spectrum implies self-similarity.  

Fractional Brownian motion is peculiar in that it has neither a well-defined derivative nor a well-defined integral.  Loosely speaking, one may say that a derivative does not exist because the limiting action of taking a derivative conflicts with the self-similarity.  Because $z(t)$ exhibits variability at infinitesimally small scales, $[z(t+\Delta)-z(t)]/\Delta$ does not have a well-defined limit as $\Delta$ tends to zero.  The integral $\int_{-\infty}^t z(u)\, \mathrm{d} u$ does not exist either, because $z(t)$ has unbounded variance as $t$ progresses toward to infinitely large negative times and is therefore not integrable.  Nevertheless, a \emph{differenced} version of fBm does exist.  This process, termed \emph{fractional Gaussian noise}, is discussed for completeness in Appendix~\ref{fgnappendix}.

\subsection{Fractal dimension}\label{fractalsection}

The property of self-similarity, which is \emph{global} in nature, was shown in the previous section to be related to the spectral slope.  The slope is also related to two \emph{local} properties, one associated with the slope at small frequencies, or the behavior of the autocovariance at large time offsets, and one associated with the slope at high frequencies, or the autocovariance at small time offsets.  The former property is the process \emph{memory} or \emph{long-range dependence} discussed in Section~\ref{diffusiveprocesssection}, while the latter is the \emph{fractal dimension}. While we view spectral slope as the more physically meaningful quantity, its relationship to fractal dimension is here discussed for completeness.
 
Fractal dimension is a measure of the dimensionality of a curve (or some higher-order surface) that accounts the effect of roughness \citep{mandelbrot85-ps,falconer}.   There are several different measures of fractal dimension in use, giving sometimes different values of dimensional measure for a particular curve \citep[see e.g.][]{mandelbrot85-ps,taylor91-jrssb,dunbar92-jmaa}.  The most well-known measure, the Hausdorff dimension, is related to the behavior of the autocovariance function or variogram at very short timescales.  One must also distinguish between the dimension of a curve as a function of the time variable, as in $u(t)=\Re\{z(t)\}$ versus $t$, and the dimension of a curve such as $z(t)=u(t)+\mathrm{i} v(t)$ in space or $u(t)$ versus $v(t)$, see e.g. \citet{qian03-pwlrc}.  In the literature, the former is known as a \emph{graph}, and the latter as a \emph{sample path}.  

The dimension of the graph is closely related to the short-time behavior of the autocovariance.  As described by \citet{gneiting04-siam}, for a univariate (or real-valued) stationary process $u(t)$ that has an autocovariance function behaving as $|\tau|^{\rho}$ for some $0<\rho\le2$ as $\tau\rightarrow 0$, the Hausdorff dimension of the graph of the process is given by  $D= 2 - \rho/2$.  The comparable result for intrinsically stationary Gaussian processes such as fBm is provided by \citet{adler77-ap}.  For fBm, $\rho=2\alpha-1$, hence the dimension of the graph of (real-valued) fBm is $D=5/2-\alpha$.  This varies from $D=1$ for the smoothest processes having $\alpha=3/2$, to $D=2$ for the roughest processes with $\alpha=1/2$, corresponding to the bottom-to-top progression seen in Fig.~\ref{randomwalk_brownian}.

As pointed out by \citet{gneiting04-siam}, the self-similarity of fBm links the behavior at very large scales and very small scales together.  Because for fBm the spectral slope is constant, the fractal behavior at small scales implies a singularity in the spectrum at the origin.  This is associated with unbounded diffusivity, and since this singularity is not integrable, with unbounded variance as well.  The Mat\'ern process examined in the next section has an additional degree of freedom compared to fBm, such that the spectrum transitions to flat values for sufficiently low frequencies.  This decouples the fractal dimension from the low-frequency behavior and permits the phenomenon of diffusivity to arise.  

\subsection{Stochastic integral equation}

Fractional Brownian motion is defined via the stochastic integral equation \citep{mandelbrot68-siam}
\begin{multline}\label{stochasticintegral}
z(t) = \frac{A}{\Gamma(\alpha)} \left\{
\int_{-\infty}^0 \left[(t-s)^{\alpha-1} - (-s)^{\alpha-1}\right] \mathrm{d} W(s) 
\right.\\\left.
+\int_{0}^t (t-s)^{\alpha-1} \mathrm{d} W(s) \right\}
\end{multline}
where $\mathrm{d} W(t)$ here are increments of the complex-valued Wiener process, the covariance of which between itself at two different times is
\begin{equation}\label{orthogonalwiener}
E\left\{\mathrm{d} W(t) \, \mathrm{d} W^*(s)\right\} = \delta(t-s)\, \mathrm{d} t\,\mathrm{d} s.
\end{equation}
The integration with respect to $\mathrm{d} W(s)$ indicates in (\ref{stochasticintegral}) that these integrals are of the Riemann-Stieltjes form, see \citet{sapa}.  The process $\mathrm{d} W(s)$ can be said to represent continuous-time white noise, thus this equation defines fBm as a weighted integral of white noise.  Because in (\ref{stochasticintegral}) one may exchange the order of the integral and the expectation operator, and $\mathrm{d} W(s)$ is zero mean, $z(t)$ is also zero mean.  As $\mathrm{d} W(s)$ is Gaussian and $z(t)$ is a linear combination of Gaussian random variables, $z(t)$ is also Gaussian. Thus $z(t)$ inherits both zero-meanness and Gaussianity from the increments of the Wiener process. Further intuitive content of (\ref{stochasticintegral}) is not initially apparent, so we will take some time to examine it in detail.

Note that standard Brownian motion, corresponding to $\alpha=1$, is defined for all $t$ as
\begin{equation}\label{standardebrownian}
z(t) = A \int_0^t \mathrm{d} W(s)
\end{equation}
in which the integral is interpreted as $z(t) = -A \int_{t}^0 \mathrm{d} W(s)$ for $t<0$.  This is simply the temporal integral of white noise.  The fBm definition (\ref{stochasticintegral}) reduces to the Brownian form with $\alpha=1$, with the first term in (\ref{stochasticintegral}) vanishing.  

The stochastic integral equation (\ref{stochasticintegral}) can be written in the somewhat more transparent form
\begin{equation}\label{stochasticintegral2}
z(t) = \frac{A}{\Gamma(\alpha)} 
\int_{-\infty}^t \left[(t-s)^{\alpha-1} - I(-s) (-s)^{\alpha-1}\right] \mathrm{d} W(s) 
\end{equation}
where $I(t)$ is the indicator, or unit step, function defined as
\begin{equation}\label{indicator}
I(t) \equiv \left\{ \begin{array}{cc}1,& \quad t\ge 0 \\
 0, & \quad t<0\end{array}\right..
\end{equation}
The purpose of the second term in (\ref{stochasticintegral2}) is now clearly seen to set the initial condition.  It is not a function of time; it is simply a random number, chosen to set $z(0)=0$ identically.  Note that the two components of (\ref{stochasticintegral2}) cannot be written as separate integrals, because writing them as two separate integrals would mean that two different realizations of $\mathrm{d} W(s)$ are involved. The two terms in (\ref{stochasticintegral2})  must be based on the {\em same} realization of $\mathrm{d} W(s)$ in order to achieve the initial condition $z(0)=0$; this is not true for the two terms in (\ref{stochasticintegral}), which correspond to two different intervals of integration.
 
The weighting factors such as $(t-s)^{\alpha-1}$ in (\ref{stochasticintegral}) may be seen as creating a \emph{fractional integral} of the Wiener process, as will now be shown.  There is a simple expression for a function $f(t)$ that is integrated $n$ times from some initial point $a$ to time $t$, an action that is the reverse of the repeated derivative $(\mathrm{d}^n/\mathrm{d} t^n) f(t)$.  This formula, known as \emph{Cauchy's formula for repeated integration}, states 
\begin{multline}\label{cauchy}
\int_{a}^t \int_{a}^{\tau_1}\cdots \int_{a}^{\tau_{n-2}}\left[\int_{a}^{\tau_{n-1}} f(\tau_n)\, 
\mathrm{d} \tau_n \right] \mathrm{d} \tau_{n-1}\cdots \mathrm{d}\tau_2 \, \mathrm{d} \tau_1 \\
=\frac{1}{(n-1)!} \int_{a}^t (t-\tau)^{n-1} f(\tau)\, \mathrm{d} \tau
\end{multline}
meaning that one may collapse an integral that is repeated $n$ times into a single integral, with a weighting to the $(n-1)$th power.  Note that applying $(\mathrm{d}^n/\mathrm{d} t^n)$ to both sides, one obtains $f(t)=f(t)$---the left-hand side by repeated applications of the fundamental theorem of calculus, and the right-hand side by repeated applications of the Leibniz integral rule.  

While the left-hand side of the Cauchy integral formula is not interpretable for non-integer $\alpha$, the right-hand side remains valid.  This allows us to \emph{define} a fractional integral of $f(t)$ by letting $n$ take on non-integer values in the right-hand-side of (\ref{cauchy}).  According to this reasoning, the quantity
\begin{equation}
\frac{1}{\Gamma(\alpha)}\int_{a}^t (t-\tau)^{\alpha-1} f(\tau) \,\mathrm{d} \tau \quad\quad \alpha>0
\end{equation}
is known as the Riemann-Liouville fractional integral, and may be said to integrate the function $f(t)$ a \emph{fractional} number of times $\alpha$.  For further details on fractional calculus, see e.g.  \citet{gorenflo97-cism}.  

Returning to the definition of fBm in (\ref{stochasticintegral}), we now see that it is simply a fractional integral of continuous-time white noise, modified to have the initial condition $z(0)=0$.  Unlike standard Brownian motion (\ref{standardebrownian}), which is integrated only from time $t=0$, for fractional Brownian motion one integrates from the infinite past in order to obtain the desired statistical behavior, and then one offsets this process by the correct amount in order to set the desired initial condition.  

\section{The Mat\'ern process}\label{maternsection}

The previous section reviewed the properties of fractional Brownian motion, including its self-similarity and fractal dimension, and showed how these are related to the spectral slope.  This section examines the Mat\'ern process in detail, with a focus on its relationship to fBm.  A simple extension, the inclusion of a `spin parameter', generalizes the Mat\'ern process to encompass a larger family of oscillatory processes that are shown to represent forced/damped fractional oscillators.  

\subsection{The Mat\'ern process and its spectrum}\label{SS:matern}

In Section~\ref{motivationsection} we showed that fractional Brownian motion is unable to capture long-time diffusive behavior, and demonstrated that this was a deficiency for the particular application to modeling particle velocities in two-dimensional turbulence.  Regarding the spectra in Fig.~\ref{spectra}, one sees a high-frequency power law slope but a low-frequency plateau.  This leads us to consider a spectrum of the form
\begin{equation}\label{maternspectrum}
 S^{M}_{zz}(\omega)=\frac{A^2}{\left(\omega^2+\lambda^2\right)^\alpha},\quad \quad \alpha>\frac{1}{2}
 \end{equation}
which is the spectrum of a type of stationary random process known as the Mat\'ern process \citep{matern60-mss,guttorp06-biometrika}. Unlike fBm, the Mat\'ern process is defined for all $\alpha>1/2$ and not just in the range $1/2<\alpha<3/2$. Compared with fBm, the Mat\'ern spectrum incorporates an additional (non-negative) parameter $\lambda$ having units of frequency, which will be shown to have the physical interpretation of a damping.  Note that the form of the Mat\'ern spectrum also generalizes that of the Ornstein-Uhlenbeck process, corresponding to the $\alpha=1$ case, to fractional orders \citep{wolpert05-sp,lim06-pla}.

\begin{figure*}[t!]
\includegraphics[width=0.99\textwidth]{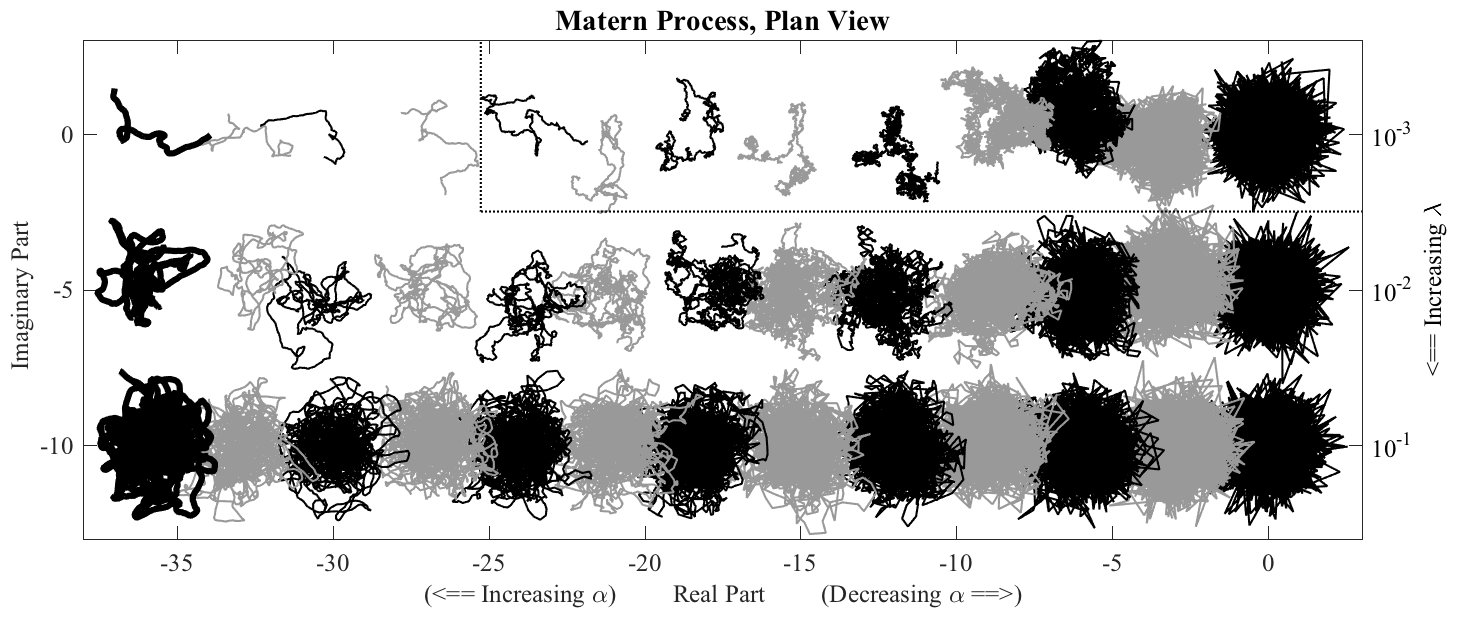}
\caption{\small  Plan view of realizations of the complex-valued Mat\'ern process, for twelve different values of the slope parameter $\alpha$ and three different values of the damping parameter $\lambda$.  Lines corresponding to successively higher values of $\alpha$ are offset by a value of $-3$ in the $x$-direction, while successively higher values of $\lambda$ are offset by a value of $-3$ in the $y$-direction.  The slope parameter $\alpha$ ranges from just greater than 1/2 to 2 with an interval of 1/8, while $\lambda$ takes the values $1/10$, $1/100$, and $1/1000$.  The various $\alpha$ values are shown as alternating black and gray lines, with largest value $\alpha=2$ shown as the heavy black line.  The dotted box corresponds to those values of $\alpha$ shown previously in Fig.~\ref{randomwalk_brownianplan}, and to the smallest of the three damping values presented here.  }
\label{maternplan}
\end{figure*}

Examples of simulated Mat\'ern processes are shown in Fig.~\ref{maternplan}, for twelve different values of $\alpha$ and three different values of $\lambda$.  The box indicates a very low-damping regime with $1/2<\alpha<3/2$, roughly corresponding to the fractional Brownian motion realizations seen in Fig.~\ref{randomwalk_brownianplan}.  There are two important differences when compared to fBm. The first is that there is no upper bound on $\alpha$, so the spectral decay can become even steeper than for the $\alpha=3/2$ case that defines the upper limit of the slope parameter for fBm. The second is the role of the additional parameter $\lambda$.  As this parameter is increased, the curves for any $\alpha$ value appear more and more like white noise. 

The damping parameter $\lambda$ thus emerges as controlling the transition between two distinct spectral regimes.  The Mat\'ern spectrum is observed to have two limits
\begin{align}
S^{M}_{zz}(\omega)\approx\frac{A^2}{|\omega|^{2\alpha}}, \quad\quad |\omega| \gg \lambda\\
S^{M}_{zz}(\omega)\approx\frac{A^2}{\lambda^{2\alpha}}, \quad\quad |\omega| \ll \lambda
\end{align}
so that, for high frequencies, an fBm-like power-law decay is recovered, while for low frequencies the spectrum approaches a constant.  The spectrum may therefore may be said to be \emph{locally white} (or constant) for small $|\omega|/\lambda$; that is, the spectrum is not constant over all frequencies, or \emph{globally white}, but it is approximately white for sufficiently low frequencies. The Mat\'ern process thus provides a continuum between the two regimes of white noise and a power-law spectrum, with a transition dictated by the value of $\lambda$. Equivalently, $\lambda$ gives the approximate timescale at which the process begins to exhibit self-similar behavior, as one zooms out from very large timescales.  It follows that in real-world applications, the sampling interval must be sufficiently small compared to $\lambda$ in order to resolve the self-similar behavior.

The theoretical spectra corresponding to the realizations in Fig.~\ref{maternplan} are shown in Fig.~\ref{threepanel}a.  When frequency is normalized by the damping parameter, the theoretical (as opposed to the sampled) spectra for the different $\lambda$ values become identical.  A transition in the vicinity of $\omega/\lambda=1$ is readily apparent.  The different spectral levels reflect the choice of normalization, which is that $\sigma^2$ has been set to unity.  Smaller values of $\alpha$, corresponding to slower decay, therefore appear with lower spectral levels in order to integrate to unit variance.  

To examine the role of $\lambda$ as a transition frequency, we take the derivative of the logarithm of the spectrum, and obtain
\begin{equation}\label{difflogspec}
\frac{\mathrm{d}}{\mathrm{d}\omega}\ln S_{zz}^M(\omega)= -\alpha \frac{2\omega}{\omega^2+\lambda^2} 
\end{equation}
and note that $\frac{\mathrm{d}^2}{\mathrm{d}\omega^2}\ln S_{zz}^M(\omega)$ vanishes at $|\omega|=\lambda$, so the rate of change (\ref{difflogspec}) obtain an extremum at that frequency.  The third derivative $\frac{\mathrm{d}^3}{\mathrm{d}\omega^3}\ln S_{zz}^M(\omega)$ is positive at $|\omega|=\lambda$, indicating that this extremum of $\frac{\mathrm{d}}{\mathrm{d}\omega}\ln S_{zz}^M(\omega)$ is a minimum.  Thus the parameter $\lambda$ gives the frequency at which $\ln S_{zz}^M(\omega)$ is decreasing most rapidly with increasing $|\omega|$, a natural choice to designate the transition between the energetic ``white'' regime at low frequencies and the decaying regime at high frequencies.  Since $\frac{d}{d\omega}\ln S_{zz}^M(\omega)=\left[\frac{d}{\mathrm{d} \omega}S_{zz}^M(\omega)\right]/S_{zz}^M(\omega)$, $|\omega|=\lambda$ is the frequency at which the \emph{fractional} decrease in $S_{zz}^M(\omega)$ is largest.

The variance and diffusivity of the Mat\'ern process are both finite, and are found to be given by
\begin{equation}
\sigma^2 =c_\alpha\frac{A^2}{\lambda^{2\alpha-1}},\quad\quad
\kappa =\frac{1}{4}\frac{A^2}{ \lambda^{2\alpha}}
\label{maternkappa}
\end{equation}
in which we have introduced the normalizing constant \begin{equation}
c_\alpha\label{cdef}
\equiv\frac{1}{2\pi} B\left(\frac{1}{2}, \alpha-\frac{1}{2}\right) =\frac{1}{2\pi} \frac{\Gamma\left(\frac{1}{2}\right) \Gamma\left(\alpha-\frac{1}{2}\right)}{\Gamma(\alpha)}
\end{equation}
where $B(x,y)\equiv\Gamma(x)\Gamma(y)/\Gamma(x+y)$ is the beta function.  The value of the diffusivity is found from $\kappa = S_{zz}(0)/4$, see (\ref{kappaspectrum}), together with the Mat\'ern spectrum form in (\ref{maternspectrum}), while the variance is 
\begin{multline}
\sigma^2
=\frac{1}{2\pi}\int_{-\infty}^\infty S_{zz}^M(\omega)\, \mathrm{d} \omega=\frac{1}{2\pi}\int_{-\infty}^\infty \frac{A^2}{\left(\omega^2+\lambda^2\right)^{\alpha}}\, \mathrm{d} \omega\\
=\frac{A^2}{2\pi \lambda^{2\alpha-1}}\int_{0}^\infty \frac{x^{-1/2}}{\left(1+x\right)^{\alpha}}\, \mathrm{d} x
\end{multline}
after the change of variables $\omega^2\mapsto x \lambda^2$.  Applying one of the defining forms of the beta function,  e.g. \citet[][3.194.3]{gradshteyn},
\begin{equation}
\int_{0}^\infty \frac{x^{\mu-1}}{(1+ x)^{\nu}}\, \mathrm{d} x= B(\mu,\nu-\mu), \quad\quad \nu>\mu>0
\end{equation}
then leads to the variance expression given in (\ref{maternkappa}).  

The Mat\'ern spectrum can be rewritten in terms of the variance $\sigma^2$ as
\begin{equation}\label{newmaternspectrum}
 S^{M}_{zz}(\omega)=\frac{\lambda^{2\alpha-1}}{c_{\alpha}}\,
\frac{\sigma^2}{\left(\omega^2+\lambda^2\right)^\alpha}
\end{equation}
so that the diffusivity becomes $\kappa=\frac{1}{4}\sigma^2/(\lambda c_\alpha)$.  In this form, the Mat\'ern spectrum becomes a function of $\sigma^2$, $\alpha$, and $\lambda$ rather than $A^2$, $\alpha$, and $\lambda$.  This will prove to be more convenient for numerical optimization during parameter fitting, because reasonable ranges for $\sigma$ are more readily determined than are ranges of $A$.  This re-parameterization also simplifies somewhat the form of the autocovariance function, presented next.  

\subsection{The autocovariance function}

The autocovariance function corresponding to the spectrum (\ref{maternspectrum}) is found to be \citep{matern60-mss,guttorp06-biometrika}
\begin{equation}
R^{M}_{zz}(\tau)=\sigma^2 \mathcal{M}_\alpha (\lambda\tau)
\label{maternacvs}
\end{equation}
where for notational convenience we have introduced the {\em Mat\'ern function} 
\begin{equation}\label{maternfunction}
\mathcal{M}_\alpha(x)\equiv  \frac{2}{\Gamma(\alpha-1/2) 2^{\alpha-1/2}} \, |x|^{\alpha-1/2}\mathcal{K}_{|\alpha-1/2|}(|x|) 
\end{equation} 
as a modified version of the $\mathcal{K}_{\alpha-1/2}(x)$, the decaying modified Bessel function of the second kind of order $\alpha-1/2$.  Integral relation 17.34.9~given on p.~1126 of \citet{gradshteyn} may be rearranged to give
\begin{equation}\label{fourierpairmatern}
 \mathcal{M}_{\alpha}(\lambda\tau)
 = \frac{1}{2\pi}\int_{-\infty}^\infty \frac{ \lambda^{2\alpha-1}}{c_\alpha}\frac{\mathrm{e}^{\mathrm{i} \omega \tau} }{\left(\omega^2+\lambda^2\right)^{\alpha}}\,\mathrm{d} \omega
\end{equation}
for $\alpha>0$ and $\lambda>0$, verifying that (\ref{maternacvs}) is the inverse Fourier transform of (\ref{newmaternspectrum}). The cosine integral version of this result is sometimes known as Basset's formula, see \citet[][p.~172]{watson}, who states the case of integer $\alpha$ is originally due to \citet[][p.~19]{basset}, and who also discusses some history of the integral on the right-hand-side.

\begin{figure*}[t!]
\includegraphics[width=0.99\textwidth]{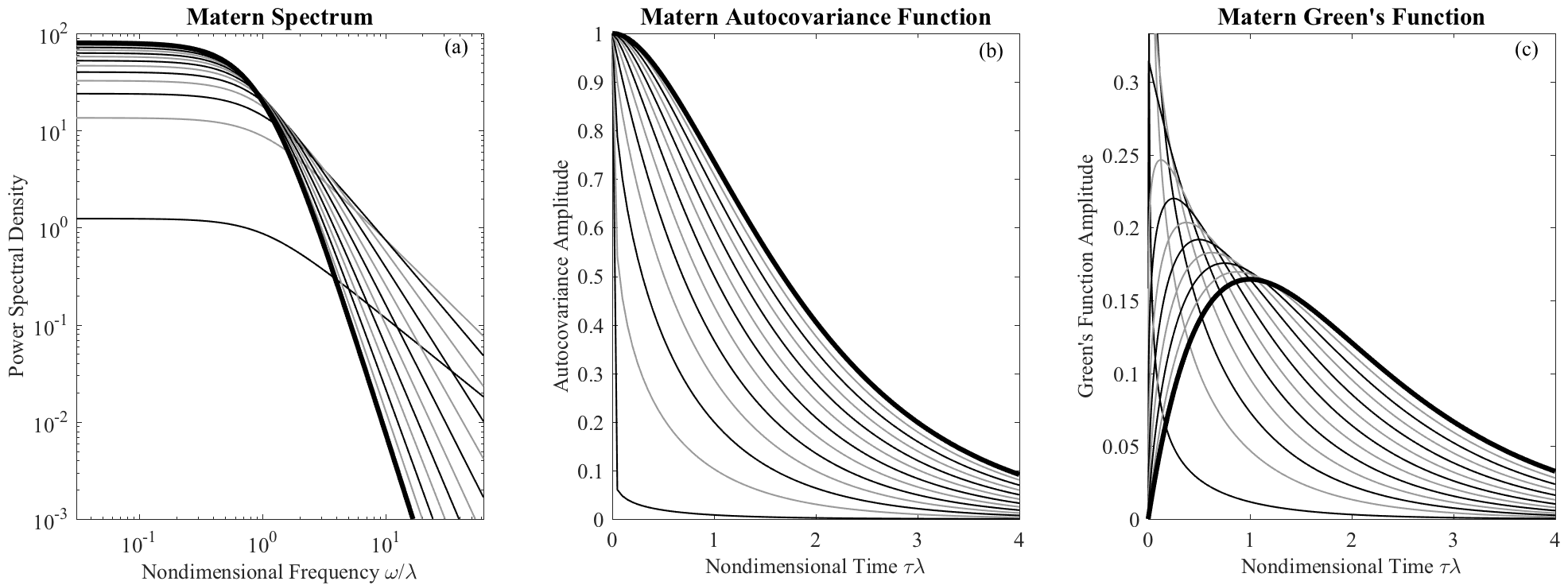}
\caption{\small Theoretical spectra (a), autocovariance functions (b), and Green's functions (c) for Mat\'ern processes corresponding to the different $\alpha$ values shown previously in Fig.~\ref{maternplan}, and with the process variance set to $\sigma^2=1$.  The corresponding expressions are (\ref{newmaternspectrum}), (\ref{maternacvs}), and (\ref{Greenfunctiontime}), respectively.  As in Fig.~\ref{maternplan}, the various $\alpha$ values are shown by alternating black and gray lines, with $\alpha=2$ shown as a heavy black line.  Time and frequency have been nondimensionalized as $\tau\lambda$ and $\omega/\lambda$, respectively; thus the transition between a flat and a sloped regime occurs in the vicinity of $\omega/\lambda=1$ in (a), while the $e$-folding time in (c) is $\tau\lambda=1$.  The autocovariance function (b) develops a strong singularity as $\alpha$ approach 1/2, which is linked to the flattening of the spectrum in (a).  The Green's function in (c) is infinite at $\tau\lambda=0$ for $\alpha<1$, and vanishes at $\tau\lambda=0$ for $\alpha>1$.}
\label{threepanel}
\end{figure*}

Examples of theoretical Mat\'ern autocovariance functions are presented in Fig.~\ref{threepanel}b, again corresponding to the realizations in Fig.~\ref{maternplan}.  As is usual with Fourier pairs, the most localized spectra correspond to the most distributed autocovariance functions, and vice-versa.  As $\alpha$ decreases, the autocovariance falls off more and more quickly from the origin, with a singularity developing at the origin as $\alpha$ approaches one-half.

The asymptotic behavior of the Mat\'ern covariance for large and small times is as follows. For  $|\tau|\gg 1/\lambda$, one has the behavior 
\begin{equation}\label{maternacvslongtime}
R^{M}_{zz}(\tau)\approx  \sigma^2\frac{\sqrt{2\pi}}{\Gamma\left(\alpha-1/2\right) 2^{\alpha-1/2}}
\left|\lambda\tau\right|^{\alpha-1}e^{-\lambda|\tau|}
\end{equation}
as follows from the asymptotic behavior of the modified Bessel function for large argument \citep[][9.7.2]{abramowitz}.  Thus the Mat\'ern process exhibits exponential decay of its covariance function, and is therefore categorized as a \emph{short-memory} process.  For time offsets that are small compared to the damping timescale, $|\tau|\ll 1/\lambda$, and for the slope parameter in the range $1/2<\alpha<3/2$, one finds
\begin{equation}\label{Rnearzerooriginal}
R^{M}_{zz}(\tau)\approx\sigma^2 
\left[1 - \left(\frac{\lambda|\tau|}{2}\right)^{2\alpha-1}\frac{\Gamma\left(\frac{3}{2}-\alpha\right)}{\Gamma\left(\alpha+\frac{1}{2}\right)}\right]
\end{equation}
as the short-time behavior of the Mat\'ern autocovariance function.  This is derived in Appendix~\ref{shorttimesection} following \citet[][their Appendix~A]{goff88-jgr}, who were apparently the first to establish it, see \citet{guttorp06-biometrika}.  It is also shown in Appendix~\ref{shorttimesection} that for $\alpha>3/2$, the lowest-order dependence of the Mat\'ern autocovariance function no longer contains a power of $\alpha$, but instead remains proportional to $\tau^{2}$ as $\alpha$ increases.

The expression (\ref{Rnearzerooriginal}) for the short-time behavior of the Mat\'ern autocovariance may be simplified by noting 
\begin{equation}
c_\alpha \frac{1}{2^{2\alpha-1}}\frac{\Gamma\left(\frac{3}{2}-\alpha\right)}{\Gamma\left(\alpha+\frac{1}{2}\right)} = \frac{1}{2}V_\alpha
\end{equation}
which relates $V_\alpha$, the coefficient of fractional Brownian motion defined in (\ref{symmetricform}), to $c_\alpha$, the normalizing constant for the Mat\'ern process defined in (\ref{cdef}).  These two definitions together with the duplication formula for the gamma function (\ref{duplication}) presented in Appendix~\ref{coefficientappendix} lead to the above result.  Substituting this into the asymptotic expansion (\ref{Rnearzerooriginal}) for small $|\tau|$, we obtain for $1/2<\alpha<3/2$
\begin{equation} \label{Mlookslikefbm}
R_{zz}^{M}(\tau)\approx \sigma^2 - \frac{1}{2} V_\alpha A^2 |\tau|^{2\alpha-1},\quad\quad |\tau|\ll 1/\lambda
\end{equation}
after making use of the expression for the Mat\'ern variance given by (\ref{maternkappa}).
This matches exactly the $\tau$-dependence inferred for a power-law spectrum inferred in Appendix~\ref{fbmspectrumappendix} using a limiting argument.  Note that the only dependence on $\lambda$ of the autocovariance  for $|\tau|\ll 1/\lambda$ is through the variance $\sigma^2$.

From this small-$\tau$ expansion, we can immediately determine the fractal dimension, as discussed in Section~\ref{fractalsection}.  One finds
\begin{equation} \label{materndimension}
D = \left\{\begin{array}{ccc} \frac{5}{2}-\alpha &&\alpha<3/2\\
1 &&\alpha\ge 3/2
\end{array}\right.
\end{equation}
so that the fractal dimension decays from $D=2$, for very rough processes with $\alpha=1/2$, to $D=1$, for smooth processes with $\alpha=3/2$, just as with fractional Brownian motion.  For slopes steeper than $\omega^{-3}$, the fractal dimension remains at unity.  This is a consequence of the fact that for $\alpha>3/2$, the highest power of $\tau$ appearing in the small-$\tau$ expansion  (\ref{Rnearzerooriginal}) is $\tau^{2}$.

\subsection{Inclusion of spin}\label{additionofspin}

A very simple modification can expand the range of possibilities of the Mat\'ern process, and also aid in the development of physical intuition.  We add a deterministic tendency for the process to spin on the complex plane at rate $\Omega$, and refer to this new process as the oscillatory Mat\'ern process or oMp.  Modulating the Mat\'ern autocovariance $R^{M}_{zz}(\tau)$ by $e^{\mathrm{i} \Omega\tau}$ gives
\begin{align}
R^{o\!\!\:M\!\!\:p}_{zz}(\tau)&\equiv e^{\mathrm{i} \Omega\tau} R^{M}_{zz}(\tau)\label{maternacvsrot}\\
S^{o\!\!\:M\!\!\:p}_{zz}(\omega)& =\frac{A^2}{\left[(\omega-\Omega)^2+\lambda^2\right]^{\alpha}}\label{maternspectrumrot}
\end{align} 
for the new autocovariance function / spectrum pair.  Note that with $\alpha=1$, these reduce to
\begin{align}\label{maternalphaoneR}
R^{o\!\!\:M\!\!\:p}_{zz}(\tau)&= \frac{A^2}{2\lambda}\,e^{\mathrm{i} \Omega\tau} e^{ -\lambda|\tau|}\\
S^{o\!\!\:M\!\!\:p}_{zz}(\omega)&=\frac{A^2}{(\omega-\Omega)^2+\lambda^2}
\label{maternalphaone}
\end{align} 
where we have made use of 10.2.17 on p.~444 of \citet{abramowitz} for the former equality.  These are observed to be the autocovariance and spectrum of the complex-valued oscillator known as the complex Ornstein-Uhlenbeck process \citep{jeffreys42-mnras,arato99-cma}.  

Thus the oscillatory Mat\'ern process subsumes the Mat\'ern process and the complex Ornstein-Uhlenbeck process into a larger family.  In this next section we will determine the stochastic integral equation of this oscillatory Mat\'ern process.  

\subsection{Stochastic integral equation}\label{sie}

Unlike fractional Brownian motion, the Mat\'ern process is not generally defined in terms of a stochastic integral equation or a stochastic differential equation. A stochastic integral equation that will generate an oscillatory Mat\'ern process is 
\begin{equation}\label{greenfunctionexpression}
z(t) = A \int_{-\infty}^{\infty} g(t-s) \,\mathrm{d} W(s)
\end{equation}
where the Green's function, or impulse response function, is 
\begin{equation}\label{Greenfunctiontime}
g(t) \equiv \left\{ \begin{array}{lll} \frac{1}{\Gamma(\alpha)} \, t^{\alpha-1} e^{\mathrm{i} \Omega t} e^{-\lambda t}, & & t \ge 0 \\ 0, &\quad & t < 0 
\end{array}\right.  .
\end{equation}
Note that the Green's function has been set to vanish before time $t=0$, thus corresponding to a causal filter.  

The Fourier transform of a Green's function $g(t)$ is an important quantity known as the \emph{transfer function}, and we find
\begin{equation}\label{Greenfourier}
G(\omega) = \int_{-\infty}^\infty g(t) e^{\mathrm{i} \omega t} \mathrm{d} t\,
=\frac{1}{\left[\mathrm{i} (\omega- \Omega) +\lambda  \right]^\alpha}
\end{equation}
for the Mat\'ern transfer function, using 3.2.3 on p.~118 of \citet{bateman}.  In terms of the Green's function, the autocovariance function is given by
\begin{multline}
R_{zz} (\tau)\equiv \mathrm{E}\left\{z(t) z^*(t-\tau) \right\}  \\=A^2\int_{-\infty}^{\infty} \int_{-\infty}^{\infty} 
 g(t-s) g^*(t-\tau-r) \,\mathrm{E}\left\{ \mathrm{d} W(s)\, \mathrm{d} W^*(r)\right\}\\
 =A^2 \int_{-\infty}^{\infty} 
 g(s) g^*(s-\tau)\, \mathrm{d} s \label{greenfunctionexpression4}
\end{multline}
with the last expression following from the orthogonality property of the Wiener increments (\ref{orthogonalwiener}), together with a change in the variable of integration. From the familiar cross-correlation theorem
\begin{equation}\label{greenfunctionexpression5}
 \int_{-\infty}^{\infty} 
 g(s) g^*(s-\tau)\, \mathrm{d} s = \frac{1}{2\pi} \int_{-\infty}^{\infty} 
 \left|G(\omega) \right|^2 e^{\mathrm{i} \omega \tau} \,\mathrm{d} \omega
 \end{equation}
it then follows that spectrum of the process generated using the Green's function (\ref{Greenfunctiontime}) matches that for the oscillatory Mat\'ern process (\ref{maternspectrumrot}).\footnote{As an aside, we point out that this result implies that with $\Omega=0$, the cross-correlation of $g(t)$ with itself as in (\ref{greenfunctionexpression4}) must recover the Bessel function form of the Mat\'ern autocovariance function, although this is not at all obvious in the time domain.}

Examples of the Green's functions for $\Omega=0$ are shown in Fig.~\ref{threepanel}c.  Note a change in behavior across $\alpha=1$.  For higher values of $\alpha$, the Green's function vanishes at $\tau=0$, thus developing a maximum that is seen to shift away from the origin as one increases $\alpha$.  For $\alpha<1$, however, a singularity develops at the origin, and the Green's function monotonically decays with increasing time.

Identifying this stochastic integral equation sheds light on the nature of the Mat\'ern process itself.  The Green's function $g(t)$ defined in (\ref{Greenfunctiontime}) is also the solution to an impulse forcing of the damped fractional oscillator equation 
\begin{equation}\label{maternoscillator}
\left[ \frac{\mathrm{d}}{\mathrm{d} t}+\lambda - \mathrm{i} \Omega \right]^\alpha g(t) = \delta(t)
\end{equation}
as shown in Appendix~\ref{materngreen}.  This establishes the physical interpretation of the oscillatory Mat\'ern process as a \emph{damped fractional oscillator} forced by continuous-time white noise.  The standard Mat\'ern process is then seen as a forced/damped fractional oscillator in which the oscillation frequency is set to zero.  

Note that here we have avoided attempting to write the Mat\'ern process as a stochastic \emph{differential} equation, as there are mathematical difficulties in ensuring that the fractional-order derivatives exist.\footnote{The expansion of the fractional-order operator in (\ref{maternoscillator}) using the generalized binomial theorem, see (\ref{binomialexpansion}), involves infinitely many higher-order derivatives; but their existence conflicts with self-similar roughness of the Mat\'ern process as one proceeds to increasingly small scales.}  The approach we have taken, comparing the impulse response function  (\ref{Greenfunctiontime})  for the Mat\'ern stochastic integral equation (\ref{greenfunctionexpression}) with that for the deterministic fractional differential equation (\ref{maternoscillator}), is intended to determine the physical nature of the system while sidestepping such mathematical difficulties.

We can also now understand why $\lambda$ is referred to as a `damping'.  In the $\alpha=1$ case, the oscillatory Mat\'ern process becomes identical to the complex Ornstein-Uhlenbeck process, as previously mentioned.  The Green's function for this process is $\mathrm{e}^{\mathrm{i}\Omega t-\lambda t}$ for non-negative $t$, and zero elsewhere.  This Green's function is also the solution to the first-order ordinary differential equation 
\begin{equation}
\frac{\mathrm{d}}{\mathrm{d} t} g(t) + \lambda g(t) -\mathrm{i} \Omega g(t)  = \delta(t) 
\end{equation}
which is the equation for a damped, one-sided oscillator forced by a delta function.  This equation appears, for example, in the study of oscillations of the ocean surface layer forced by the wind \citep{pollard70-dsr}, in which $\lambda$ parameterizes a physical drag.  In the Green's function, $\lambda$ sets the timescale of the decay of the oscillations, and it therefore also controls the decorrelation time in the autocovariance function (\ref{maternalphaoneR}).  In the spectrum (\ref{maternalphaone}), $\lambda$ removes the singularity at $\omega=\Omega$, replacing it with a `bump' that becomes more spread out as $\lambda$ increases.

All of these factors support interpreting $\lambda$ as a damping for $\alpha=1$.  For other values of $\alpha$, we see that $\lambda$ still controls the decay of the Green's function (\ref{Greenfunctiontime}), the long-term decay (\ref{maternacvslongtime}) of the autocovariance function (\ref{maternacvsrot}), and the spreading out of the singular peak at $\omega=\Omega$ in the spectrum (\ref{maternspectrumrot}).   In the fractional differential equation (\ref{maternoscillator}) as well, $\lambda$ appears as a quantity that can trade off against the rate of change. Thus, for $\alpha\ne 1$, the parameter $\lambda$ still acts in a way that supports its identification as a damping.

As shown in the next section, if the damping vanishes, the stochastic integral equation for the Mat\'ern process becomes identical to that for fractional Brownian motion, apart from a modification that sets the initial condition for fBm.

\subsection{Relationship to fractional Brownian motion}\label{relationshiptofBm}

Having identified the stochastic integral equation for the Mat\'ern process, we now examine its relationship with fractional Brownian motion.  The Green's function of the oscillatory Mat\'ern process (\ref{Greenfunctiontime}) can be rewritten as
\begin{equation}\label{Greenfunctiondetail}
g_{\alpha,\lambda,\Omega}(t) \equiv  \frac{I(t)}{\Gamma(\alpha)} t^{\alpha-1} e^{\mathrm{i} \Omega t} e^{-\lambda t}
\end{equation}
where $I(t)$ is the indicator function defined in (\ref{indicator}), and where we explicitly specify the dependence of $g(t)$ upon the Mat\'ern parameters.  In terms of this Green's function, the stochastic integral equation defining fBm (\ref{stochasticintegral2}) becomes
\begin{align}
z(t) &= \frac{A}{\Gamma(\alpha)} 
\int_{-\infty}^t \left[(t-s)^{\alpha-1} - I(-s) (-s)^{\alpha-1}\right] \mathrm{d} W(s)\\
 &= A\int_{-\infty}^t \left[g_{\alpha,0,0}(t-s) - g_{\alpha,0,0}(-s)\right] \mathrm{d} W(s)\label{modifiedmatern}
\end{align}
in which $g_{\alpha,0,0}(t) = \frac{1}{\Gamma(\alpha)} I(t)\, t^{\alpha-1}$.  The only difference between this and the equation for the undamped, non-oscillatory Mat\'ern process (\ref{greenfunctionexpression}) is the second term in the integral, which as shown earlier, serves the function of enforcing the initial condition $z(0)=0$. This confirms that the standard Mat\'ern process with $\lambda>0$, and consequently with a Green's function of the form $g_{\alpha,\lambda,0}(t) = \frac{1}{\Gamma(\alpha)} I(t)\, t^{\alpha-1} \mathrm{e}^{-\lambda t}$, is rightly thought of as \emph{damped} fractional Brownian motion.

If fractional Brownian motion and the standard Mat\'ern processes are essentially facets of the same process, one should be able to see this directly from their autocovariances.  This is indeed the case.  For time shifts $\tau$ that are very small compared to the global time $t$, the fBm autocovariance (\ref{fbmautocovariance}) is approximately given by
\begin{equation}
R_{zz}^{f\!Bm}(t,\tau)\approx \sigma^2(t) - \frac{1}{2} V_\alpha A^2 |\tau|^{2\alpha-1}, \quad\quad|\tau|\ll |t|
\end{equation}
where $\sigma^2(t)\equiv R_{zz}^{f\!Bm}(t,0)=V_\alpha \, A^2 |t|^{2\alpha-1}$ is the time-varying fBm variance encountered earlier in (\ref{fbmvariance}).  This matches (\ref{Mlookslikefbm}) for the Mat\'ern autocovariance at small $|\tau|/\lambda$.  

The intuitive interpretation of this result is that a Mat\'ern process has a second-order structure that behaves for small time offsets $\tau$ in the same way as does fractional Brownian motion, considered for offsets $\tau$ that are small compared with the current global time $t$.  Or, even more succinctly, the \emph{local} behaviors of the Mat\'ern process and fBm are the same; they differ from each other only for sufficiently large time offsets.  

To look at this another way, imagine that a modified Mat\'ern process were constructed with an integral matching the form of that for fractional Brownian motion (\ref{modifiedmatern}).  In other words, we define $z(t)$ as in (\ref{modifiedmatern}) but for arbitrary values of $\lambda$.  Such a process would then by definition have $z(0)=0$, and would therefore not be stationary.  For nonzero $\lambda$, after a sufficiently long time this initial condition is `forgotten' on account of the decaying exponential in the Green's function, and the process will eventually behave as if it were stationary.  For $\lambda=0$, however, this initial condition is never forgotten.  

 The qualitatively significant difference between the Mat\'ern process and fBm---that the former is stationary, while the latter is non-stationary---can be seen as a consequence of the lack of damping in the latter case.  In applications, we believe it would be unphysical to observe a process that remains nonstationary for all timescales.  Rather, for sufficiently long observational periods, it is more likely that the process will eventually settle into stationary behavior.  For the Mat\'ern process, this occurs when the observational window is sufficiently long compared with the decay timescale $\lambda^{-1}$.  Another difference is that the value of fBm at time $t=0$ is fixed to an exact value of zero, while that of the Mat\'ern process is random.  However, since it is common practice to remove the sample mean prior to analyzing a data time series, and/or to add a constant offset to a generated process, this distinction makes little practical difference for applications such as the one presented here.

\section{Generation}\label{Generation}

This section addresses means to simulate realizations of fractional Brownian motion and the Mat\'ern process numerically.  The main contribution is a new approach to simulating a diffusive process such as the Mat\'ern in $O(N \log N)$ operations, by relying on the knowledge of its Green's function.  Readers not interested in these numerical details may feel free to proceed to the application in Section~\ref{Application}.

\subsection{The Cholesky decomposition}

The standard approach to simulating a Gaussian random process with a known covariance matrix is a method called the Cholesky decomposition, which we discuss here.  In this section, as we will be dealing with vectors and matrices, a change of notation is called for.  We now let $z_n\equiv z(n\Delta)$ with integer $n$ denote a discretely sampled random process, sampled at $N$ times separated by the uniform interval $\Delta$.  

This sequence is arranged into a length $N$ random column vector denoted $\mathbf{z}$.  We define the expected $N\times N$ covariance matrix of $\mathbf{z}$ as $\mathbf{R}\equiv \mathrm{E}\left\{\mathbf{z}\mathbf{z}^H\right\}$, where the superscript ``$H$'' denotes the conjugate transpose, having components 
\begin{multline}
R_{m,n} = \mathrm{E}\left\{z_m z^*_{n}\right\} = 
\mathrm{E}\left\{z\left(m\Delta \right)\,z^*\!\left(n\Delta \right)\right\}
\\=R_{zz}\left(n\Delta ,(m-n)\Delta \right).
\end{multline}
Here $n\Delta $ plays the role of global time $t$, and $(m-n)\Delta $ that of the time offset $\tau$, in the evaluation of the nonstationary covariance function $R_{zz}(t,\tau)=\mathrm{E}\left\{z(t+\tau)\,z^*\!(t)\right\}$.  Thus variation in $\mathbf{R}$ of the time offset $\tau$ with fixed global time $t$ occurs in the direction perpendicular to the main diagonal, while variation of $t$ with fixed $\tau$ occurs along the main diagonal.  In the case of a stationary process, there is no variation parallel to the main diagonal, and $\mathbf{R}$ is then said to be a Toeplitz matrix.

The Cholesky decomposition factorizes the covariance matrix as $\mathbf{R}=\mathbf{L} \mathbf{U}$, where $\mathbf{L}$ is lower triangular and $\mathbf{U}$ is upper triangular.  It follows from the Hermitian symmetry of $\mathbf{R}$ that $\mathbf{L}=\mathbf{U}^H\!\!$.  Now let $\mathbf{w}$ be an $N$-vector of unit-variance, independent, complex-valued Gaussian random variables.  Forming the sequence $\hat{\mathbf{z}}=\mathbf{L} \mathbf{w}$, we find the covariance matrix $\widehat{\mathbf{R}}\equiv \mathrm{E}\left\{\hat{\mathbf{z}}\hat{\mathbf{z}}^H\right\}$ associated with $\hat{\mathbf{z}}$ is given by 
\begin{equation}
\widehat{\mathbf{R}}\equiv \mathbf{L}\, \mathrm{E}\left\{ \mathbf{w}\mathbf{w} ^H\right\} \mathbf{L}^H =
 \mathbf{L} \mathbf{I} \mathbf{L}^H = \mathbf{R}
\end{equation}
where $\mathbf{I}$ is the $N\times N$ identity matrix.  Note while we could have also chosen to use $\mathbf{U}$ to generate the random sequence, the use of $\mathbf{L}$ is more natural as it corresponds to a causal filter.  

Thus to simulate a length $N$ sequence of a possibly nonstationary Gaussian random process, one simply populates an $N\times N$ matrix with the known values from the autocovariance function, applies the Cholesky decomposition to generate a lower triangular matrix, and multiplies the result by a vector of white noise.  The resulting sequence has the \emph{identical} covariance structure to a length $N$ sample of the original random process.  

A limitation of this approach is that the Cholesky decomposition requires, in its most straightforward implementation, $O(N^3)$ operations.  Computational costs therefore increase steeply with increasing $N$.  However, it is the case that many realizations of sequences of a fixed length can be generated quickly, because one only needs to form the Cholesky decomposition once for a given autocovariance matrix.  For simulation of stationary processes, the Toeplitz matrix structure can in principle be used to accelerate the Cholesky decomposition to $O(N^2)$ or even $O(N\log N)$, see \citet{yagle85-aam} and \citet{dietrich97-sjsc} respectively.  The latter method, termed circulant embedding, while $O(N\log N)$, involves embedding the covariance matrix of interest within a larger matrix, and may lead to somewhat unpredictable tradeoffs between minimizing error and increasing the matrix size \citep{percival06-sp}.  The method presented here has the advantages that it is very straightforward to implement, and that the error terms are well understood provided the Green's function is known.  

\subsection{Discretization effects in fast generation}

To devise our generation method, we will first renormalize the Green's function so that we may use $\sigma$ rather than $A$ to parameterize the process amplitude.  A modified Green's function is defined as 
\begin{equation}\label{Greenfunctiondetail2}
\tilde g(t)=\tilde g_{\alpha,\lambda,\Omega}(t) \equiv  \frac{\lambda^{\alpha-1/2}}{\sqrt{c_\alpha}}\frac{I(t)}{\Gamma(\alpha)} t^{\alpha-1} e^{\mathrm{i} \Omega t} e^{-\lambda t}
\end{equation}
where the subscripts on $\tilde g(t)$ will be dropped unless explicitly needed.  The stochastic integral equation for the Mat\'ern process (\ref{greenfunctionexpression}) then becomes
\begin{equation}\label{approxgreen1pt5}
z(t) = \sigma \int_{-\infty}^\infty \tilde g(t-s) \,\mathrm{d} W(s) 
\end{equation}
recalling that $\sigma$ and $A$ related by $\sigma^2=c_\alpha A^2/\lambda^{2\alpha-1}$.  Next we introduce a temporal spacing $\tilde\Delta\equiv \Delta/k$ that is finer than the sampling interval $\Delta$, where $k$ is a positive integer termed the \emph{oversampling parameter}.  We then have 
\begin{equation}\label{approxgreen2}
z(t) = \sigma  \sum_{p=0}^{\infty}\int_{t-(p+1)\tilde\Delta}^{t-p\tilde\Delta} \tilde g(t-s) \,\mathrm{d} W(s) 
\end{equation}
by splitting the integral in (\ref{approxgreen1pt5}) into contributions from smaller integrals over segments of duration $\tilde \Delta$.  Here we have replaced the upper limit of integration with $t$, as $\tilde g(t-s)$ vanishes for negative values of its argument.

For each of these integrals over a short segment, we approximate the Green's function by a constant, namely the value of the Green's function at the segment midpoint, which occurs when $t-s=(p+1/2)\tilde\Delta$.  Employing this approximation and evaluating the result at the discrete times $t=n\Delta$ \emph{defines} a discrete series
\begin{equation}\label{approxgreen1}
\tilde z_n \equiv \sigma \sum_{p=0}^{\infty} \tilde g\left((p+1/2)\tilde\Delta\right) \int_{n\Delta-(p+1)\tilde\Delta}^{n\Delta-p\tilde\Delta} \,\mathrm{d} W(s)
\end{equation}
for all integers $n=-\infty,\ldots,-2,-1,0,1,2,\ldots \infty$.  Because $\int_{a}^{b} \,\mathrm{d} W(s)$ is a zero-mean Gaussian random variable with variance $(b-a)$, the integral in the above expression simplifies~to 
\begin{equation}\label{approxgreen2}
 \int_{n\Delta-(p+1)\tilde\Delta}^{n\Delta-p\tilde\Delta} \,\mathrm{d} W(s)=
 \int_{(nk-p-1)\tilde\Delta}^{(nk-p)\tilde\Delta} \,\mathrm{d} W(s)= w_{nk-p} \,\sqrt{\tilde\Delta}
\end{equation}
where $w_n$, defined for integer $n$, is a sequence of complex-valued, unit variance, independent Gaussian random variables.  Introducing an oversampled version of the discrete Green's function as
\begin{equation}
\tilde g_n^{\{k\}}\equiv \tilde g\left((n+1/2)\Delta/k\right) 
\end{equation}
our expression (\ref{approxgreen1}) for $\tilde z_n$ becomes
\begin{equation}\label{approxgreen3}
\tilde z_n=\sigma\sqrt{\frac{\Delta}{k}} \sum_{p=0}^{\infty}  \tilde g_{p}^{\{k\}} w_{nk-p}.
 \end{equation}
This is a discrete convolution, but modified by the fact that the output will have a temporal resolution that is $k$ times more coarse than that of the two input series. 

The numerical evaluation of the oversampled Green's function can be simplified by noting the behavior of $\tilde g(t)$ with respect to a rescaling of the time axis by some factor $r$,
\begin{multline}\label{Greenfunctiondetail3}
\tilde g_{\alpha,\lambda,\Omega}(rt) = \frac{\lambda^{\alpha-1/2}}{\sqrt{c_\alpha}}\frac{I(rt)}{\Gamma(\alpha)} (rt)^{\alpha-1} e^{\mathrm{i} \Omega rt} e^{-\lambda rt}\\= \frac{1}{\sqrt{r}}\,\tilde g_{\alpha,\lambda r,\Omega r}(t).
\end{multline}
Then the Green's function $\tilde g_n^{\{k\}}$ in (\ref{approxgreen3}) can be rewritten as
\begin{multline}
\tilde g_n^{\{k\}}= \tilde g_{\alpha,\lambda,\Omega}\left((n+1/2)\Delta/k\right) 
\\=\sqrt{\frac{k}{\Delta}\,} \tilde g_{\alpha,\lambda\Delta/k,\Omega\Delta/k}\left(n+1/2\right)
\end{multline}
which replaces the time rescaling with a rescaling of the damping $\lambda$ and frequency shift $\Omega$, resulting in a cancellation of the factor $\sqrt{\Delta/k}$.

The autocovariance function of $\tilde z_n$ is very close to the sampled autocovariance function of the Mat\'ern process, and can be made arbitrary close by a suitable choice of oversampling rate $k$, as will now be shown.  The autocovariance sequence associated with $\tilde z_n$ is found to be 
\begin{multline}\label{Greenfunctiontimezz}
\widetilde R_{n}\equiv \mathrm{E}\left\{\tilde z_m \tilde z_{m-n}^* \right\}
= \sigma^2 \frac{\Delta}{k}\times \\
\sum_{p=0}^\infty\sum_{q=0}^\infty \tilde g_{p}^{\{k\}}\,\left[\tilde g_{q}^{\{k\}}\right]^* \,\mathrm{E}\left\{w_{mk-p} \,w_{(m-n)k-q}^*\right\}
\end{multline}
and since $\mathrm{E}\{w_m w_n^*\} = \delta_{m,n}$ where $\delta_{m,n}$ is the Kronecker delta function, all terms in the summation vanish except for when $mk-p=(m-n)k-q$ or equivalently $q=p-nk$.  Thus 
\begin{equation}\label{Greenfunctiontimezz}
\widetilde R_{n}=\mathrm{E}\left\{\tilde z_m \tilde z_{m-n}^* \right\} 
=\sigma^2 \frac{\Delta}{k}\sum_{p=0}^{\infty} \tilde g_{p}^{\{k\}}\,\left[\tilde g_{p-nk}^{\{k\}}\right]^*
\end{equation}
which is clearly an approximation to (\ref{greenfunctionexpression4}) for an autocovariance function in terms of its Green's function.  The discretely sampled autocovariance sequence can therefore be approximated to arbitrary precision by a choosing a suitable degree of oversampling.  However, notice that the summations in (\ref{approxgreen3}) and (\ref{Greenfunctiontimezz}) extend to infinity, which is not possible in practice.  In the next subsection we examine the impact of additional errors resulting from finite sample size effects.

\subsection{Sample size effects in fast generation}\label{finitesize}

In practice, the summations over the duration of the Green's function must be truncated at some point.  It is tempting to truncate the Green's function after a relatively short time.  However, for spectra having a large dynamic range, this truncation leads to undesirable leakage effects, just as in spectral analysis, that degrade the spectrum of the generated sequences.  Instead, we will utilize a Green's function that is longer than entire length of the time series. 

Firstly we need to determine a suitable cutoff for limiting the long-term influence of the Green's function.  We denote by $T_\epsilon$ the time such that the magnitude of the Green's function, integrated to this time, rises to within a fraction $\epsilon$ of the value it obtains when integrated over all times:
\begin{equation}\label{Greenfunctiontime1}
\frac{\int_{0}^{\,T_\epsilon} \left|\tilde g(s)\right| \,ds}{\int_{0}^{\infty} \left|\tilde g(s) \right|\,ds } = 1-\epsilon.
\end{equation}
Using the definition of the Mat\'ern Green's function (\ref{Greenfunctiontime}), one may readily show that this occurs when 
\begin{equation}\label{Greenfunctiontime2}
\frac{\gamma\left(\alpha,\lambda T_\epsilon\right)}{\Gamma(\alpha)} = 1-\epsilon, \quad\quad \gamma(\alpha,t)\equiv \int_{0}^{\,t} s^{\alpha-1} e^{-s} \,ds.
\end{equation}
where $\gamma(\alpha,t)$ is the incomplete gamma function of order $\alpha$ evaluated at time $t$.  

 Anticipating transforming to the Fourier domain, we will define sequences that are periodized.  Because we intend to employ a periodic convolution, yet wish to prevent noise values at the end of the time series from influencing the beginning, we will create a longer sequence of length $\widehat N \equiv N+N_\epsilon$ where $N_\epsilon\equiv \mathrm{ceil}(T_\epsilon/\Delta)$ with $\mathrm{ceil}(\cdot)$ being the ceiling function.  Let $ \widehat w_{n}$ be a version of the noise that is periodic with period $\widehat N$, and $ \widehat g_{n}^{\{k\}}$ be a version of $ \tilde g_{n}^{\{k\}}$ that is set to zero for $n>\widehat N-1$.  Form a length-$\widehat N$ vector $\hat{\mathbf{z}}$ with entries given by
\begin{equation}\label{periodiccorrelation}
\hat z_n \equiv \sigma\sqrt{\frac{\Delta}{k}\,} \sum_{p=0}^{\widehat N-1} \widehat g_{p}^{\{k\}} \widehat w_{nk-p}
 \end{equation}
and now decompose this vector into two parts, $\hat{\mathbf{z}} = [ \hat{\mathbf{z}}_\epsilon \,\,\,\, \hat{\mathbf{z}}_o]^T$ where the superscript ``$T$'' is the transpose operator.  In the initial portion $ \hat{\mathbf{z}}_\epsilon$, of length $N_\epsilon$, the decaying Green's function is interacting with noise wrapped around from the end of the periodic noise sequence.  This portion is discarded, while the second portion $ \hat{\mathbf{z}}_o$ is of length $N$ and is the simulated series we desire.  

The $N\times N$ covariance matrix associated with the latter sequence, $\widehat{\mathbf{R}} = \mathrm{E}\left\{\hat{\mathbf{z}}_o \hat{\mathbf{z}}_o^H\right\}$, has components given by
\begin{multline}
\widehat R_{m,n}=
\sigma^2\frac{\Delta}{k} \times \\
\sum_{p=0}^{\widehat N-1} \sum_{q=0}^{\widehat N-1} \widehat g_{p}^{\{k\}} \left[\widehat g_{q}^{\{k\}}\right]^* \mathrm{E}\left\{ \widehat w_{mk-p+kN_\epsilon }\label{Rmnmatrix}
\widehat w_{nk-q+kN_\epsilon }^*\right\}.
 \end{multline}
To simplify this expression, observe that the covariance of the periodized noise sequence~$\widehat w_n$~is 
 \begin{equation}\label{ellsum}
 \mathrm{E}\{\widehat w_{m} \widehat w_{n}^*\}=\sum_{\ell=-\infty}^\infty\delta_{m,n+\ell\widehat N}
 \end{equation}
 with the sum indicating that the periodized noise is correlated with copies of itself from the future and the past.  Thus in (\ref{Rmnmatrix}), all terms vanish except for when $mk-p=(nk-q)+\ell\hat N$ or equivalently $q=p-(m-n)k+\ell\hat N$.  We then have
 \begin{equation}
\widehat R_{m,n} \label{rhat}
=\sigma^2\frac{\Delta}{k} \sum_{p=0}^ {\widehat N-1} \widehat g_{p}^{\{k\}} \left[\widehat g_{p-(m-n)r}^{\{k\}} + \widehat g_{p-(m-n)r+\hat N}^{\{k\}}\right]^* 
 \end{equation}
for the terms in the $N \times N$ covariance matrix $\widehat{\mathbf{R}}$. Note that this consists only of the $\ell=0$ and $\ell=1$ terms from (\ref{ellsum}).  The first term in (\ref{rhat}) is due to the $\ell=0$ term.  The second ($\ell=1$) term arises from the Green's function interacting with a copy of itself shifted by $\hat N$ due to the periodization of the noise, and is expected to be much smaller than the first term. Note that contributions from negative $\ell$ do not appear due to the fact that $\widehat g_{n}^{\{k\}}$ vanishes for negative $n$; but all contributions from $\ell>1$ also vanish because $\widehat g_{n}^{\{k\}}$ has been truncated to vanish for $n>\widehat N-1$.
 
\subsection{Comparison of fast and Cholesky methods}

The advantage to the Green's function approach is that (\ref{periodiccorrelation}) is a discrete, periodic convolution that can be implemented using a Fast Fourier Transform in $O(\widehat N \log \widehat N)$ operations; if $\widehat N\approx N$, this is approximately $O(N \log N)$.  In the numerical implementation described in Appendix~\ref{jlabappendix}, we set $\epsilon=0.01$, such that $T_\epsilon$ gives the time at which the time-integrated Green's function reaches one percent of its total time-integrated magnitude.  We also set the oversampling parameter $k$ such that there will be at least 10 points per damping timescale $\lambda^{-1}$, which is accomplished by choosing $k=\mathrm{ceil}\left(10\times\lambda\Delta\right)$ since $1/(\lambda\Delta)$ is the number of sampled points in one damping timescale.  These settings are observed to give fast but accurate performance for a broad range of parameters. 

If desired, the matrix $\widehat R_{m,n}$ in (\ref{rhat}) can be computed in order to explicitly check the errors in computing the covariance matrix, although this will necessarily slow down the algorithm.  The terms in the true, discretely sampled autocovariance matrix are given exactly~by 
\begin{multline}\label{rmnsplit}
R_{m,n} 
 = \sigma^2 \int_{0}^{\hat T}
 \tilde g(s)  \tilde g^*\left(s-(m-n)\Delta\right) ds \\
  + \sigma^2 \int_{\hat T}^\infty
 \tilde g(s)  \tilde g^*\left(s-(m-n)\Delta\right) ds 
 \end{multline}
where $\hat T=(\hat N-1) \Delta$; this follows from the form of the Mat\'ern autocovariance function in terms of the Green's function (\ref{greenfunctionexpression4}).  We may observe that discretizing the first integral corresponds to the first summation in (\ref{rhat}).  There are therefore three error terms between $R_{m,n}$ and $\widehat R_{m,n}$: errors associated with this discretization, which are minimized by choosing the oversampling rate $k$ to be sufficiently large; and errors from the second integral in (\ref{rmnsplit}) and the second summation in (\ref{rhat}), both of which are minimized by choosing $N_\epsilon$ sufficiently large.  Thus error can be computed by comparing the difference between the true discretely sampled autocovariance matrix $R_{m,n}$ and the autocovariance matrix $\widehat R_{m,n}$ that is satisfied by the process generated through the Green's function method.  While this is numerically expensive, it need only be computed one time for a given set of parameters $\alpha$, $\lambda$, $N$, $k$, and~$\epsilon$.

\begin{figure*}[t!]
	\includegraphics[width=\textwidth,angle=0]{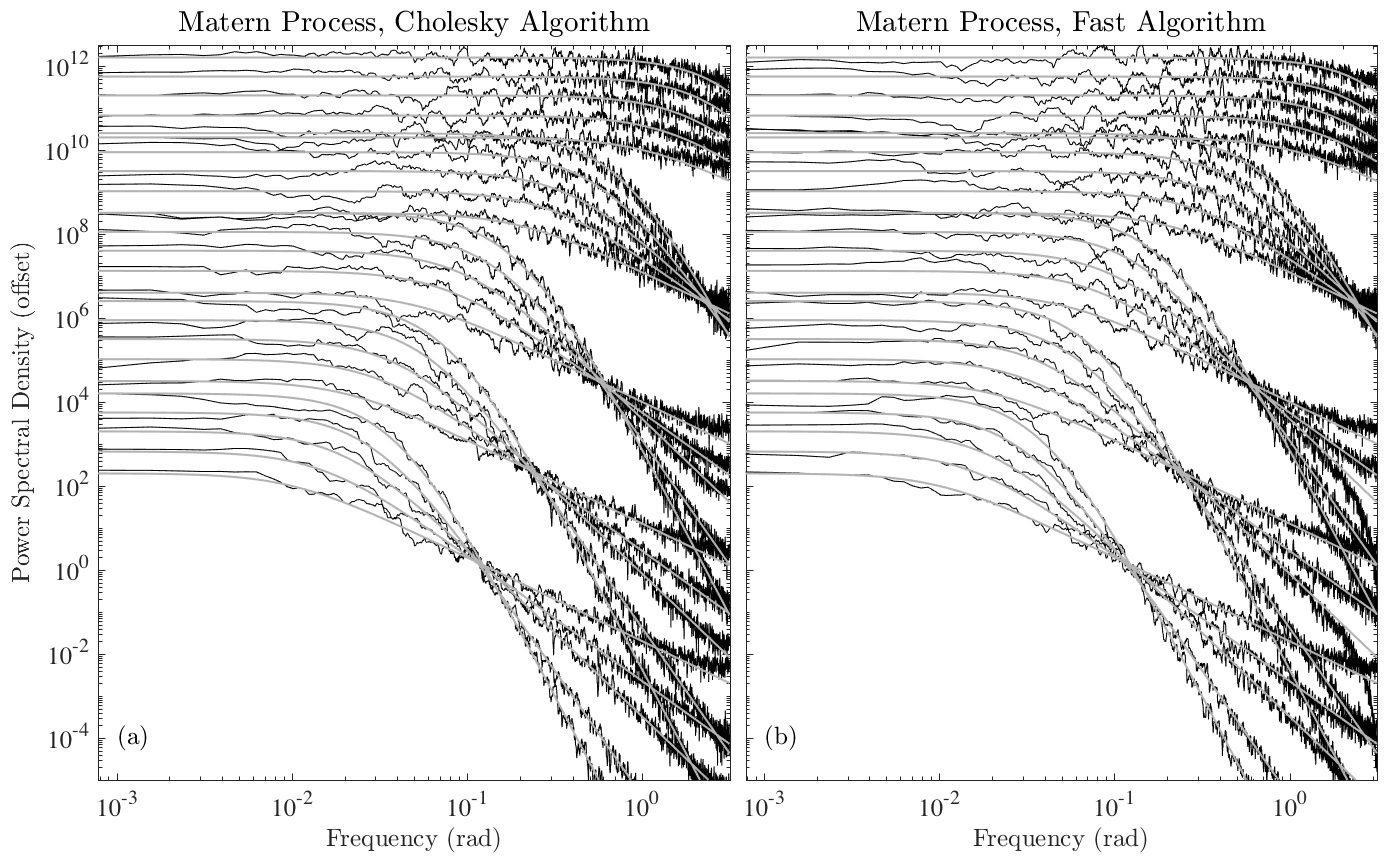}
	\caption{\small A comparison of the spectra of simulated unit-variance Mat\'ern process having twenty-five different $(\alpha,\lambda)$ values for (a) the Cholesky decomposition algorithm and (b) the fast generation algorithm presented based on the Green's function.  The process samples are each 8000 points long, with the sample interval $\Delta$ set to unity.  Black curves show the multitaper spectral estimates, as described in the text, while gray curves are the theoretical spectral forms.  Successive spectral plots have been offset in the vertical by a factor of $\sqrt{10}$ for presentational clarity.  The five lines within each group correspond to the five $\alpha$ values 1, 1.5, 2, 3, and 4.  The five groups correspond to different values of $\lambda$, with $\lambda$ equal to 0.01, 0.02, 0.05 0.2, or 1 times the value of $\alpha$ for each curve, proceeding from bottom to top.  Only positive frequencies are shown, as the theoretical spectra at negative frequencies are identical.  Simulated spectra from the $O(N \log N)$ fast algorithm and those from the $O(N^3)$ Cholesky algorithm are found to be virtually identical.}
	\label{algorithms}
\end{figure*}

As an example, in Fig.~\ref{algorithms} we present spectra of 25 samples of Mat\'ern processes generated using both the Cholesky decomposition and the fast Green's function algorithm.  The estimated spectrum for each realization is computed using Thomson's adaptive multitaper algorithm \citep{thomson82-ieee,park87b-jgr} using 15 orthogonal Slepian tapers having a time-bandwidth product of eight.  The adaptive algorithm employs frequency-domain smoothing only to the extent that it can be achieved without the expense of broadband bias.  

No substantial difference between spectra computed with the two different algorithms is seen over many decades of structure, indicating that fast algorithm is able to simulate the Mat\'ern process to a very high degree of accuracy.  In generating this plot for time series of length 1000, 2000, 4000, and 8000 points (as shown here), the Green's function method executes respectively 3, 7, 11, and 45 times faster than the Cholesky algorithm on a Mac desktop.  Note that the Green's function method does not depend on any special properties of the Mat\'ern process, apart from the particular definition of the cutoff time $T_\epsilon$ for the initial time period (\ref{Greenfunctiontime1}).  The method is therefore suitable for any Gaussian random process having a decaying and sufficiently smooth autocovariance for which the Green's function has an analytic expression.  A more detailed comparison between the Green's function method of generation, and other methods such as circulant embedding \citep{dietrich97-sjsc,percival06-sp}, is outside the scope of this paper, and is a natural direction for further work.

\section{Application}\label{Application}

This section presents the details of an application of the Mat\'ern process to modeling particle velocities in a numerical simulation of two-dimensional fluid turbulence, a preview of which was presented in Section~\ref{applicationpreview}. Details of the numerical model are given in Section~\ref{modelsection}, the estimation of parameter values is discussed in Section~\ref{likelihoodsection}, and 
the means by which realizations of the stochastic models are obtained is described in Section~\ref{stochasticrealizations}.

\subsection{Numerical simulation of 2D turbulence}\label{modelsection}

A system called \emph{forced-dissipative quasigeostrophic turbulence} is created by integrating an equation for the streamfunction $\Phi(x,y,t)$.  For nondivergent flows, the streamfunction is a scalar-valued quantity at each point giving the velocity components through $U(x,y,t)=-\frac{\partial}{\partial y}\Phi$ and $V(x,y,t)=\frac{\partial}{\partial x}\Phi$.  The equation to be integrated is 
\begin{equation}\label{vorticityequation}
\frac{\partial}{\partial t} \left(\nabla^2\Phi - \Phi/L_D^2 \right) + J(\Phi,\nabla^2\Phi) = F -D
\end{equation}
where $J(a,b)\equiv\frac{\partial a}{\partial x} \frac{\partial b}{\partial y} -\frac{\partial b}{\partial x}\frac{\partial a}{\partial y} $ is the Jacobian operator, $L_D$ is a spatial scale termed the deformation radius, $F$ is a forcing function, and $D$ is a damping.  This equation is derived from a conservation law following particle trajectories.  This simple system is considered an idealized representation of turbulence in planetary fluid dynamics, on scales large enough that the rotation of the planet is important, but not so large that the planet's curvature needs to be taken into account.

An integration of (\ref{vorticityequation}) is carried at $1024^2$ resolution in a doubly periodic domain of dimension 2500 $\times$ 2500~km.  As is typical in such problems, the forcing $F$ consists of random fluctuations of a particular spatial scale imposed everywhere in the domain at each time step.  A characteristic forcing scale of 117~km is chosen here such that the scale of the forcing is intermediate between the grid scale and the domain scale.  The damping is chosen to take the form $D=r\nabla^2\Phi$ where $r$ is set to $1.5 \times 10^{-8}$~s$^{-1}$. After an initial spin-up period, during which an equilibration of the energy level is achieved, the simulation is run for three years or 3*365=1095 days.  

A snapshot of current speed from the first day of the simulation after the end of the spin-up period is shown in the left panel of Fig.~\ref{snapshot}.  As mentioned previously, the circular areas of high speed represent long-lived vortices  \citep[see e.g][]{mcwilliams90b-jfm,scott13-jfm}, which are not the subject of this study.  Instead we are interested in the behavior of particles that inhabit the spaces between the vortices.  

The analysis here is based on a set of 1024 particle trajectories that are tracked throughout this experiment, shown in the right panel of Fig.~\ref{snapshot}.  The trajectories are output at high temporal resolution, decimated to a six hour sampling interval, and first central differenced to produce velocities.  Position and velocity records are then decimated again to daily resolution, which we find to be sufficient to capture meaningful variability.  One-half of the trajectories are then discarded in order to exclude those most directly effected by vortices, as described next, leaving 512 trajectories of length 1095 to be analyzed.  
 
The simplest way to remove the effects of vortices is simply to discard those trajectories which conspicuously exhibit the effects of vortex trapping.  A common measure of the impact of vortices on a trajectory is the so-called spin parameter \citep{sawford99-blm,veneziani05b-jmr,veneziani05a-jmr}, defined as
\begin{equation}
 \overline\Omega \equiv \frac{\overline{u(t) \frac{\mathrm{d}}{\mathrm{d} t}v(t) - v(t) \frac{\mathrm{d}}{\mathrm{d} t}u(t)}}{\overline{u^2(t) +v^2(t)}} = 
 \frac{\overline{\Im\left\{z^*(t) \frac{\mathrm{d}}{\mathrm{d} t}z(t) \right\}}}{\overline{|z(t)|^2}}
\end{equation}
in which ``$\Im$'' is the imaginary part.  In our implementation, these time derivatives are adequately approximated by first central differences at daily resolution.  The overbar here is a temporal average over the extent of a trajectory; note that since the mean velocity is zero, the denominator is the velocity variance along the trajectory.  

We take the modulus of the time-averaged spin, $|\overline\Omega|$, as a measure of the overall impact of vortices.  Because of the long-term persistence of particles within vortices \citep[see e.g.][]{pasquero02-prl}, it is unlikely that a small value of $|\overline\Omega|$ would result from cancellation of positive and negative contributions within the same time series for the three-year lengths we consider.  Conservatively, we keep the half of the 1024 trajectories having the lower values of spin magnitude.  The resulting 512 trajectories, offset to begin at the origin in Fig.~\ref{dispersion}a, exhibit a meandering character in addition to their dispersion.  The omitted trajectories  typically present a dense and regular looping structure, some of which may be seen in the right-hand panel of Fig.~\ref{snapshot}.  

\subsection{Frequency-domain maximum likelihood}\label{likelihoodsection}

This section describes the method by which the Mat\'ern parameters are estimated from a finite data sample, which necessitates some new notation.  In reality one only observes a random process $z(t)$ at a finite set of discrete times $z[n]=z(n\Delta)$ separated by the fixed time interval $\Delta$, and with $n=0,1,2,\ldots,N-1$.  In this subsection, we use square brackets for time series which take discrete arguments, thereby distinguishing a discretely sampled time series $z[n]$ from its continuous-time analogue $z(t)$.  Based on this sample $z[n]$, one wishes to estimate the parameters of stochastic model, conventionally denoted by the vector $\bm \theta$, which in the case of the Mat\'ern model is $\bm\theta=(\sigma,\alpha,\lambda)$.

A standard approach would be to form a parametric estimate using the maximum likelihood method implemented in the time domain.  However, this method involves a computationally expensive matrix inversion, which becomes a limiting factor when analyzing large datasets.  An alternative approach to estimating the parameters is to do so in the frequency domain using a method called the Whittle likelihood \citep{whittle53-afm}.  This approach is considerably faster than time-domain maximum likelihood, with $O(N\log N)$ versus $O(N^2)$ computational cost, yet is known to give approximately the same results.  It also has the advantage of letting us only fit the parametric model over a specified band of frequencies.  The Whittle likelihood method proceeds as follows. The discrete Fourier transform of the length $N$ sequence $z[n]$ is given by
 \begin{equation}\label{Zdef}
Z[m]\equiv\sum_{n=0}^{N-1} z[n]\, e^{-\mathrm{i} 2\pi m n/N}
\end{equation}
for $m=0, 1, 2, \ldots,(N-1)$.  The squared modulus of this sequence of $N$ Fourier coefficients, renormalized by $1/N$, defines a spectral estimate known as the \emph{periodogram} 
\begin{equation}\label{periodogram}
\widehat S_{zz}[m]\equiv \frac{1}{N}\left|Z[m]\right|^2.
\end{equation}
This is to be compared with the discretely sampled theoretical spectrum (\ref{newmaternspectrum}) for a particular value of the parameters $\bm\theta$
\begin{equation}
S_{zz}^{\bm\theta}[m]= S^{M}_{\!zz}\left(\frac{2\pi m}{N\Delta}\right)
=\frac{\lambda^{2\alpha-1}}{c_{\alpha}}\,\frac{\sigma^2}{
\left[\left(\frac{2 \pi m}{N\Delta}\right)^2+\lambda^2\right]^{\alpha}}
\end{equation}
where $2\pi m /(N\Delta)$ is recognized as the $m$th Fourier frequency.  

The model parameters are estimated by finding the value of $\bm\theta$ that maximizes the so-called Whittle log-likelihood 
\begin{equation}\label{whittle}
\ell(\bm\theta) = -\sum_{m\in\mathcal{F}} \left\{ \ln S_{zz}^{\bm\theta}[m]
+\frac{\widehat S_{zz}[m]}{ S_{zz}^{\bm\theta}[m]} \right\}
 \end{equation}
in which $\mathcal{F}$ is a set of integers indicating the Fourier frequencies over which the fit is to be applied.  For example, $\mathcal{F}$ could be chosen to be $m=0, 1, 2, \ldots,(N-1)$, in which case the fit will be applied to all frequencies.  

In turns out to be the case that in the inference of parameters for a steep spectrum, such as we are dealing with here, this approach is inadequate as it ignores potentially significant effects associated with the finite sample size.  In particular, \emph{spectral blurring} associated with the periodogram can lead to quite incorrect slopes at high frequencies.  Instead we use the \emph{de-biased} Whittle likelihood method recently developed by \citet{sykulski16-arxiv}.  In that approach, the periodogram $\widehat S_{zz}[m]$ in (\ref{whittle}) is replaced with a \emph{tapered} spectral estimate, and the theoretical spectrum $S_{zz}^{\bm\theta}[m]$ is replaced with the \emph{expected} tapered estimate for a Mat\'ern process characterized by the particular value of $\bm\theta$.  The de-biased Whittle likelihood allows the parameters $\bm\theta$ to be more accurately estimated, as it correctly accounts for the effect of spectral leakage as well as aliasing.  

\subsection{Stochastic model realizations}\label{stochasticrealizations}

 
Here we give details on how the realizations shown in Fig.~\ref{dispersion}b--d have been created.  First, in preparing Fig.~\ref{spectra}, tapered spectral estimates as well as periodogram estimates are formed.  As discussed in Section~\ref{applicationpreview}, for data tapers we use the lowest-order Slepian taper \citep{slepian78-bell,thomson82-ieee,park87b-jgr,sapa} with the time-bandwidth product set to  10.  The average over all time series, and over both sides of the frequency spectrum, are shown for both estimates.  In contrast with the tapered estimates, the periodogram (not shown) is seen to accurately estimate the spectrum over only about half of the dynamic range.  This fact illustrates the potentially severe problems with using the standard Whittle likelihood for parameter inference involving steep spectra, and motivates our use of the de-biased method.

After forming the tapered spectral estimate for each of the 512 turbulence velocity time series, we apply the de-biased Whittle likelihood to infer the best fit Mat\'ern parameters for each time series.  Here the frequency set $\mathcal{F}$ is chosen to include frequencies up to 1.5~radians per day, as this corresponds to the upper limit of apparent structure in the spectra.  For each set of parameters, we generate a realization of  a Mat\'ern process having these properties as described in Section~\ref{Generation}, and then cumulatively sum these velocity time series to produce the trajectories shown in Fig.~\ref{dispersion}b.  Estimation of the spectra for these Mat\'ern realizations in the same manner as for the turbulence data leads to the black dashed line shown in Fig.~\ref{spectra}, which is seen to be a very close match to the velocity spectra for the particle trajectories from the turbulence simulation.

To generate the trajectories shown in Fig.~\ref{dispersion}c and Fig.~\ref{dispersion}d, we proceed as follows.  The parameter values from the fit to the Mat\'ern form are converted to a diffusivity through $\kappa=\frac{1}{4}\sigma^2/(\lambda c_\alpha)$, which is then used to scale realizations of white noise.  The spectra of the associated velocities in Fig.~\ref{spectra}c are seen as matching the low-frequency values of the Lagrangian velocity spectra from our turbulence simulation.  Cumulatively summing these white noise velocities produces the trajectories in Fig.~\ref{dispersion}c; note that these trajectories therefore consist of discrete samples of standard Brownian motion.  These are seen to match well the dispersion characteristics of the turbulence trajectories, but to have far too high a degree of small-scale roughness.

For the power-law realizations, we cannot employ fractional Brownian motion because the observed slopes---which in this simulation is steeper than those found in the ocean---are outside the fBm range.  Instead we use the implied spectral amplitudes $A^2=\sigma^2\lambda^{2\alpha-1}/c_\alpha$ and slope parameters $\alpha$ from the Mat\'ern fit to fix the properties of a different Mat\'ern process having a very small damping value, chosen as $\lambda=2\pi/T$ where $T$ is the record duration.  Realizations are then generated and cumulatively summed to give the trajectories shown in Fig.~\ref{dispersion}d. As mentioned before, these have vastly too much energy on account of extending the high-frequency slope to very low frequencies.  The flattening of the estimated spectrum for these realizations seen in Fig.~\ref{spectra} is a result of the extreme dynamic range hitting the limit of numerical precision. 

The point of the application is to show that Mat\'ern process provides an excellent match to the turbulence data.  This opens the door to investigating a number of interesting physical questions regarding the distributions and interpretations of those parameters, which must, however, be left to the future. 

\section{Discussion}\label{S:Conclusion}

This paper has examined the Mat\'ern process as a stochastic model for time series, which we have shown to be equivalent to damped fractional Brownian motion (fBm).  The damping is shown to be essential for permitting the phenomenon of diffusivity to arise in the temporal integral of the process, referred to here as the \emph{trajectory}, which disperses from its initial location at a constant rate.  The rate of diffusion of the trajectory is given by the value of the spectrum of the process at zero frequency.  At higher frequencies, the spectrum transitions to a power-law slope, like fBm, with the location of this transition being controlled by the damping parameter.  

Because damping is a common feature in physical systems, the Mat\'ern process is expected to be valuable in describing time series which, when observed over shorter time intervals, appear to consists of fractional Brownian motion.  The addition of a spin parameter leads to a still more general process that satisfies the stochastic integral equation for a damped fractional oscillator forced by continuous-time white noise, and that encompasses the standard Mat\'ern process as well as the complex \citep{jeffreys42-mnras,arato99-cma} and standard \citep{uhlenbeck30-pr} Ornstein-Uhlenbeck processes within a single larger family.  A simple algorithm for generating approximate realizations of this `oscillatory Mat\'ern' process in $O(N\log N)$ operations was presented.

A categorization of stochastic processes as \emph{diffusive}, \emph{subdiffusive}, and \emph{superdiffusive} was proposed, depending upon their value at zero frequency.  These categorizations refer to the nature of the dispersion experienced by the trajectory associated with the process, assuming that the integral of the process is well defined.  This categorization is related to, yet distinct from, the conventional designation of a random process as short-memory or long-memory \citep{beran}.  We have argued that the diffusivity categorization may prove to be a powerful way to describe stochastic processes in general.  

The Mat\'ern process was found to provide an excellent match to velocity time series from particle trajectories in forced/dissipative two-dimensional fluid turbulence that are not directly influenced by the presence of vortices.  This is an important contribution, since we show that a power-law process such as fBm cannot hope to capture the diffusive behavior.  Despite its simple three-parameter form, trajectories associated with the Mat\'ern process were seen to be visually virtually indistinguishable from those from the numerical model.  This suggests that the Mat\'ern form may prove useful for describing similar trajectories taken by instruments tracking the actual ocean currents.  Such `Lagrangian data' is one of the main windows into observing the ocean circulation, yet surprisingly little work has been done to analyze the velocity spectra in major Lagrangian datasets \citep{rupolo96-jpo,elipot08-grl}.  Apart from \citet{rupolo96-jpo}, the spectral slope in oceanographic Lagrangian data is almost completely unexplored, although it is implicit in several fractal dimension studies \citep{osborne89-tellus,sanderson90-tellus,sanderson91-tellus,summers02-npg}.

In this paper, we have taken essentially an observational approach, and sought to fit a parametric model to the trajectories as a descriptive analysis, without requiring a physical justification.  A next step is  to attempt to understand this model on physical grounds.  A number of researchers have attempted to derive forms for the Lagrangian velocity spectrum (or, equivalently, the autocovariance function) under simplified dynamical assumptions \citep{griffa96-smpo,weiss98-pf,majda99-pr,berloff02b-jpo,veneziani05a-jmr,majda13-ptrsla}.  One promising avenue of comparison is with the work of \citet{berloff02b-jpo}, who derive dynamical models roughly equivalent to integral orders of the Mat\'ern process.  Another is with \citet{majda99-pr}, see their Section 3.1.2, who construct idealized velocity fields that give rise to the diffusive, subdiffusive, and superdiffusive regimes of Lagrangian behavior.  Exploring the relationship of the Mat\'ern form to these dynamical models is a promising direction for future research.

\bigskip

\section*{Acknowledgements}
The work of J.~M.  Lilly and J.~J.  Early was supported by award \#1235310 from the Physical Oceanography program of the United States National Science Foundation.  The work of A.  M.  Sykulski was supported by a Marie Curie International Outgoing Fellowship. The work of S.~C.  Olhede was supported by awards \#EP/I005250/1 and \#EP/L025744/1 from the Engineering and Physical Sciences Research Council of the United Kingdom, and by award \#682172 from the European Research Council. 

The authors are grateful to an anonymous referee and to Peter Ditlevsen for their comments during the review process, which led to an improved paper.  Helpful and inspiring interactions with Alfred Hanssen, Tim Garrett, Joe LaCasce, Shane Elipot, Rick Lumpkin, and Brendon Lai at various stages in the preparation of this work are also gratefully acknowledged.  
\appendix

\section{A freely available software package}\label{jlabappendix}
All software needed to carry out the analyses described in this paper, and to generate all figures, is distributed as a part of a freely available toolbox of Matlab functions. This toolbox, called \texttt{jLab}, is available at \url{http://www.jmlilly.net} and is distributed under a Creative Commons license. The package to implement the Mat\'ern analysis, called \texttt{jMatern}, includes the following functions: \texttt{materncov}, \texttt{maternspec}, and \texttt{maternimp}, which implement the Mat\'ern autocovariance function, spectrum, and impulse response or Green's function, respectively; \texttt{maternoise}, which generates realizations of the Mat\'ern process using either the standard Cholesky decomposition method, or the fast generation method described in Section~\ref{Generation}; \texttt{maternfit}, which performs a parametric spectral fit for the Mat\'ern process and a number of variations, using the de-biased Whittle likelihood method discussed in Section~\ref{likelihoodsection}; and \texttt{blurspec}, which accounts for the blurring and/or aliasing of the theoretical spectrum associated with truncation of a continuous random process or the tapering of a finite sample.  All functions support the oscillatory Mat\'ern process as well as the standard Mat\'ern process.  Finally, \texttt{makefigs\_matern} generates all figures in this paper based on model output that can be downloaded from \url{http://www.jmlilly.net/ftp/pub/materndata.zip}.

\section{Diffusivity in terms of the spectrum}\label{diffusivityderivation}

Here we show that for a second-order stationary process, the diffusivity $\kappa$ is the value of the spectrum at zero frequency, as stated in (\ref{kappaspectrum}).  This is done by beginning with the nonstationary case. The \emph{time-dependent} diffusivity $\kappa(t)$ of a nonstationary process can be expressed in terms of the nonstationary autocovariance function $R_{zz}(t,\tau)$ as
\begin{align}
\kappa(t) &=\frac{1}{4} \frac{\mathrm{d}}{\mathrm{d} t} \int_{0}^t \int_{0}^t \mathrm{E}\left\{z(t_1)\,z^*(t_2) \right\} \mathrm{d} t_1\, \mathrm{d} t_2\\
&=\frac{1}{4} \frac{\mathrm{d}}{\mathrm{d} t} \int_{0}^t \left[\int_{0}^t R_{zz}(t_2,t_1-t_2)\,\mathrm{d} t_1\right]\, \mathrm{d} t_2\label{kappa2}
\end{align}
after substituting (\ref{rintermsofz}) into (\ref{kappadef}) and making use of (\ref{autocovariance}).  Applying the Leibniz rule for differentiation of an integral, in the form 
 \begin{equation}\label{leibniz}
\frac{\mathrm{d}}{\mathrm{d} t} \int_{0}^t f(\tau,t)\, \mathrm{d} \tau = f(t,t) + \int_{0}^t \frac{\partial}{\partial t} f(\tau,t) \,\mathrm{d} \tau 
\end{equation}
the expression for the time-dependent diffusivity simplifies to
\begin{align}
\kappa(t) &=\frac{1}{4} \int_{0}^t R_{zz}(t,t_1-t) \,\mathrm{d} t_1
+ \frac{1}{4} \int_{0}^t R_{zz}(t_2,t-t_2) \,\mathrm{d} t_2\nonumber\\
&=  \frac{1}{2}\int_{0}^t \Re\left\{R_{zz}(t,\tau-t)\right\} \,\mathrm{d} \tau\label{timevaryingkappa}
\end{align}
where in applying (\ref{leibniz}), $f(\tau,t)$ is taken to be the entire quantity in square brackets in (\ref{kappa2}).  The second line in (\ref{timevaryingkappa}) follows from the symmetry $R_{zz}(t,\tau)=R_{zz}^*(t+\tau,-\tau)$, with $\Re\{\cdot\}$ denoting the real part.

The time-dependent diffusivity can be understood in several different ways, see also \citet{lacasce08-pio}.  Substituting the definition of the autocovariance (\ref{autocovariance}), the last expression in (\ref{timevaryingkappa}) becomes
\begin{equation}
\kappa(t) = \frac{1}{2}\int_{0}^t \Re\left\{\mathrm{E}\left[z(\tau) z^*(t)\right]\right\} \,\mathrm{d} \tau 
\end{equation}
which states that the time-dependent diffusivity $\kappa(t)$ is the integral of the covariance between the velocity at time $t$ and the velocity at all times between $0$ and $t$.  However, $z^*(t)$ can be pulled outside the integral, leading to
\begin{equation}
\kappa(t) = \frac{1}{2}\Re\left\{\mathrm{E}\left[z^*(t) \int_{0}^t z(\tau)\,\mathrm{d} \tau\right]\right\} = \frac{1}{2}\Re\left\{\mathrm{E}\left[z^*(t) r(t)\right]\right\}
\end{equation}
so that $\kappa(t)$ can equivalently be seen as the \emph{inner product} of the velocity at time $t$ and the displacement at time $t$.

In the case that $z(t)$ is stationary, $R_{zz}(t,\tau)=R_{zz}(\tau)$, and the long-time limiting diffusivity value $\kappa$ is given by
\begin{align}\label{changeofvariables}
\kappa &=\lim_{t\longrightarrow\infty} \frac{1}{2} \int_{0}^t
\Re \left\{R_{zz}(\tau-t)\right\} \,\mathrm{d} \tau\\ &=\lim_{t\longrightarrow\infty} \frac{1}{2}\int_{-t}^0
\Re\left\{R_{zz}(\tau)\right\} \,\mathrm{d} \tau=\frac{1}{4} \int_{-\infty}^\infty R_{zz}(\tau)\, \mathrm{d} \tau \label{diffusivityintermsofR}
\end{align}
after a change of variables.  One may invert the inverse Fourier transform (\ref{fourierpair}) to give $S_{zz}(\omega)= \int_{-\infty}^\infty e^{-\mathrm{i} \omega \tau } R_{zz} (\tau) \,\mathrm{d}\tau$, and we then see that $\kappa=S_{zz}(0)/4$, as claimed in  (\ref{kappaspectrum}). Thus, while diffusivity is generally thought of as a time-domain quantity, it may also be expressed in the frequency domain.  

\section{Diffusiveness and memory}\label{memorysection}

In this appendix we examine the relationship between the properties of memory and diffusiveness, by constructing examples of processes with different combinations of these two properties through modifying the Mat\'ern process.  Here we will make use of a number of quantities that are not defined until the Mat\'ern process is examined in Section~\ref{maternsection}.

Spectra of stationary processes corresponding to different combinations of memory and diffusiveness are given in Table~\ref{table:memory}.  These processes can be generated through the stochastic integral equation (\ref{greenfunctionexpression}), and are most simply described by specifying modifications to the transfer function $G(\omega)$ defined in (\ref{Greenfourier}), with attendant changes for its Fourier transform, the time-domain Green's function $g(t)$. As discussed in Section~\ref{diffusiveprocesssection}, the classification of a process as `diffusive' means that its spectrum takes on a finite nonzero value at zero frequency, such that the integrated version of the process exhibits diffusive dispersion, with the expected squared distance from an initial location increasing at a constant rate.

\begin{table*}[t!]
\begin{center}
\caption{Examples of spectra for short-and long-memory processes of subdiffusive, diffusive, and superdiffusive types.  The term in the box is the spectrum of the Mat\'ern process, as given in (\ref{maternspectrum}).  Note that the two spectra corresponding to diffusive processes have been normalized such that $\kappa=S_{zz}(0)/4=1$. $\lambda$ is a nonnegative constant, while $\Omega$ is here a nonzero constant of either sign.  } \label{table:memory}\vspace{0.2in}
\begin{tabular}{||l|l|lc|lc||}
 \hline\hline
&$\kappa $ & Short Memory & Stationarity & Long Memory & Stationarity \\\hline & & & &  &\vspace{-0.05in}\\
Superdiffusive & $\infty$& (not possible) & --- & $S_{zz}(\omega)=\displaystyle{\frac{1}{\omega^{2\beta}\left(\omega^2+\lambda^2\right)^\alpha}}$ & $\alpha+\beta>\frac{1}{2},\,\,\,\,\beta<\frac{1}{2} $\\& & & & &\vspace{-0.1in}\\
Diffusive & 1 & $S_{zz}(\omega)=\boxed{\frac{4\lambda^{2\alpha}}{\left(\omega^2+\lambda^2\right)^\alpha}}\quad$ & $\alpha>\frac{1}{2}$ & $S_{zz}(\omega)=\displaystyle{\frac{4\Omega^{2\beta}\left(\Omega^{2}+\lambda^2\right)^\alpha}{|\omega-\Omega|^{2\beta}\left(|\omega-\Omega|^{2}+\lambda^2\right)^\alpha}}$& $\alpha+\beta>\frac{1}{2},\,\,\,\,\beta<\frac{1}{2}$  \\& & & & &\vspace{-0.1in}\\
Subdiffusive & 0 &$S_{zz}(\omega)=\displaystyle{\frac{\omega^2}{\left(\omega^2+\lambda^2\right)^\alpha}}$ & $\alpha>\frac{3}{2}$ & $S_{zz}(\omega)=\displaystyle{\frac{\omega^{2}}{|\omega-\Omega|^{2\beta}\left(|\omega-\Omega|^{2}+\lambda^2\right)^\alpha}}$ & $\alpha+\beta>\frac{3}{2},\,\,\,\,\beta<\frac{1}{2} $  \\ & & & & &\vspace{-0.05in}\\
\hline\hline
\end{tabular} 
\end{center}
\end{table*}
 
Multiplying the Mat\'ern transfer function given by (\ref{Greenfourier}), with the spin $\Omega$ set to zero, by $\omega$ multiplies the spectrum by $\omega^2$ and thus leads to a short-memory subdiffusive process, with a spectrum shown at the lower left of Table~\ref{table:memory}.  This process has finite variance provided we choose $\alpha>3/2$.  Dividing the  Mat\'ern transfer function by $|\omega|^\beta$, corresponding to a fractional integration, divides the spectrum by $|\omega|^{2\beta}$.  This gives a process that is both long-memory and superdiffusive, with a spectrum shown at the upper right.  Adding a spin to this latter process, by shifting the transfer function frequency by $\Omega$ as in (\ref{Greenfourier}), also shifts the spectrum as $\omega\mapsto\omega-\Omega$. The resulting spectrum, shown at the center right of  Table~\ref{table:memory}, has a finite value at frequency zero but a singularity off zero, and is therefore diffusive but long-memory; we note that this continuous-time process is related to the discrete-time Gegenbauer process \citep{gray89-jtsa,baillie96-jecon}.  Finally, multiplying the transfer function of the previous process by $\omega$ multiplies the spectrum by $\omega^2$, causing the spectrum at zero frequency to vanish; however this does not remove the singularity at $\omega=\Omega$, leading to a long-memory subdiffusive process, the spectrum of which is at the lower right in the table.

These results show that diffusiveness and memory, while related, are distinct properties that can be varied independently.  In this table we have also noted the parameter ranges required for the process spectrum to integrate to a finite variance, and therefore for the process to be stationarity.  In general for a spectrum of the form $|\omega|^{-2\alpha}$, the behavior of the singularity at zero contributes to unbounded variance for $\alpha>\frac{1}{2}$, while the behavior at large frequencies contributes to unbounded variance for $\alpha<\frac{1}{2}$.  Ensuring that neither the singularities nor the large-frequency decay will contribute to unbounded variance leads to the parameter ranges for stationarity shown in the table.  

\section{The fBm Rihaczek distribution}\label{fbmspectrumappendix}

Here we derive (\ref{fbmvariogramtransform}) for the Rihaczek distribution of fractional Brownian motion, an expression that was previously presented by \citet{oigard06-pre}, adding some additional details. For fBm, there arises a complication in defining the Rihaczek distribution as in (\ref{fbmvariogramtransformgeneral}), because the integral in (\ref{fbmvariogramtransformdefinition}) is divergent.  Despite this, (\ref{fbmvariogramtransform}) may be derived by interpreting this integral in a limiting sense, as is now shown.  For $\alpha>1/2$, consider the integral
\begin{equation}\label{interpretasable}
\int_{-\infty}^\infty |\tau|^{2\alpha-1} e^{-\mathrm{i} \omega \tau } \mathrm{d}\tau =
2\Re\left\{ \int_{0}^\infty \tau^{2\alpha-1} e^{\mathrm{i} \omega \tau } \mathrm{d}\tau \right\}
\end{equation}
which does not exist in the usual sense, since the integral is divergent.  However, a limiting form does exist, given by 
\begin{equation}\label{abellimit}
\lim_{\epsilon\longrightarrow 0}
\int_{0}^\infty \tau^{2\alpha-1} e^{-\epsilon \tau +\mathrm{i} \omega \tau} \mathrm{d}\tau = e^{\mathrm{i} \alpha \pi }\frac{\Gamma(2\alpha)}{\omega^{2\alpha}},\quad\quad \omega\ne 0
\end{equation}
which is an example of what is termed an \emph{Abel limit},  see \citet[p.~407]{wong80-siam}.  Thus interpreting (\ref{interpretasable}) as an Abel limit leads to
\begin{equation} \label{abeltransform}
\frac{1}{ 2 \cos\left(\pi\alpha\right) \Gamma(2\alpha)}\int_{-\infty}^\infty |\tau|^{2\alpha-1} e^{- \mathrm{i}\omega \tau } \mathrm{d}\tau = 
\frac{1}{\left|\omega\right|^{2\alpha}}
\end{equation}
such that a decaying power law in the frequency domain is associated with a growing power law, of one lower order, in the time domain.  Here we have noted that changing the sign of $\omega$ in (\ref{abellimit}) is equivalent to a complex conjugation, since $(-1)^{2\alpha}=e^{ 2\mathrm{i}\pi\alpha}$, this leading to the absolute value of $\omega$. 

The coefficient of the integral in (\ref{abeltransform}) simplifies to $-V_{\alpha}/2$, as shown in Appendix~\ref{coefficientappendix}.  One then finds
\begin{equation}\label{variogramtransform}
 -\int_{-\infty}^\infty \frac{V_\alpha}{2} A^2 |\tau|^{2\alpha-1} \,e^{-\mathrm{i} \omega \tau} \,\mathrm{d} \tau 
= \frac{A^2}{|\omega|^{2\alpha}}=\widetilde{S}_{zz}^{f\!Bm}(\omega)
\end{equation}
which shows that $A^2/|\omega|^{2\alpha}$ is the Fourier transform, in the Abel limit sense, of that part of the nonstationary autocovariance function $R_{zz}^{f\!Bm}(t,\tau)$ depending only on $\tau$.  The Fourier transformed quantity on the left-hand side of (\ref{variogramtransform}) is also recognized from (\ref{fbmvariogram}) as the negative of the fBm variogram $\gamma_{zz}^{f\!Bm}(\tau)$.  A change of variables gives
\begin{equation}\label{variogramtransformshifted}
 -\int_{-\infty}^\infty\frac{V_\alpha}{2} A^2 |t+\tau|^{2\alpha-1} \,e^{-\mathrm{i} \omega \tau} \,\mathrm{d} \tau 
 = e^{\mathrm{i} \omega t}\frac{A^2}{|\omega|^{2\alpha}}
\end{equation}
and substituting (\ref{variogramtransform}) and (\ref{variogramtransformshifted}) into (\ref{fbmvariogramtransformdefinition}), and making use of $ \int_{-\infty}^\infty \,e^{-\mathrm{i} \omega \tau} \,\mathrm{d} \tau = 2\pi\delta(\omega)$, one obtains (\ref{fbmvariogramtransform}).  From left to right in (\ref{fbmvariogramtransform}), we have the inverse Fourier transforms of the $|\tau|$ term, the $|t+\tau|$ term, and the $|t|$ term from the fBm autocovariance function (\ref{fbmautocovariance}).  

\section{The form of the fBm coefficient $V_\alpha$}\label{coefficientappendix}
The usual form of the coefficient for fractional Brownian motion, in terms of the Hurst parameter $H=\alpha-1/2$, is 
\begin{equation}
V_H\equiv\frac{\Gamma(1-2H)\cos(\pi H)}{\pi H}
\end{equation}
see \citet{barton88-itit}.  In terms of the slope parameter $\alpha$, this becomes 
\begin{equation}
V_\alpha\equiv\frac{\Gamma(2-2\alpha)\sin(\pi\alpha)}{\pi (\alpha-1/2)}
\end{equation}
which can be expressed in a more symmetric form as follows.  First we expand the denominator using $\Gamma(1+\nu)=\nu\Gamma(\nu)$ or $\nu=\Gamma(1+\nu)/\Gamma(\nu)$ with $\nu=\alpha-1/2$, giving
\begin{equation}\label{intermediatealpha}
V_\alpha=\frac{\Gamma(2-2\alpha)\Gamma\left(\alpha-\frac{1}{2}\right)\sin(\pi\alpha)}{\pi\Gamma\left(\alpha+\frac{1}{2}\right)}.
\end{equation}
The so-called \emph{reflection} and \emph{duplication} theorems for the gamma function are, respectively,
 \begin{align}
 \sin(\pi\nu) &= \frac{\pi}{\Gamma(\nu) \Gamma(1-\nu)} \label{reflection}\\
 \Gamma(2\nu)&= \frac{1}{\sqrt{\pi}}\,2^{2\nu-1}\Gamma(\nu)\Gamma\left(\nu+\frac{1}{2}\right) \label{duplication}
\end{align}
see 6.1.17 and 6.1.18 on p.~256 of \citet{abramowitz}.  Applying the later to both $\Gamma(2\alpha)$ and $\Gamma(2-2\alpha)$ gives their product as
\begin{multline}
\Gamma(2\alpha)\Gamma(2-2\alpha) \\= 
\frac{1}{\pi} \Gamma(\alpha)\Gamma\left(\alpha+\frac{1}{2}\right)
\Gamma(1-\alpha)\Gamma\left(\frac{3}{2}-\alpha\right)
\end{multline} 
in which all powers of two exactly cancel.  Employing the reflection theorem with $\nu=\alpha$, this becomes
\begin{equation}\label{intermediategamma}
\Gamma(2\alpha)\Gamma(2-2\alpha) =
\frac{\Gamma\left(\alpha+\frac{1}{2}\right)\Gamma\left(\frac{3}{2}-\alpha\right)}{\sin(\pi\alpha)}
\end{equation}
and substituting this into (\ref{intermediatealpha}) leads to (\ref{symmetricform}), as claimed.

Now, using the reflection formula (\ref{reflection}) together with a trigonometric identity we find 
\begin{equation}\label{trigreflection}
-\cos(\pi\alpha)=\sin\left(\pi\alpha-\frac{\pi}{2}\right) = \frac{\pi}{\Gamma\left(\alpha-\frac{1}{2}\right)\Gamma\left(\frac{3}{2}-\alpha\right)} 
\end{equation}
and therefore
\begin{equation}\label{abelfbm}
-\frac{1}{\cos\left(\pi\alpha\right) \Gamma(2\alpha)}
=\frac{1}{\pi}\frac{\Gamma\left(\alpha-\frac{1}{2}\right)\Gamma\left(\frac{3}{2}-\alpha\right)}{ \Gamma(2\alpha)}=V_\alpha.
\end{equation}
This establishes that the coefficient of the integral in (\ref{abeltransform}) is the same as $-V_{\alpha}/2$.

\section{Fractional Gaussian noise} \label{fgnappendix}
  
Define the difference of a fractional Brownian motion process at one time and itself a different time as
\begin{equation} \label{definefgn}
 z_\Delta(t) \equiv z(t + \Delta)-z(t)
\end{equation}
which will be explicitly labeled by the time interval $\Delta$ for clarity.  The resulting process is called \emph{fractional Gaussian noise} or fGn \citep{mandelbrot68-siam,mandelbrot69-wrr,sapa}.  While it is more usual to sample the process defined by (\ref{definefgn}) at regular intervals, here we will examine the properties of the continuous-time process.  

The autocovariance function for continuous-time fractional Gaussian noise will be denoted as
\begin{equation}
 R_{zz,\Delta}^{f\!Gn}(t,\tau)\equiv\mathrm{E}\left\{ z_\Delta(t+\tau)\, z^*_\Delta(t)\right\}
\end{equation}
and this expands to give
\begin{multline}
R_{zz,\Delta}^{f\!Gn}(t,\tau)=
R_{zz}^{f\!Bm}(t,\tau)+R_{zz}^{f\!Bm}(t+\Delta,\tau)
\\-R_{zz}^{f\!Bm}(t,\tau+\Delta) - R_{zz}^{f\!Bm}(t+\Delta,\tau-\Delta).
\end{multline}
Substituting the form of the fBm autocovariance (\ref{fbmautocovariance}), cancellations occur, leading to
\begin{multline}\label{fbmincrement}
R_{zz,\Delta}^{f\!Gn}(\tau)\equiv
R_{zz,\Delta}^{f\!Gn}(t,\tau)=\frac{V_\alpha}{2} A^2 \times\\
\left[\left|\tau+\Delta\right|^{2\alpha-1}+\left|\tau-\Delta\right|^{2\alpha-1}-2\left|\tau\right|^{2\alpha-1}\right]
\end{multline}
where our notation is modified to reflect the fact that the autocovariance is independent of $t$.  Fractional Gaussian noise is therefore a stationary process.  On account of the self-similar scaling of the fBm autocovariance function (\ref{originalform}), one finds 
\begin{equation}
R_{zz,\Delta}^{f\!Gn}(\tau)= \Delta^{2\alpha-1}R_{zz,1}^{f\!Gn}(\tau/\Delta)
\end{equation}
so that we may without loss of generality set $\Delta=1$.  For convenience we let $\tilde \tau\equiv\tau/\Delta$ be a nondimensional time offset.

The expression (\ref{fbmincrement})  may be compared with (5.2) of \citet{mandelbrot68-siam}, who permitted the durations of the two increments to differ.  Our expression differs from that in \citet{mandelbrot68-siam} because we have chosen to apply the similarity scaling to remove the increment \emph{duration} rather than the \emph{separation}, for reasons to become apparently shortly; ``$T$'' in \citet{mandelbrot68-siam} refers to what we call $\tau$ here.  

\begin{figure}[t!]
\begin{center}
\includegraphics[width=0.5\textwidth]{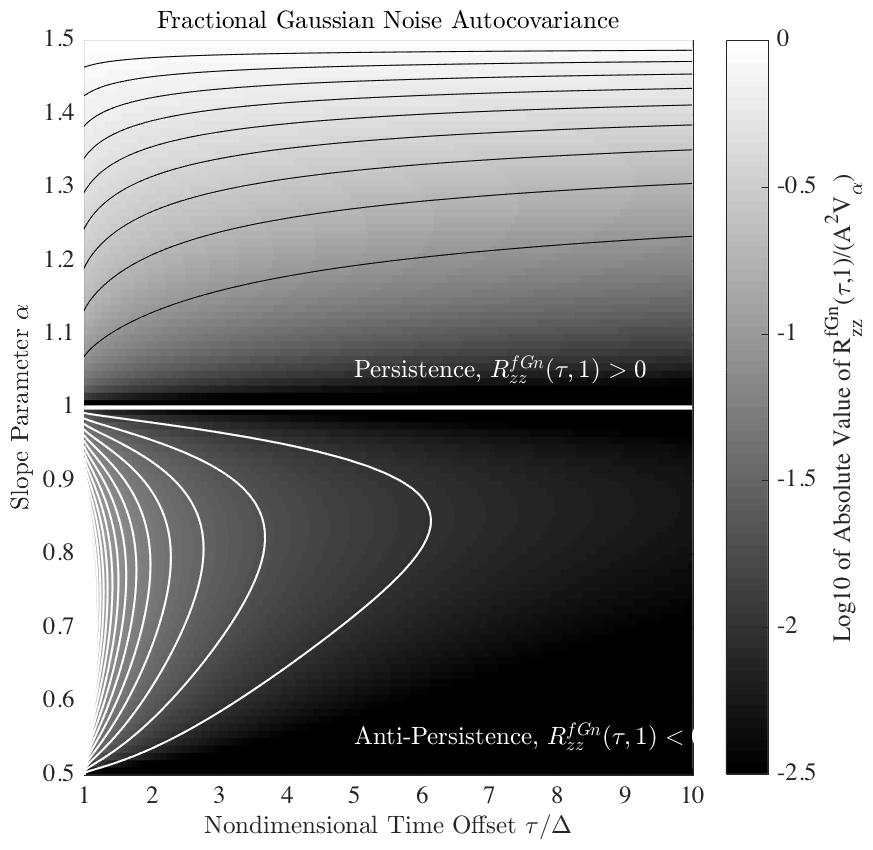}\end{center}
\caption{\small The fractional Gaussian noise autocovariance function $R_{zz,1}^{f\!Gn}(\widetilde\tau)$, as defined in (\ref{fbmincrement}), here normalized by dividing by $V_\alpha$ and $A^2$.  The time axis is interpreted as the normalized time $\tilde\tau=\tau/\Delta$.  The shading shows $\log_{10}$ of the magnitude of the normalized autocovariance function, which obtains a maximum of unity at $\alpha=3/2$ for all $\widetilde \tau$.  A sign change occurs at $\alpha=1$, with positive values at higher $\alpha$ and negative values at lower $\alpha$.  Black lines are contours of positive values, with a contour interval of 0.1 beginning at zero, while thin white lines are contours of negative values with an interval of 0.01.  The heavy white curve is the zero contour at $\alpha=1$.  
}
\label{increment}
\end{figure}

The normalized fGn covariance function $R_{zz,1}^{f\!Gn}(\tilde\tau)/(A^2V_\alpha)$ is shown in Fig.~\ref{increment}.  Because fGn will generally be sampled, we are typically interested only in time offsets $\tau$ that exceed the sample interval $\Delta$, corresponding to $\tilde \tau>1$.  Analyzing $R_{zz,1}^{f\!Gn}(\tilde\tau)/(A^2V_\alpha)$ using (\ref{fbmincrement}), one sees that for $\tilde \tau>1$ it obtains a maximum value of unity at $\alpha=3/2$, while it vanishes both for $\alpha=1/2$ and $\alpha=1$.  It is found that $R_{zz,1}^{f\!Gn}(\tilde\tau)$ is positive for $\alpha>1$, and negative for $\alpha<1$, see \citet{mandelbrot68-siam}.  The maximum positive value is at $\alpha=3/2$ for all $\tilde\tau$, but the maximum negative value occurs at some intermediate value of $\alpha$ in the range $(1/2,1)$.  For any fixed $\alpha$, increasing $\tilde\tau$ leads to absolute values of $R_{zz,1}^{f\!Gn}(\tilde\tau)$ that decay toward zero.  

The behavior of the fractional Gaussian noise covariance function allows us to discuss the property of \emph{persistence}.  For $\alpha>1$, fGn exhibits positive correlations, such that positive values will tend to be followed by positives value and negative values by negative values.  However, for $\alpha<1$, fGn is \emph{anti-persistent}, and positive values will tend to be followed by negative values and vice-versa.  Note that $R_{zz,1}^{f\!Gn}(\tilde\tau)$ is not symmetric about $\alpha=1$: the most positive correlations occur at $\alpha=3/2$, but the most negative correlations do not occur at $\alpha=1/2$.  This may perhaps be seen as reflecting a difference between persistence and anti-persistence.  Values of the same sign can follow one another indefinitely, for any timescale; but the same cannot be true for values of the opposite sign.

The persistence transition in fractional Gaussian noise at $\alpha=1$ is reflected in the behavior of fractional Brownian motion seen in Fig.~\ref{randomwalk_brownian}.  Values of $\alpha>1$ coincide with the tendency for the process to systematically drift away from an initial value, as differenced versions of the process will tend to keep contributing perturbations of one particular sign.  Similarly, for $\alpha<1$, the anti-correlations of the differenced process tend to act to restore fBm toward a baseline, and therefore these process are more closely distributed around the mean value of zero.  The important point is that for fractional Brownian motion, the spectral slope can be seen as being linked to the degree of persistence or anti-persistence associated with a differenced version of the process.  

The memory of fractional Gaussian noise may be determined as follows.  The fGn autocovariance (\ref{fbmincrement}) can be rewritten as
\begin{multline}\label{fbmincrement2}
R_{zz,\Delta}^{f\!Gn}(\tau)=\frac{V_\alpha}{2} A^2 \tau^{2\alpha-1}\times\\
\left[\left|1+\Delta/\tau\right|^{2\alpha-1}+\left|1-\Delta/\tau\right|^{2\alpha-1}-2\right]
\end{multline}
after pulling out the factor of $\tau^{2\alpha-1}$.  Employing the binomial expansion, $(1+x)^\gamma=1+\gamma x + \frac{1}{2} \gamma(\gamma-1) x^2 +O(x^3)$ for small $x$, cancellations occur and we find
\begin{multline}
R_{zz,\Delta}^{f\!Gn}(\tau)=\frac{V_\alpha}{2} A^2 \Delta^2 (2\alpha-1)(2\alpha-2)\tau^{2\alpha-3} \\
+ O\left(\tau^{2\alpha-2}\right)
\end{multline}
for the asymptotic behavior at large $\tau$.  Recall from Section~\ref{memorysection} that a \emph{long-memory} stationary process is one for which the long-time behavior of the autocovariance function behaves as $R_{zz}(\tau) \sim |\tau|^{-\mu}$ for 
$0<\mu\le1$.  For fGn we have $\mu=3-2\alpha$, thus  $1\le \alpha<3/2$ corresponds to $0<\mu\le 1$, and fractional Gaussian noise is a long-memory process in this range of $\alpha$.  

\section{The Mat\'ern autocovariance for small $\tau$} \label{shorttimesection}

In this appendix we derive the form of the small-$\tau$ behavior of the Mat\'ern autocovariance function, as was apparently first done by \citet{goff88-jgr}, their p. 13,606.  Here we follow those authors, paying particularly close attention to the $\alpha$ range over which the result is valid.  For this we will make use of the identity 9.6.2 of \citet{abramowitz}
 \begin{equation}\label{KintermsofI}
\mathcal{K}_{\nu}(\tau) = \frac{1}{2}\pi \frac{\mathcal{I}_{-\nu}(\tau)-\mathcal{I}_{\nu}(\tau)}{\sin(\nu\pi)}
\end{equation}
together with the series expansion 9.6.10 of \citet{abramowitz}
 \begin{equation}
\mathcal{I}_{\nu}(\tau) = \left(\frac{1}{2} \tau\right)^{\!\nu} \sum_{n=0}^\infty \frac{ \left(\frac{1}{2} \tau\right)^{2n}}{n! \,\Gamma(n+1+\nu)}.
\end{equation} 
Employing the reflection formula (\ref{reflection}), these combine to give
 \begin{multline}\label{Kexpansion}
\tau^\nu\mathcal{K}_{\nu}(\tau) =
 \frac{1}{2} \Gamma(1-\nu) \Gamma(\nu)
\left[ 2^\nu \sum_{n=0}^\infty \frac{ \left(\frac{1}{2} \tau\right)^{2n}}{n! \,\Gamma(n+1-\nu)} \right.\\\left.
- \frac{ \tau^{2\nu}}{2^\nu} \sum_{n=0}^\infty \frac{ \left(\frac{1}{2} \tau\right)^{2n}}{n! \,\Gamma(n+1+\nu)}\right] 
\end{multline}
and gathering the terms for $n=0$, one finds
 \begin{multline}\label{Knearzero}
\mathcal{M}_{\nu+1/2}(\tau)=\frac{1}{\Gamma(\nu) 2^{\nu-1}}  |\tau|^\nu\mathcal{K}_{\nu}(|\tau|) \\= 1 - \left(\frac{1}{2}|\tau|\right)^{\!2\nu}\frac{\Gamma(1-\nu)}{\Gamma(1+\nu)}
+\sum_{n=1}^\infty\tau^{2n}\left[c_n + d_n |\tau|^{2\nu}\right] 
\end{multline}
where $c_n$ and $d_n$ are constants describing the behavior proportional to $\tau^{2n}$ and $|\tau|^{2n+2\nu}$, respectively.  Here $\mathcal{M}_{\alpha}(\tau)$ is the Mat\'ern function introduced in (\ref{maternfunction}).

Provided that $\nu>0$, we have $\mathcal{M}_{\nu+1/2}(\tau)\approx 1$ for $\tau$ sufficiently close to zero.  For $0<\nu<1$ and small $\tau$, the term outside the summation in (\ref{Knearzero}), which is proportional to $|\tau|^{2\nu}$, dominates the first term in the summation, which is proportional to $\tau^2$;  all other terms are then smaller still. The range of $\nu$ for which this result is valid does appear to have been mentioned by \citet{goff88-jgr}. Since $\nu$ in these expressions is related to $\alpha$ in the Mat\'ern autocovariance function through $\nu=\alpha-1/2$, this domination occurs for $1/2<\alpha<3/2$, and we obtain the asymptotic behavior (\ref{Rnearzerooriginal}) for $|\tau|\ll1/\lambda$.   For larger values of $\alpha$, the smallest power of $\tau$ in (\ref{Knearzero}) is the $\tau^2$ term on the second line of (\ref{Knearzero}), which therefore dominates.

\section{The Mat\'ern oscillator equation} \label{materngreen}

The Green's function (\ref{Greenfunctiontime}) for the oscillatory Mat\'ern process is also the solution the fractional differential equation 
(\ref{maternoscillator}), which describes a damped fractional oscillator forced by a delta function at the origin, as we now show. We expand the operator in (\ref{maternoscillator}) as
\begin{multline}\label{binomialexpansion}
\left[\frac{\mathrm{d}}{\mathrm{d} t}+ \lambda -\mathrm{i} \Omega\right]^\alpha =\\
\sum_{n=0}^\infty \frac{\alpha(\alpha-1)\cdots(\alpha-n+1)}{n!} \left[\frac{\mathrm{d}^n}{\mathrm{d} t^n}+ \left(\lambda -\mathrm{i} \Omega\right)^{\alpha-n}\right]
\end{multline}
using Newton's generalization of the binomial theorem to non-integral orders.  Substituting $g(t)=
\frac{1}{2\pi}\int_{-\infty}^{\infty} G(\omega) e^{\mathrm{i} \omega t} \mathrm{d} \omega$ into the left-hand side of the differential equation (\ref{maternoscillator}), applying (\ref{binomialexpansion}), and carrying out the indicated derivatives, leads to
\begin{multline}\label{maternoscillator2}
\left[ \frac{\mathrm{d}}{\mathrm{d} t}+\lambda - \mathrm{i} \Omega \right]^\alpha g(t)= \\
\frac{1}{2\pi}\int_{-\infty}^{\infty} G(\omega) \left[ \mathrm{i}\left(\omega- \Omega\right) +\lambda \right]^\alpha e^{\mathrm{i} \omega t} \mathrm{d} \omega
\end{multline}
after collapsing the summation using a second application of the generalized binomial theorem.  Now a cancellation occurs, and the right-hand side of (\ref{maternoscillator2}) becomes simply $\frac{1}{2\pi}\int_{-\infty}^{\infty} e^{\mathrm{i} \omega t} \mathrm{d} \omega$, which is equal to $\delta(t)$, thus verifying (\ref{maternoscillator}).


\begin{thebibliography}{106}
\providecommand{\natexlab}[1]{#1}
\providecommand{\url}[1]{{\tt #1}}
\providecommand{\urlprefix}{URL }
\expandafter\ifx\csname urlstyle\endcsname\relax
  \providecommand{\doi}[1]{doi:\discretionary{}{}{}#1}\else
  \providecommand{\doi}{doi:\discretionary{}{}{}\begingroup
  \urlstyle{rm}\Url}\fi

\bibitem[{Abramowitz and Stegun(1972)}]{abramowitz}
Abramowitz, M. and Stegun, I.~A.: Handbook of Mathematical Functions with
  Formulas, Graphs, and Mathematical Tables, National Bureau of Standards,
  Washington, D. C., tenth printing edn., 1972.

\bibitem[{Adler(1977)}]{adler77-ap}
Adler, R.~J.: {H}ausdorff dimension and {G}aussian fields, Ann. Probab., 5,
  145--151, 1977.

\bibitem[{Arat\'o et~al.(1999)Arat\'o, Baran, and Isp\'any}]{arato99-cma}
Arat\'o, M., Baran, S., and Isp\'any, M.: Functionals of complex
  {O}rnstein-{U}hlenbeck processes, Comput. Math. Appl., 37, 1--13, 1999.

\bibitem[{Baillie(1996)}]{baillie96-jecon}
Baillie, R.~T.: Long memory processes and fractional integration in
  econometrics, J. Econometrics, 73, 5--59, 1996.

\bibitem[{Barton and Poor(1988)}]{barton88-itit}
Barton, R.~J. and Poor, H.~V.: Signal detection in fractional {G}aussian noise,
  IEEE T. Inform. Theory, 34, 943--959, 1988.

\bibitem[{Basset(1888)}]{basset}
Basset: A Treatise on Hydrodynamics, with Numerous Examples, Cambridge Univ
  Press, 1888.

\bibitem[{Bateman(1954)}]{bateman}
Bateman, H.: Tables of Integral Transforms, McGraw-Hill Book Company, Inc,
  1954.

\bibitem[{Beran(1992)}]{beran92-ss}
Beran, J.: Statistical methods for data with long-range dependence, Stat. Sci.,
  7, 404--416, 1992.

\bibitem[{Beran(1994)}]{beran}
Beran, J.: Statistics for Long-Memory Processes, vol.~61 of {\em Monographs on
  Statitics and Applied Probability\/}, Chapman \& Hall / CRC, 1994.

\bibitem[{Berloff and McWilliams(2002)}]{berloff02b-jpo}
Berloff, P. and McWilliams, J.: Material transport in oceanic gyres. {P}art
  {II}: {H}ierarchy of stochastic models, J. Phys. Oceanogr., 32, 797--830,
  2002.

\bibitem[{Bracco and McWilliams(2010)}]{bracco10-jfm}
Bracco, A. and McWilliams, J.~C.: Reynolds-number dependency in homogeneous,
  stationary two-dimensional turbulence, J. Fluid Mech., 646, 517--526, 2010.

\bibitem[{Cressie(1988)}]{cressie88-jasa}
Cressie, N.: A graphical procedure for determining nonstationarity in time
  series, J. Acoust. Soc. Am., 83, 1108--1116, 1988.

\bibitem[{Davis(1983)}]{davis83-jmr}
Davis, R.~E.: Oceanic property transport, {L}agrangian particle statistics, and
  their prediction, J. Mar. Res., 41, 163--194, 1983.

\bibitem[{Dietrich and Newsam(1997)}]{dietrich97-sjsc}
Dietrich, C.~R. and Newsam, G.~N.: Fast and exact simulation of stationary
  {G}aussian processes through circulant embedding of the covariance matrix,
  SIAM J. Sci. Comput., 18, 1088--1107, 1997.

\bibitem[{Dritschel et~al.(2008)Dritschel, Scott, Gottwald, and
  Tran}]{dritschel08-prl}
Dritschel, D.~G., Scott, R.~K., Gottwald, G.~A., and Tran, C.~V.: Unifying
  scaling theory for vortex dynamics in two-dimensional turbulence, Phys. Rev.
  Lett., 101, 94\,501, 2008.

\bibitem[{Dunbar et~al.(1992)Dunbar, Douglass, and Camp}]{dunbar92-jmaa}
Dunbar, S.~R., Douglass, R.~W., and Camp, W.~J.: The divider dimension of the
  graph of a function, J. Math. Anal. Appl., 167, 403--413, 1992.

\bibitem[{Elipot and Lumpkin(2008)}]{elipot08-grl}
Elipot, S. and Lumpkin, R.: Spectral description of oceanic near-surface
  variability, Geophys. Res. Lett., 35, L05\,606, \doi{10.1029/2007GL032874},
  2008.

\bibitem[{Emery and Thomson(2014)}]{emery}
Emery, W.~J. and Thomson, R.~E.: Data Analysis Methods in Physical
  Oceanography, Elsevier, third edn., 2014.

\bibitem[{Falconer(1990)}]{falconer}
Falconer, K.: Fractal Geometry: Mathematical Foundations and Applications, John
  Wiley \& Sons, 1990.

\bibitem[{Flandrin(1989)}]{flandrin89-itit}
Flandrin, P.: On the spectrum of fractional {B}rownian motion, IEEE T. Inform.
  Theory, 35, 197--199, 1989.

\bibitem[{Flandrin(1999)}]{flandrin}
Flandrin, P.: Time-Frequency / Time-Scale Analysis, Academic Press, San Diego,
  1999.

\bibitem[{Fofonoff(1969)}]{fofonoff69-dsr}
Fofonoff, N.~P.: Spectral characteristics of internal waves in the ocean,
  Deep-Sea Res., 16, 59--71, (Supplement), 1969.

\bibitem[{Gneiting and Schlather(2004)}]{gneiting04-siam}
Gneiting, T. and Schlather, M.: Stochastic models that separate fractal
  dimension and the {H}urst effect, SIAM Rev., 46, 269--282, 2004.

\bibitem[{Gneiting et~al.(2010)Gneiting, Kleiber, and
  Schlather}]{gneiting10-jasa}
Gneiting, T., Kleiber, W., and Schlather, M.: Mat\'ern cross-covariance
  functions for multivariate random fields, J. Acoust. Soc. Am., 105,
  1167--1177, 2010.

\bibitem[{Goff and Jordan(1988)}]{goff88-jgr}
Goff, J.~A. and Jordan, T.~H.: Stochastic modeling of seafloor morphology:
  Inversion of sea beam data for second-order statistics, J. Geophys. Res., 93,
  13\,589--13\,608, 1988.

\bibitem[{Gonella(1972)}]{gonella72-dsr}
Gonella, J.: A rotary-component method for analyzing meteorological and
  oceanographic vector time series, Deep-Sea Res., 19, 833--846, 1972.

\bibitem[{Gorenflo and Mainardi(1997)}]{gorenflo97-cism}
Gorenflo, R. and Mainardi, F.: Fractals and Fractional Calculus in Continuum
  Mechanics, vol. 378 of {\em {CISM International Centre for Mechanical
  Sciences Series}\/}, chap. Fractional calculus: Integral and differential
  equations of fractional order, pp. 223--276, Springer-Verlag Wien, 1997.

\bibitem[{Gradshteyn and Ryzhik(2000)}]{gradshteyn}
Gradshteyn, I.~S. and Ryzhik, I.~M.: The Table of Integrals, Series and
  Products, 6th Edition, Academic Press, 2000.

\bibitem[{Gray et~al.(1989)Gray, Zhang, and Woodward}]{gray89-jtsa}
Gray, H.~L., Zhang, N.-F., and Woodward, W.~A.: On generalized fractional
  processes, J. Time Ser. Anal., 10, 233--257, 1989.

\bibitem[{Griffa(1996)}]{griffa96-smpo}
Griffa, A.: Stochastic Modelling in Physical Oceanography, chap. Applications
  of stochastic particle models to oceanographic problems, pp. 113--140,
  Springer, Boston, MA, 1996.

\bibitem[{Guttorp and Gneiting(2006)}]{guttorp06-biometrika}
Guttorp, P. and Gneiting, T.: Studies in the history of probability and
  statistics {XLIX}. {O}n the {M}at\'ern correlation family, Biometrika, 93,
  989--995, 2006.

\bibitem[{Handcock and Stein(1993)}]{handcock93-technometrics}
Handcock, M.~S. and Stein, M.~L.: A {B}ayesian analysis of kriging, Technom,
  35, 403--410, 1993.

\bibitem[{Hanssen and Scharf(2003)}]{hanssen03-itsp}
Hanssen, A. and Scharf, L.~L.: A theory of polyspectra for nonstationary
  stochastic processes, IEEE T. Signal Proces., 51, 1243--1252, 2003.

\bibitem[{Hartikainen and S\"arkk\"a(2010)}]{hartikainen10-mlsp}
Hartikainen, J. and S\"arkk\"a, S.: Kalman filtering and smoothing solutions to
  temporal {G}aussian process regression models, in: Proceedings of the {IEEE}
  International Workshop on Machine Learning for Signal Processing {(MLSP)},
  2010.

\bibitem[{Hedevang and Schmiegel(2014)}]{hedevang14-ijnsns}
Hedevang, E. and Schmiegel, J.: A {L}\'{e}vy based approach to random vector
  fields: with a view towards turbulence, Int. J. Nonlin. Sci. Num., 15,
  411--435, 2014.

\bibitem[{Hindberg and Hanssen(2007)}]{hindberg07-itsp}
Hindberg, H. and Hanssen, A.: Generalized spectral coherences for
  complex-valued harmonizable processes, IEEE T. Signal Proces., 55,
  2407--2413, 2007.

\bibitem[{Hunt(1951)}]{hunt51-tams}
Hunt, G.~A.: Random {F}ourier transforms, Trans. Amer. Math. Soc., 71, 38--69,
  1951.

\bibitem[{Jeffreys(1942)}]{jeffreys42-mnras}
Jeffreys, H.: The variation of latitude, Mon. Not. R. Astron. Soc., 100,
  139--155, 1942.

\bibitem[{Kadoch et~al.(2011)Kadoch, del Castillo-Negrete, Bos, and
  Schneider}]{kadoch11-pre}
Kadoch, B., del Castillo-Negrete, D., Bos, W. J.~T., and Schneider, K.:
  Lagrangian statistics and flow topology in forced two-dimensional turbulence,
  Phys. Rev. E, 83, 036\,314, \doi{https://doi.org/10.1103/PhysRevE.83.036314},
  2011.

\bibitem[{{Kamp\'e de F\'eriet}(1939)}]{kampedeferiet39-assb}
{Kamp\'e de F\'eriet}, J.: Les fonctions al\'eatoires stationnaires et la
  th\'eorie statistique de la turbulence homog\'ene, Ann. Soc. Sci. Brux., 59,
  145--194, 1939.

\bibitem[{Kirkwood(1933)}]{kirkwood33-pr}
Kirkwood, J.~G.: Quantum statistics of almost classical assemblies, Phys. Rep.,
  44, 31--37, 1933.

\bibitem[{Koszalka and LaCasce(2010)}]{koszalka10-od}
Koszalka, I.~M. and LaCasce, J.~H.: Lagrangian analysis by clustering, Ocean
  Dyn., 60, 957--972, 2010.

\bibitem[{La{C}asce(2008)}]{lacasce08-pio}
La{C}asce, J.~H.: Statistics from {L}agrangian observations, Prog. Oceanogr.,
  77, 1--29, 2008.

\bibitem[{Li et~al.(2010)Li, Lu, Li, and Chen}]{li10-wtc}
Li, J.-Y., Lu, X., Li, M., and Chen, S.: Data simulation of {M}at\'ern type,
  {WSEAS Transactions on Computers}, 9, 696--705, 2010.

\bibitem[{Lilly and Gascard(2006)}]{lilly06-npg}
Lilly, J.~M. and Gascard, J.-C.: Wavelet ridge diagnosis of time-varying
  elliptical signals with application to an oceanic eddy, Nonlinear Proc.
  Geoph., 13, 467--483, 2006.

\bibitem[{Lilly and Olhede(2009)}]{lilly09-asilomar}
Lilly, J.~M. and Olhede, S.~C.: Wavelet ridge estimation of jointly modulated
  multivariate oscillations, in: 2009 Conference Record of the Forty-Third
  Asilomar Conference on Signals, Systems, and Computers, pp. 452--456, 2009.

\bibitem[{Lilly et~al.(2011)Lilly, Scott, and Olhede}]{lilly11-grl}
Lilly, J.~M., Scott, R.~K., and Olhede, S.~C.: Extracting waves and vortices
  from {L}agrangian trajectories, Geophys. Res. Lett., 38, 1--5, 2011.

\bibitem[{Lim and Eab(2006)}]{lim06-pla}
Lim, S.~C. and Eab, C.~H.: Riemann-{L}iouville and {W}eyl fractional oscillator
  processes, Phys. Lett. A, 355, 87--93, 2006.

\bibitem[{Lin(1972)}]{lin72-jas}
Lin, J.-T.: Relative dispersion in the enstrophy-cascading inertial range of
  homogeneous two-dimensional turbulence, J. Atmos. Sci., 29, 394--396, 1972.

\bibitem[{Lindgren et~al.(2011)Lindgren, Rue, and
  Lindstr{\"o}m}]{lindgren11-jrssb}
Lindgren, F., Rue, H., and Lindstr{\"o}m, J.: An explicit link between
  {G}aussian fields and {G}aussian {M}arkov random fields: the stochastic
  partial differential equation approach, J. Roy. Stat. Soc. B Met., 73,
  423--498, 2011.

\bibitem[{Lumpkin and Pazos(2007)}]{lumpkin07-lapcod}
Lumpkin, R. and Pazos, M.: Lagrangian Analysis and Prediction in Coastal and
  Ocean Processes, chap. Measuring surface currents with {S}urface {V}elocity
  {P}rogram drifters: the instrument, its data, and some recent results, pp.
  39--67, Cambridge University Press, 2007.

\bibitem[{Ma(2004)}]{ma04-jap}
Ma, C.: The use of the variogram in the construction of stationary time series
  models, J. Appl. Probab., 41, 1093--1103, 2004.

\bibitem[{Majda and Gershgorin(2013)}]{majda13-ptrsla}
Majda, A.~J. and Gershgorin, B.: Elementary models for turbulent diffusion with
  complex physical features: eddy diffusivity, spectrum and intermittency,
  Philos. T. Roy. Soc. A, 371, 20120\,184, 2013.

\bibitem[{Majda and Kramer(1999)}]{majda99-pr}
Majda, A.~J. and Kramer, P.~R.: Simplified models for turbulent diffusion:
  {T}heory, numerical modelling, and physical phenomena, Phys. Rep., 1999.

\bibitem[{Mandelbrot(1985)}]{mandelbrot85-ps}
Mandelbrot, B.~B.: Self-affinity and fractal dimension, Phys. Scripta, 32,
  257--260, 1985.

\bibitem[{Mandelbrot and {Van Ness}(1968)}]{mandelbrot68-siam}
Mandelbrot, B.~B. and {Van Ness}, J.~W.: Fractional {B}rownian motions,
  fractional noises and applications, SIAM Rev., 10, 422--437, 1968.

\bibitem[{Mandelbrot and Wallis(1969)}]{mandelbrot69-wrr}
Mandelbrot, B.~B. and Wallis, J.~R.: Computer experiments with fractional
  {G}aussian noises: {P}art 3, mathematical appendix, Water Resour. Res., 5,
  260--267, 1969.

\bibitem[{Mat{\'e}rn(1960)}]{matern60-mss}
Mat{\'e}rn, B.: Spatial variation: stochastic models and their applications to
  some problems in forest surveys and other sampling investigations,
  Meddelanden fr{\aa}n Statens Skogsforskningsinstitut, 49, 1--144, 1960.

\bibitem[{Matheron(1963)}]{matheron63-eg}
Matheron, G.: Principles of geostatistics, Econ. Geol., 58, 1246--1266, 1963.

\bibitem[{McWilliams(1990{\natexlab{a}})}]{mcwilliams90a-jfm}
McWilliams, J.~C.: The vortices of two-dimensional turbulence, J. Fluid Mech.,
  219, 361--385, 1990{\natexlab{a}}.

\bibitem[{McWilliams(1990{\natexlab{b}})}]{mcwilliams90b-jfm}
McWilliams, J.~C.: The vortices of geostrophic turbulence, J. Fluid Mech., 219,
  387--404, 1990{\natexlab{b}}.

\bibitem[{Metzner(2007)}]{metzner07-thesis}
Metzner, P.: Transition path theory for {M}arkov processes, Ph.D. thesis,
  Freien Universit\"at Berlin,
  \urlprefix\url{http://www.diss.fu-berlin.de/diss/servlets/MCRFileNodeServlet/FUDISS_derivate_000000003512/},
  2007.

\bibitem[{Molz et~al.(1997)Molz, Liu, and Szulga}]{molz97-wrr}
Molz, F.~J., Liu, H.~H., and Szulga, J.: Fractional {B}rownian motion and
  fractional {G}aussian noise in subsurface hydrology: A review, presentation
  of fundamental properties, and extensions, Water Resour. Res., 33,
  2273--2286, 1997.

\bibitem[{Monin(1958)}]{monin58-tpa}
Monin, A.~S.: The structure of atmospheric turbulence, Theor. Probab. Appl., 3,
  266--296, 1958.

\bibitem[{Monin and Yaglom(2007)}]{moninandyaglomII}
Monin, A.~S. and Yaglom, A.~M.: Statistical Fluid Mechanics, {V}olume {II}:
  {M}echanics of Turbulence, Dover Publications, Inc., 2007.

\bibitem[{Mooers(1973)}]{mooers73-dsr}
Mooers, C. N.~K.: A technique for the cross spectrum analysis of pairs of
  complex-valued time series, with emphasis on properties of polarized
  components and rotational invariants, Deep-Sea Res., 20, 1129--1141, 1973.

\bibitem[{Neeser and Massey(1993)}]{neeser93-itit}
Neeser, F.~D. and Massey, J.: Proper complex random processes with applications
  to information theory, IEEE T. Inform. Theory, 39, 1293--1302, 1993.

\bibitem[{{\O}ig{\aa}rd et~al.(2006){\O}ig{\aa}rd, Hanssen, and
  Scharf}]{oigard06-pre}
{\O}ig{\aa}rd, T.~A., Hanssen, A., and Scharf, L.~L.: Spectral correlations of
  fractional {B}rownian motion, Phys. Rev. E, 74, 1--6, 2006.

\bibitem[{Osborne et~al.(1989)Osborne, Jr., Provenzale, and
  Bergamasco}]{osborne89-tellus}
Osborne, A.~R., Jr., A.~K., Provenzale, A., and Bergamasco, L.: Fractal drifter
  trajectories in the {K}uroshio extension, Tellus, 41, 416--435, 1989.

\bibitem[{Park et~al.(1987)Park, {Vernon III}, and Lindberg}]{park87b-jgr}
Park, J., {Vernon III}, F.~L., and Lindberg, C.~R.: Frequency-dependent
  polarization analysis of high-frequency seismograms, J. Geophys. Res., 92,
  12,664--12,674, 1987.

\bibitem[{Pasquero et~al.(2002)Pasquero, Provenzale, and
  Weiss}]{pasquero02-prl}
Pasquero, C., Provenzale, A., and Weiss, J.~B.: Vortex statistics from
  {E}ulerian and {L}agrangian time series, Phys. Rev. Lett., 89, 284\,501,
  2002.

\bibitem[{Percival(2006)}]{percival06-sp}
Percival, D.~B.: Exact simulation of complex-valued {G}aussian stationary
  processes via circulant embedding, Signal Process., 86, 1470--1476, 2006.

\bibitem[{Percival and Walden(1993)}]{sapa}
Percival, D.~B. and Walden, A.~T.: Spectral Analysis for Physical Applications,
  Cambridge University Press, New York, 1993.

\bibitem[{Picinbono and Bondon(1997)}]{picinbono97a-itsp}
Picinbono, B. and Bondon, P.: Second-order statistics of complex-valued time
  series, IEEE T. Signal Proces., 45, 411--420, 1997.

\bibitem[{Pollard and {Millard, Jr.}(1970)}]{pollard70-dsr}
Pollard, R.~T. and {Millard, Jr.}, R.: Comparison between observed and
  simulated wind-generated inertial oscillations, Deep-Sea Res., 17, 813--821,
  1970.

\bibitem[{Qian(2003)}]{qian03-pwlrc}
Qian, H.: Processes with Long-Range Correlations, chap. Fractional {B}rownian
  motion and fractional {G}aussian noise, pp. 22--33, Springer, 2003.

\bibitem[{Rihaczek(1968)}]{rihaczek68-itit}
Rihaczek, A.~W.: Signal energy distribution in time and frequency, IEEE T.
  Inform. Theory, 14, 369--374, 1968.

\bibitem[{Rogers(1997)}]{rogers97-mf}
Rogers, L. C.~G.: Arbitrage with fractional {B}rownian motion, Math. Financ.,
  7, 95--105, 1997.

\bibitem[{Rossby(2007)}]{rossby07-lapcod}
Rossby, H.~T.: Lagrangian Analysis and Prediction in Coastal and Ocean
  Processes, chap. Evolution of {L}agrangian methods in oceanography, pp.
  1--38, Cambridge University Press, 2007.

\bibitem[{Rupolo et~al.(1996)Rupolo, Artalea, Huab, and
  Provenzale}]{rupolo96-jpo}
Rupolo, V., Artalea, V., Huab, B.~L., and Provenzale, A.: Lagrangian velocity
  spectra at 700 m in the western {N}orth {A}tlantic, J. Phys. Oceanogr., 26,
  1591--1607, 1996.

\bibitem[{Sanderson and Booth(1991)}]{sanderson91-tellus}
Sanderson, B.~G. and Booth, D.~A.: The fractal dimension of drifter
  trajectories and estimates of horizontal eddy-diffusivity, Tellus, 43,
  334--349, 1991.

\bibitem[{Sanderson et~al.(1990)Sanderson, Goulding, and
  Okubo}]{sanderson90-tellus}
Sanderson, B.~G., Goulding, A., and Okubo, A.: The fractal dimension of
  relative {L}agrangian motion, Tellus, 42, 550--556, 1990.

\bibitem[{Sawford(1999)}]{sawford99-blm}
Sawford, B.~L.: Rotation of trajectories in {L}agrangian stochastic models of
  turbulent dispersion, Bound.-Lay. Meteorol., 93, 411--424, 1999.

\bibitem[{Schlather(2012)}]{schlather12-stmne}
Schlather, M.: Advances and Challenges in Space-time Modelling of Natural
  Events, vol. 207 of {\em Lecture Notes in Statistics\/}, chap. Construction
  of covariance functions and unconditional simulation of random fields, pp.
  25--54, Springer Berlin Heidelberg, 2012.

\bibitem[{Schreier and Scharf(2003)}]{schreier03b-itsp}
Schreier, P.~J. and Scharf, L.~L.: Stochastic time-frequency analysis using the
  analytic signal: why the complementary distribution matters, IEEE T. Signal
  Proces., 51, 3071--3079, 2003.

\bibitem[{Scott and Dritschel(2013)}]{scott13-jfm}
Scott, R.~K. and Dritschel, D.~G.: Halting scale and energy equilibration in
  two-dimensional quasigeostrophic turbulence, J. Fluid Mech., 721, 1--12,
  2013.

\bibitem[{Slepian(1978)}]{slepian78-bell}
Slepian, D.: Prolate spheriodal wave functions, {F}ourier analysis, and
  uncertainty-- {V}: {T}he discrete case, Bell Syst. Tech. J., 57, 1371--1430,
  1978.

\bibitem[{Solo(1992)}]{solo92-sjam}
Solo, V.: Intrinsic random functions and the paradox of l/f noise, SIAM J.
  Appl. Math., 52, 270--291, 1992.

\bibitem[{Summers(2002)}]{summers02-npg}
Summers, D.~M.: Impulse exchange at the surface of the ocean and the fractal
  dimension of drifter trajectories, Nonlinear Proc. Geoph., 9, 11--23, 2002.

\bibitem[{Sykulski et~al.(2016{\natexlab{a}})Sykulski, Olhede, Lilly, and
  Danioux}]{sykulski16-jrssc}
Sykulski, A.~M., Olhede, S.~C., Lilly, J.~M., and Danioux, E.: Lagrangian time
  series models for ocean surface drifter trajectories, J. Roy. Stat. Soc. C
  App., 65, 29--50, 2016{\natexlab{a}}.

\bibitem[{Sykulski et~al.(2016{\natexlab{b}})Sykulski, Olhede, Lilly, and
  Early}]{sykulski16-arxiv}
Sykulski, A.~M., Olhede, S.~C., Lilly, J.~M., and Early, J.~J.: The {W}hittle
  likelihood for complex-valued time series, in revision; draft available at
  \url{http://arxiv.org/pdf/1605.06718}., 2016{\natexlab{b}}.

\bibitem[{Sykulski et~al.(2017)Sykulski, Olhede, Lilly, and
  Early}]{sykulski17-itsp}
Sykulski, A.~M., Olhede, S.~C., Lilly, J.~M., and Early, J.~J.:
  Frequency-domain stochastic modeling of stationary bivariate or
  complex-valued signals, IEEE T. Signal Proces., 65, 3136--3151, 2017.

\bibitem[{Taylor and Taylor(1991)}]{taylor91-jrssb}
Taylor, C.~C. and Taylor, S.~J.: Estimating the dimension of a fractal, J. Roy.
  Stat. Soc. B Met., pp. 353--364, 1991.

\bibitem[{Taylor(1921)}]{taylor21-plms}
Taylor, G.~I.: Diffusion by continuous movements, P. Lond. Math. Soc., 20,
  196--212, 1921.

\bibitem[{Thomson(1982)}]{thomson82-ieee}
Thomson, D.~J.: Spectrum estimation and harmonic analysis, Proc. IEEE, 70,
  1055--1096, 1982.

\bibitem[{Uhlenbeck and Ornstein(1930)}]{uhlenbeck30-pr}
Uhlenbeck, G.~E. and Ornstein, L.~S.: On the theory of the {B}rownian motion,
  Phys. Rep., 36, 823--841, 1930.

\bibitem[{Vallis(2006)}]{vallis}
Vallis, G.~K.: Atmospheric and Oceanic Fluid Dynamics: Fundamentals and
  Large-Scale Circulation, Cambridge University Press, 2006.

\bibitem[{Veneziani et~al.(2005{\natexlab{a}})Veneziani, Griffa, Garraffo, and
  Chassignet}]{veneziani05a-jmr}
Veneziani, M., Griffa, A., Garraffo, Z., and Chassignet, E.: Lagrangian spin
  parameter and coherent structures from trajectories released in a
  high-resolution ocean model, J. Mar. Res., 63, 753--788, 2005{\natexlab{a}}.

\bibitem[{Veneziani et~al.(2005{\natexlab{b}})Veneziani, Griffa, Reynolds,
  Garraffo, and Chassignet}]{veneziani05b-jmr}
Veneziani, M., Griffa, A., Reynolds, A.~M., Garraffo, Z.~D., and Chassignet,
  E.~P.: Parameterizations of {L}agrangian spin statistics and particle
  dispersion in the presence of coherent vortices, J. Mar. Res., 63,
  1057--1083, 2005{\natexlab{b}}.

\bibitem[{{Von Karman}(1948)}]{vonkarman48-pnas}
{Von Karman}, T.: Progress in the statistical theory of turbulence, P. Natl.
  Acad. Sci. USA, 34, 530--539, 1948.

\bibitem[{Watson(1922)}]{watson}
Watson, G.~N.: A Treatise on the Theory of {B}essel Functions, Cambridge Univ
  Press, 1922.

\bibitem[{Weiss et~al.(1998)Weiss, Provenzale, and McWilliams}]{weiss98-pf}
Weiss, J.~B., Provenzale, A., and McWilliams, J.~C.: Lagrangian dynamics in
  high-dimensional point-vortex systems, Phys. Fluids, 10, 1929--1941, 1998.

\bibitem[{Whittle(1953)}]{whittle53-afm}
Whittle, P.: Estimation and information in stationary time series, Ark. Mat.,
  2, 423--434, 1953.

\bibitem[{Wolpert and Taqqu(2005)}]{wolpert05-sp}
Wolpert, R.~L. and Taqqu, M.~S.: Fractional {O}rnstein-{U}hlenbeck {L}\'evy
  processes and the telecom process: upstairs and downstairs, Signal Process.,
  85, 1523--1545, 2005.

\bibitem[{Wong(1980)}]{wong80-siam}
Wong, R.: Error bounds for asymptotic expansions of integrals, SIAM Rev., 22,
  401--435, 1980.

\bibitem[{Yagle and Levy(1985)}]{yagle85-aam}
Yagle, A.~E. and Levy, B.~C.: The {S}chur algorithm and its applications, Acta
  Appl. Math., 3, 255--284, 1985.

\end{thebibliography}

\end{document}